\documentclass[trackchanges,twocolumn]{aastex701}

\submitjournal{ApJ}
\usepackage{amsmath}
\usepackage{comment}
\usepackage{cleveref}

\newcommand{\ohb}{{[O{\small III}]/H$\beta$}}

\newcommand{\otot}{[O{\small III}]/[O{\small II}]}

\shortauthors{Cooper et al.}

\begin{document}

\title{The Physical Conditions of Low-Mass Galaxies at $z=3.5-7.0$ from JWST Spectroscopy: The Behaviour of Spectral Line Ratios with Burstiness}

\shorttitle{The Physical Conditions of Low-Mass Galaxies at $z=3.5-7$ from JWST Spectroscopy}

\author[0009-0000-0413-5699]{Ryan A. Cooper}
\affiliation{Kapteyn Astronomical Institute, University of Groningen,
P.O. Box 800, 9700AV Groningen,
The Netherlands
}
\email[show]{cooper@astro.rug.nl}

\author[0000-0001-8183-1460]{Karina I. Caputi}
\affiliation{Kapteyn Astronomical Institute, University of Groningen,
P.O. Box 800, 9700AV Groningen,
The Netherlands
}
\email{karina@astro.rug.nl}

\author[0000-0002-9889-4238]{Alessandro Marconi}
\affiliation{Dipartimento di Fisica e Astronomia, Universit\`a degli Studi di
Firenze, Via G. Sansone 1, I-50019 Sesto Fiorentino, Firenze,
Italy
}
\affiliation{INAF – Osservatorio Astrofisico di Arcetri, Largo E. Fermi 5, I-
50125 Firenze, Italy}
\email{alessandro.marconi@unifi.it}

\author[0000-0001-6865-2871]{Anna Feltre}
\affiliation{INAF – Osservatorio Astrofisico di Arcetri, Largo E. Fermi 5, I-
50125 Firenze, Italy}
\email{anna.feltre@inaf.it}

\author[0000-0001-8386-3546]{Edoardo Iani}
\affiliation{Institute of Science and Technology Austria (ISTA), Am Campus 1, 3400 Klosterneuburg, Austria
}
\email{Edoardo.Iani@ist.ac.at}

\author[0000-0002-5104-8245]{Pierluigi Rinaldi}
\affiliation{Department of Astronomy, The University of Texas at Austin, Austin, TX 78712, USA
}
\email{prinaldi@stsci.edu}

\author[0000-0001-8325-1742]{Guillaume Desprez}
\affiliation{Kapteyn Astronomical Institute, University of Groningen,
P.O. Box 800, 9700AV Groningen,
The Netherlands
}
\email{desprez@astro.rug.nl}

\author[0000-0003-1561-3814]{Harley Katz}
\affiliation{Department of Astronomy \& Astrophysics, University of Chicago, 5640 S Ellis Avenue, Chicago, IL 60637, USA
}
\affiliation{Kavli Institute for Cosmological Physics, University of Chicago, Chicago IL 60637, USA}
\email{harley.b.katz@gmail.com}

\begin{abstract}

We present a study of the physical conditions of galaxies at redshifts $z=3.5-7.0$, based on emission-line diagnostics of JWST NIRSpec spectra for 1889 star-forming sources spanning more than 4~dex in stellar mass.  We find that the median H$\alpha$/H$\beta$ and [OIII]/H$\beta$ are  constant throughout these redshifts, indicating a slow evolution of the interstellar medium conditions from the Epoch of Reionization to Cosmic Noon. The stellar-mass evolution of these line ratios follow the expected trend for chemical enrichment and dust build-up, with \ohb~being maximum for galaxies with $\rm M_{\star} \approx 10^9 \, \rm M_\odot$, and the medians of H$\alpha$/H$\beta$ and [OII]/H$\beta$ monotonically increasing from $\rm{M}_{\rm \star} \approx 10^7$ through $ \approx 10^{10}\, \rm M_\odot$. Focusing on the 1552 low stellar-mass ($<10^9 \, \rm M_\odot$) galaxies, we particularly investigate the behaviour of the main spectral-line ratios with burstiness. We find a significantly higher degree of scatter in [OIII]/H$\beta$ and [OII]/H$\beta$ for high-burstiness than for low-burstiness galaxies. Values of log(\ohb)~$\gtrsim 0.8$ are almost exclusively seen amongst those with high burstiness. Meanwhile, log(\ohb)~$\lesssim 0.3$ are virtually not found amongst high-burstiness galaxies, except at the lowest stellar masses $\mathrm{log(M_\star/M_\odot)}<7.5$. Our results are consistent with a scenario in which star-formation bursts produce temporary chemical enrichment that is quickly lost via stellar winds. They also suggest that metallicity measures in bursting galaxies are unstable, and that the search for the most metal-poor objects should optimally be conducted amongst low-burstiness or very low-stellar mass sources.

\end{abstract}


\section{Introduction} 

Unveiling the physical conditions and chemical compositions of galaxies through cosmic time is crucial for a complete understanding of galaxy formation and evolution. These parameters are encoded into a galaxy's spectrum through discrete emission lines produced by the diversity of elements present across the galactic environment. Historically, huge efforts have been made to observe these lines and calibrate them to a galaxy's underlying properties, see e.g., \citet{BPT81,Brinchmann04,Feltre16};  and \citet{kewley_2019,MM19} for recent reviews.

Typically, the comparative ratios of emission line combinations are the most effective, and often most accessible, measures of a galaxy's physiochemical conditions. For example, the [OII]$\rm \lambda\lambda3726,3729$ and [SII]$\rm \lambda\lambda6716,6731$ doublets effectively probe gas density \citep{os_and_fer, kaasinen_2017, topping_2025}, whilst ratios of the same elements in different ionisation states (such as [OIII]$\rm \lambda 5007$/[OII]) trace the hardness of ionising radiation (from stellar or AGN sources) and the strength of the radiation field which is quantified by the ionisation parameter log(U) \citep{kaasinen_2018,hayes_2025,cleri_2026}. 

However, the most revealing constraints on a galaxy's interstellar medium (ISM) are the amount of dust and heavy elements present. Metals are synthesised in stars before being released, alongside dust, through mass-loss processes in post-main-sequence stars, such as thermally pulsating asymptotic-giant-branch stars and supernovae \citep{Dwek98}. As such, the dust and metal content of a galaxy is expected to be linked to its star formation history. Understanding the dust content of a galaxy is also crucial for modelling the reddening of both nebular and stellar emission to produce a galaxy's intrinsic spectrum. 

Constraints on nebular reddening can be readily obtained through the relative strength of Balmer lines which, under typical (case B) conditions, have physically defined ratios \citep{os_and_fer}. Metallicity meanwhile is less trivial to estimate precisely, with the most direct method relying on the ratio of the typically faint [OIII]$\lambda4363$ auroral line with that of [O{\small III}]$\lambda5007$ \citep[e.g.][]{moustakas_2011,andrews_2013,perez-montero_2017,laseter24,khostovan26}. The observational challenges of observing [OIII]$\lambda4363$ have lead to a huge collection of work calibrating gas-phase metallicities to other emission line diagnostics both in the local Universe \citep[e.g.,][]{pagel_1979,dopita_1986,Tremonti04,marino_2013} and towards Cosmic Noon \citep[e.g.,][]{Erb06,maiolino_2008,Maier14,curti_2020}.

JWST, and in particular its spectroscopic instruments, have been conducive to observing complete sets of UV and optical emission lines beyond Cosmic Noon. As a result, spectroscopic studies of metal abundances have been extended to higher redshifts \citep{cameron_2023,curti_2024,sarkar_2025}, including efforts to recalibrate line ratio based metallicity diagnostics \citep{sanders23,cataldi25, Scholte25, Isobe26}. Initial metallicity studies showed a clear offset from the mass-metallicity (MZR) relation found in the local Universe \citep{maiolino_2008,Mannucci10}, whilst more recent works have additionally described minimal evolution in this relation beyond Cosmic Noon \citep[e.g.,][]{nakajima_2023}.

Up to now, the bulk of these studies have focussed on high and intermediate-mass galaxies at $\mathrm{log(M_\star/M_\odot) \gtrsim 9.0}$ where fainter emission lines are inherently more readily available. However, it is crucial for this analysis to be extended to the low-mass ($\mathrm{log(M_\star/M_\odot) < 9.0}$) regime. These low-mass systems are the progenitors of the typical galaxies that we see at present times, so understanding their conditions is imperative for understanding galaxy evolution. 

At low redshift, low-mass galaxies are well studied and show significant diversity in their properties. For example, whilst many are metal poor \citep{berg_2019, romano_2023}, populations of metal rich dwarf galaxies have been observed \citep{peeples_2008}, and their overall location on the MZR is subject to significant scatter \citep{Mannucci2011,calabro_2017}. Such diversity is to some extent credited to the bursty star formation commonplace in these systems \citep[e.g][]{mcquinn_2010, bose_2019, emami_2019, ting_2025}. Violent star forming episodes drive strong outflows that can overcome the shallow potential wells of these systems, stripping them of dust \citep{romano_2024}, metals, and fuel for further star formation \citep{bland-hawthorn_2015, hirai_2024,outflows_1}. This effect is expected to be exacerbated at higher redshifts, where galaxies are frequently observed to have bursty or stochastic star formation \citep{Harshan24,Endsley25,mcclymont25,navarro-carrera_2026}. However, to date, no systematic study has investigated the effect of burstiness on the spectral line ratios and inferred metallicities. 

To this end, we endeavour to provide a large statistical study into the ISM conditions of low-mass galaxies across a significant portion of cosmic time, encompassing the end of Cosmic Reionisation and the following billion years until Cosmic Noon ($z=3.5-7$). We leverage a large collection of deep NIRSpec PRISM spectroscopy from a variety of public observations to analyse the full suite of optical emission lines now available at such redshifts. In particular, we incorporate spectroscopy from lensing clusters to dramatically increase the number of low-mass galaxies in our sample, allowing for statistically rigorous descriptions down to $\mathrm{log(M_\star/M_\odot)\sim7.0}$.

The sample analysed in this work contains 1889 sources at $3.5 \leq z \leq 7.5$, spanning 4 dex in stellar mass and $\sim$1~Gyr of cosmic time. We exploit this large dynamic range to describe the evolution and diversity of emission line ratios and the physical parameters they encode. These results are framed in context with higher-mass systems, as well as previous observations and results from the local Universe. We then isolate the low-mass galaxies in our sample and investigate how their star formation histories interact with observed line ratios, focussing in particular on their burstiness.

We describe the various datasets used throughout this work in Section \ref{sec_obs}. In Section \ref{sec_method} we outline the approaches for estimating line fluxes and physical parameters, as well as describe the photoionisation modelling used as part of this work. In Section \ref{sec_fullsample} we study the evolution of line ratios across the full range of masses and redshifts in our sample, whilst Section \ref{sec_lowmass} discusses the low-mass galaxies in our sample, with considerations made on their ISM conditions and star formation histories (SFH). We conclude with a summary and discussion of all the findings of this work in Section \ref{sec_summary}.

This work uses the cosmological parameters defined in \cite{planck_2016} with values H$_{0}=67.8~$km s$^{-1}$Mpc$^{-1}$, $\Omega_{m}=0.308$ and $\Omega_{\Lambda}=0.69$. Common line ratios used throughout this work are $\mathrm{R3=[O{\small III}]/H\beta}$, $\mathrm{R2=[O{\small II}]/H\beta}$, and $\mathrm{O32=[O{\small III}]/[O{\small II}]}$ and we assume a Chabrier IMF \citep{chabrier_2003}.

\section{Datasets}\label{sec_obs}

The sample used throughout this work is built from a collection of publicly available JWST NIRCam and NIRSpec observations, as well as substantial archival photometry from HST. Here we group observations by spectroscopic program and outline the sources of accompanying imaging.
\subsection{Spectroscopy and Line Fitting}

In this work, we incorporate publicly available low-resolution (R$\sim30-300$) JWST NIRSpec PRISM spectroscopy from several observing programs. We opt for PRISM spectroscopy to allow for an increased sample size and more consistent detections of galaxy continua necessary for this study. We will provide a brief discussion of the caveats associated with using low-resolution spectra, specifically concerning the blending of emission lines, as relevant for our results.

For self-consistency, we perform line fitting (see Section \ref{ssec_linefluxcal}) on all the spectroscopy used in this work with \texttt{msaexp} \citep{msaexp_zenodo}. We assign each object a redshift prior of $\pm0.1~z_{\mathrm{spec}}$ using either the spectroscopic redshift provided by the associated public data release or from the best-fit redshift provided by the Dawn JWST Archive (DJA). In all instances, we find redshifts and line fluxes in excellent agreement with those in public datasets.

\subsection{CANUCS}\label{ssec_canucs}

CANUCS is a Cycle 1 GTO program (P.I: C. Willott, PID: 1208) that provided NIRCam and NIRISS imaging, as well as NIRISS slitless spectroscopy and follow-up NIRSpec PRISM spectroscopy. For this study, we make use of the first CANUCS public data release (DR1) which covers 5 lensing cluster fields previously observed extensively by HST as part of the CLASH \citep{CLASH} and Frontier Fields \citep{F_Fields} programs. Spectra are retrieved from the DJA whilst photometry, spectroscopic redshifts and lensing magnifications are obtained from the DR1 release catalogues. Details for the photometric reduction can be found in \cite{canucs_phot}, whilst the spectroscopic reduction is described in \cite{canucs_spec}.

\subsection{UNCOVER}\label{ssec_uncover}

UNCOVER is a Cycle 1 Treasury program (P.I: I. Labbe, PID: 2561) that collected PRISM spectroscopy in the lensing cluster Abell 2744 whilst also providing NIRCam photometry in 7 NIRCam filters. Follow up medium band NIRCam photometry was obtained in the Cycle 2 program MegaScience (P.I: K. A. Suess, PID: 4111) and legacy HST photometry is again available from the CLASH and Frontier Fields surveys. 

Throughout this work, we use the photometry provided by the UNCOVER team as part of Data Release 3 which incorporates imaging from UNCOVER \citep{UNCOVER}, MegaScience \citep{MegaScience} and the previously mentioned HST programs. Spectroscopy is obtained from Data Release 4 \citep{UNCOVER_spec} and objects are demagnified following the lensing models and magnifications obtained by the UNCOVER team \citep{UNCOVER_magnif, UNCOVER_magnif_2}.

\subsection{CAPERS}\label{ssec_capers}

CAPERS is a Cycle 3 GO program (P.I: M. Dickinson, PID: 6368) providing NIRSpec observations across several CANDELS fields \citep{CANDELS, CANDELS_2}. In particular, CAPERS acquired spectra in the COSMOS \citep{COSMOS}, UDS \citep{UDS} and EGS \citep{EGS} legacy fields where NIRCam imaging has been acquired by the PRIMER (P.I: J. Dunlop, PID: 1837) and CEERS surveys (P.I: S. Finkelstein, PID: 1345, \citealt{CEERS}).

For this work we obtain both spectra and photometry from the DAWN JWST Archive. Spectroscopic data is reduced through the \texttt{msaexp} pipeline following the procedures outlined in \cite{DJ_Spec_1, DJ_Spec_2}. Photometry for both HST and JWST is reduced with the \texttt{grizli} \citep{grizli} pipeline with details of the reduction procedure found in \citep{DJ_Phot}.

\subsection{JADES}\label{ssec_jades}

JADES \citep{JADES_overview} is a Cycle 1 GTO program providing NIRCam and NIRSpec observations across the GOODS-N (P.Is: D. Eisenstein, N. Luetzgendorf, PIDs: 1180, 1210, 1286, 3215) and GOODS-S (P.I: D. Eisenstein, PID: 1181) legacy fields \citep{GOODS_legacy, GOODS_legacy_2}.

This work incorporates photometry from JADES Data Release 5 (DR5, \citealt{JADES_DR5_1, JADES_DR5_2}) which includes JWST imaging from the primary JADES observations as well as a collection of programs further targeting the GOODS fields (P.Is (PIDs): R. Windhorst (1176), G. Oestlin (1283), P. Oesch (1895), C. Williams (1963), S. Finkelstein (2079), C. Williams (2514), T. Morishita (3990), E. Egami (6434), G. Oestlin (6511)). Both GOODS fields are further covered by HST imaging as part of the Hubble Legacy Fields program \citep{GOODS_HST_1, GOODS_HST_2}, which has similarly been included in DR5 (see also \citealt{Hainline26}). 

PRISM spectroscopy is obtained from JADES Data Release 4, where details on the sample selection and observation strategy are detailed in \cite{JADES_DR4_1}. Information regarding the data reduction process can be found in \cite{JADES_DR4_2}.

\section{Methodology}\label{sec_method}

\subsection{Sample Selection}\label{ssec_sample}

For this study, we primarily require robust estimations of H$\alpha$, and [O{\small III}]$\lambda$5007 emission line fluxes. To allow for this, we select objects in the redshift range $3.5 < z < 7.0$, where H$\alpha$ is still detectable whilst [O{\small III}] and H$\beta$ are sufficiently de-blended in the PRISM spectra. We further impose a signal-to-noise ratio (S/N) of S/N$>3$ for both H$\alpha$ and [O{\small III}]. We note this step may exclude objects with prominent Balmer lines but little to no [O{\small III}] emission, in particular galaxies with the lowest metallicities and/or highest gas densities (e.g., \citealt{vanzella23,cliff,morishita26,PLRD, qso1}). These initial cuts produce a total sample of 2164 galaxies. We apply no cut based on the H$\beta$ S/N, but only about 15\% of the final sample have no significant H$\beta$ detection. These objects are not removed from our work as they potentially represent the dustiest sources in our sample (See Figure \ref{fig:o3hb_bd}), but their inclusion does not alter the conclusions of this work.

\begin{figure}
    \centering
    \includegraphics[width=1\linewidth]{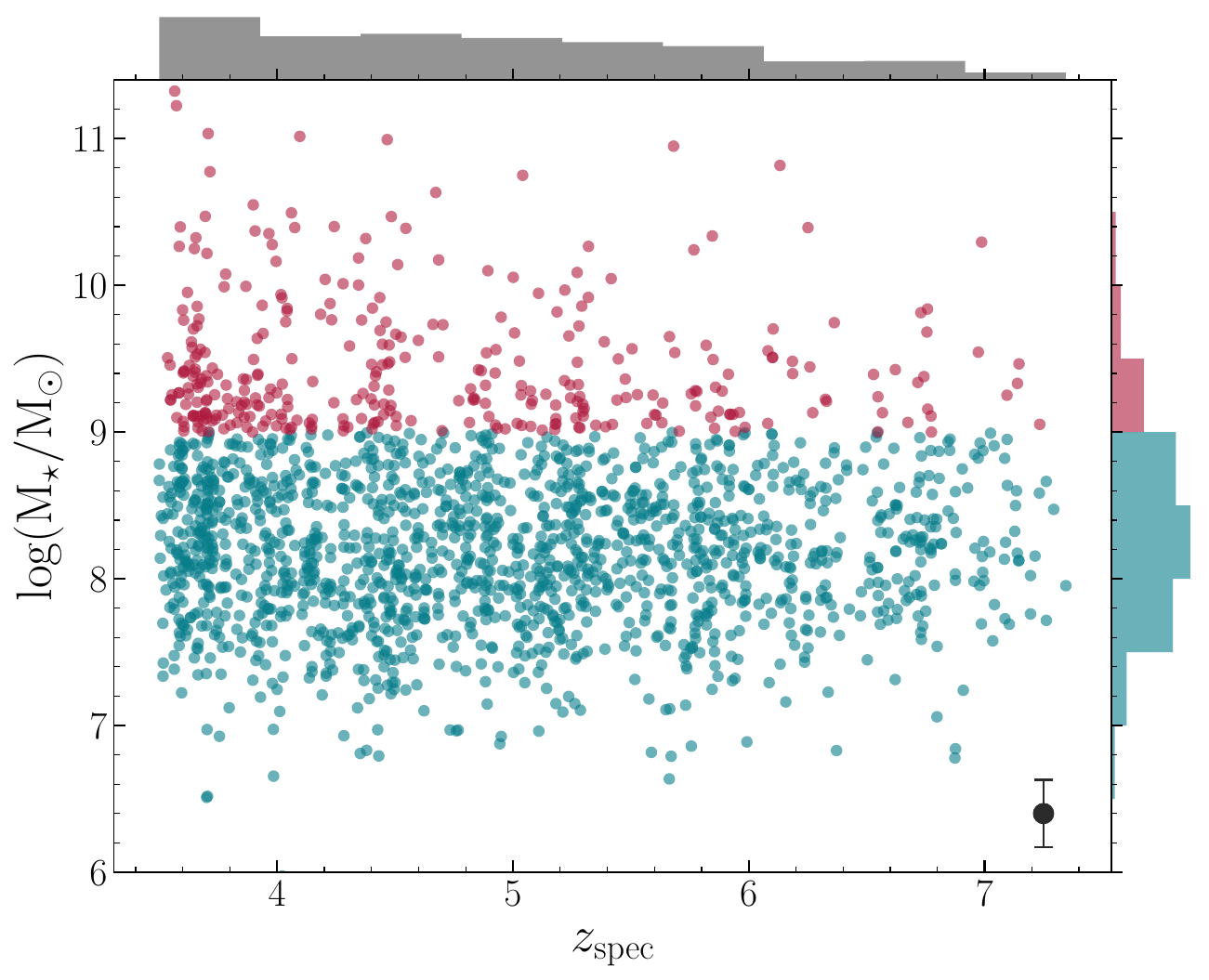}
    \caption{Stellar mass distribution with redshift for the sample considered in this work. Stellar masses are derived from the SED fitting procedure described in Section \ref{ssec_physparam}. The final sample produced by our selection procedure of 1889 galaxies is shown, with the sample of 1552 galaxies with $\mathrm{M_\star \leq 10^9M_\odot}$ highlighted in blue.}
    \label{fig:mass_dist}
\end{figure}

Before proceeding with the analysis of any sample, it is important to consider potential sources of error in line flux estimations from the PRISM spectra. This is especially crucial for objects with redshifts $3.5 \leq z < 4$. At this redshift, the [O{\small III}] doublet and H$\beta$ are typically resolved, with each line (or doublet) showing a distinct peak. However, the presence of a Type-1 AGN and outflows or high dispersion in the ionised-gas kinematics can broaden H$\beta$ and [O{\small III}] respectively (e.g., \citealt{outflows_2, outflows_1}), resulting in incorrect estimations for these line fluxes. As such, we perform a visual inspection to ensure that $H\beta$ shows a distinct peak in the spectrum. This additional cut produces a final sample of 1889 sources, with 1552 galaxies having $\mathrm{M_\star \leq 10^9M_\odot}$ (see Section \ref{ssec_physparam}). We present the full stellar-mass distribution of our galaxy sample versus redshift in Figure \ref{fig:mass_dist}. 

\subsection{Line Flux Calculation}\label{ssec_linefluxcal}

As mentioned previously, rather than using line fluxes provided by the various public data releases, the spectra in our sample are modelled with \texttt{msaexp} to ensure a consistent procedure across the full dataset. For each object, the continuum is first estimated using a combination of splines whilst line fluxes are estimated through pixel-integrated Gaussians. Line fluxes are then corrected for dust attenuation with $\rm E(B-V)$ values calculated from the Balmer decrement (see Section \ref{ssec_dust}) assuming the standard case B recombination value of 2.86 that arises from typical density ($\mathrm{n_e}=100\mathrm{~cm^{-3}}$) and temperature (T$_e$$=10^4$K) conditions \citep{os_and_fer}. Throughout this work we employ the dust extinction curve from \cite{calzetti_2000}. Through repeated testing (See Appendix \ref{ap:diff_dust}), we find that even the most dust sensitive quantities used throughout this work remain consistent (within errors) regardless of the choice of extinction curve up to stellar masses $\mathrm{log(M_\star/M_\odot) \sim 9.5}$.

Due to the low resolution of PRISM spectra, sets of nearby emission lines are typically blended into what appears as a single peak. Of significance to this work are the [O{\small III}]$\lambda$4959,$\lambda$5007 doublet and H$\alpha+$[N{\small II}]. In these cases, we correct by factors of [O{\small III}]$\lambda$5007/[O{\small III}]$\lambda$4959=2.98 \citep{oiii_ratio} and H$\alpha$/(H$\alpha$+[N{\small II}])=0.96 \citep{sandles_bd}.

\subsection{Photoionisation Modelling}\label{ssec_photmod}
To better constrain the ISM conditions of our sample, we run a grid of photoionisation models using version C25 of \texttt{CLOUDY} \citep{cloudy_2025}. We choose a stellar ionising source and employ stellar population synthesis models from \texttt{BPASS v2.3} \citep{BPASS_1, BPASS_2, BPASS_3} with a Kroupa IMF \citep{kroupa_2001},  an upper mass limit of 300~$\mathrm{M_\odot}$, a single burst in star formation, and stellar binaries. We assume a spherical cloud geometry with constant density as well as solar chemical abundance \citep{solar_abundance}, and stop all calculations when the electron density ($n_e$) reaches 1/100 of the original value. The complete parameter space probed by our models is presented in Table \ref{tab:cigale_params}.

To ensure that we span the full space of physical parameters that our galaxies can exhibit, stellar ages are allowed to increase up to 1 Gyr. However, across most of this work this age is not used. Our SED fitting (see Section \ref{ssec_physparam}) finds that $\sim98\%$ of our galaxies have light-weighted ages $< 300~$Myr and no galaxies with ages $\geq 1$~Gyr, with the oldest galaxy in our sample having a predicted age of $\sim800~$Myr.

For each model, we output the relative intensity of a set of relevant emission lines as well as the final transmitted spectrum. We then attenuate the emitted spectrum in steps of $\rm E(B-V)$=0.1 from 0 to 1 using the \cite{calzetti_2000} dust-attenuation curve. This approach allows us to predict line ratios over a wide combination of gas and dust conditions.

\begin{table}
\centering
\begin{tabular}{l|r}
\tablewidth{0pt}
\hline
\hline
\\
\textbf{Parameter} &  \textbf{Parameter Space}\\
\hline
Stellar Age [log(yr)] & [6, 6.5, 7, 7.5, 8, 8.5, 9] \\
Z$_\star$ [log(Z)] & [-1.7, -2.4, -2.7, -3.7] \\
log(U) & [-4, -3, -2, -1]\\ 
log($n_H$) & [2, 3, 4, 5, 5.5, 6]\\
Z$_{\mathrm{gas}}$ [log(Z/Z$_\odot$)] & [-3, -2.5, -2, -1.5, -1, -0.5, 0] \\
\hline
\end{tabular}
\caption{Parameters for the stellar and gas conditions of the \texttt{CLOUDY} models used throughout this work.} 
\label{tab:cigale_params}
\end{table}

\subsection{Physical Parameter Estimation}\label{ssec_physparam}

Various galaxy physical parameters are estimated from the SED fitting code \texttt{CIGALE} \citep{cigale_2019}. We make use of the stochastic star formation history defined in \cite{cigale_stoch} with a moderate `burstiness' parameter of $\sigma$=0.3, as well as a broad range of galaxy ages and star formation timescales. Models are built with stellar templates from \cite{BC03} with a set grid of stellar metallicities, whilst nebular emission templates are generated with the photoionisation code \texttt{CLOUDY 13.01} \citep{cloudy_13}. Dust attenuation follows the \cite{calzetti_2000} attenuation law, and we permit a linear grid of $\rm E(B-V)$ values ranging from 0-1.

Star formation rates (SFR) are estimated for all sources using both UV and Balmer line tracers measured from photometry and spectroscopy respectively. Recent works have highlighted the importance of metallicity specific conversion factors when considering the SFR of high redshift galaxies (see e.g. \citealt{dicesare_2026, kramarenko_2026}). As such we follow the relations defined in \cite{theios_2019}:

\begin{equation}
    \log(\mathrm{SFR_{H\alpha/UV}})=\log(\mathrm{L_{H\alpha/UV}})-C,
\end{equation}

\noindent where $\mathrm{L_{UV}}$ is evaluated at 1500\AA~and C takes the value of 41.59 and 43.46 for the H$\alpha$ and UV conversions respectively. These values correspond to an initial mass function with a mass cut-off at 100$\rm M_\odot$ and metallicity $1/5~\rm Z_\odot$. Whilst lower metallicity calibrations are available \citep[e.g][]{reddy_2022}, we find the difference in SFR estimate remains consistent within errors irrespective of the calibration chosen. UV luminosities are estimated through a median of the appropriate photometric bands and are corrected for dust attenuation with the $\rm E(B-V)$ values obtained from SED fitting.

\section{Spectral Properties of the Entire Sample}\label{sec_fullsample}

\subsection{AGN Identification}\label{ssec_agn}

The presence of a central black hole as a galaxy's primary source of ionising radiation introduces additional complexity to the interpretation of galaxy physical properties and line ratios \citep[][]{Marconi24,Moreschini26}. It is therefore crucial to identify any AGN present in our data and observe their line ratios in comparison to star forming galaxies. We inspect our sample for both Type-1 and Type-2 AGN candidates and remove both sub-sets. 

To identify Type-1 AGN, we make use of the line fitting code \texttt{unite} \citep{hviding_unite} to fit the H$\alpha$ line for every object in our sample and determine in which cases a broad component is significant. For this, we first fit each line with a single narrow-line Gaussian  component defined with a $\sigma$ prior of $100-800~\mathrm{km~s^{-1}}$. This upper bound is chosen to account for typical dynamical effects whilst also factoring in a varying line spread function (LSF, as adopted by \texttt{unite} from \citealt{degraaff_2024}).

\begin{figure}
    \centering
    \includegraphics[width=1.0\linewidth]{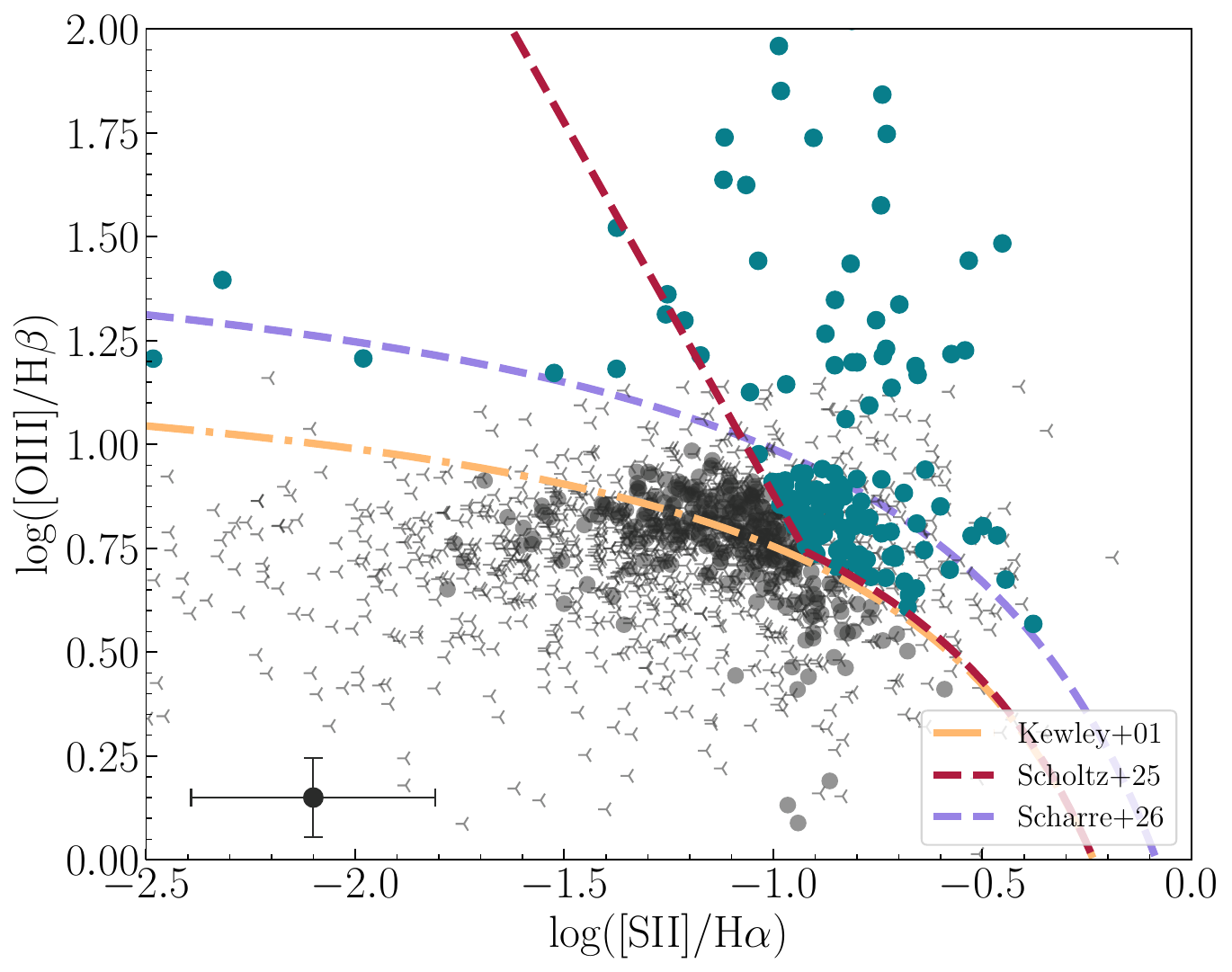}
    \caption{The S2-VO87 \citep{VO87} diagram for all sources in our sample. Selected Type-2 AGN candidates are presented as blue circles. Grey circles show objects with $3\sigma$ [S{\small II}] detections whilst grey arrows are upper limits. Over-plotted are the demarcations defined by \cite{kewley_2001} (orange), \cite{scholtz_2025} (red) and \cite{scharre_2026} (purple).}
    \label{fig:vo87}
\end{figure}

Following the single, narrow Gaussian fit, we repeat the fitting with a two component Gaussian model. As before, we define a narrow component with matching $\sigma$ priors to the individual fits and introduce a broad component with $\sigma=600-5000~\mathrm{km~s^{-1}}$. We then compare the Widely Applicable Information Criterion (WAIC) from \texttt{unite} for each model and use the criteria defined in \cite{hviding_2025} of $\mathrm{WAIC_{narrow}-WAIC_{broad}} \geq 11.8$ and FWHM$\mathrm{_{broad}~ > 1000~km~s^{-1}}$ to identify Type-1 AGN candidates. From this process we identify 124 potential broad-line AGN.
 
Whilst broad-line AGN have clear signatures readily available in their spectra, narrow-line AGN are considerably more difficult to differentiate without access to specific faint emission lines (e.g. \citealt{scholtz_2025, mazzolari_2025}). This problem is exacerbated at high redshift, where the ISM can host denser gas, and younger, metal poor stellar populations which can produce harder ionising spectra (e.g. \citealt{steidel_2014, hirschmann_2023, cameron_2023, scharre_2026}), resulting in typical AGN diagnostics becoming unreliable. However, efforts have been made to update these diagnostics by accounting for the change in ISM conditions between the local Universe and early times \citep{scholtz_2025, scharre_2026}.

As this work relies on PRISM spectra, where [N{\small II}] and H$\alpha$ remain blended at all redshifts, the most accessible AGN diagnostic is the S2-VO87 \citep{VO87} diagram which we reproduce in Figure \ref{fig:vo87}. Considering the caveats described above, we select Type-2 AGN candidates based primarily on the demarcation defined in \cite{scholtz_2025}, but also impose an R3 selection of $\mathrm{R3 > 1.16}$ motivated in part by the findings of \cite{scharre_2026} (although many AGN candidates lie below their R3 limit). Objects are included in the AGN selection only if they have a $3\sigma$ [S{\small II}] detection, unless they lie above the R3 cut where they are included regardless. In total, this produces a selection of 124 (158) Type-1 (2) AGN candidates, with 57 (59) of these having stellar mass $\mathrm{\log(M_\star/M_\odot)} < 9.0$. These sources are excluded from all results presented in this work (unless explicitly stated), but will be subject to a more detailed future study.

\subsection{Dust attenuation}\label{ssec_dust}

\begin{figure}
    \centering
    \includegraphics[width=1.0\linewidth]{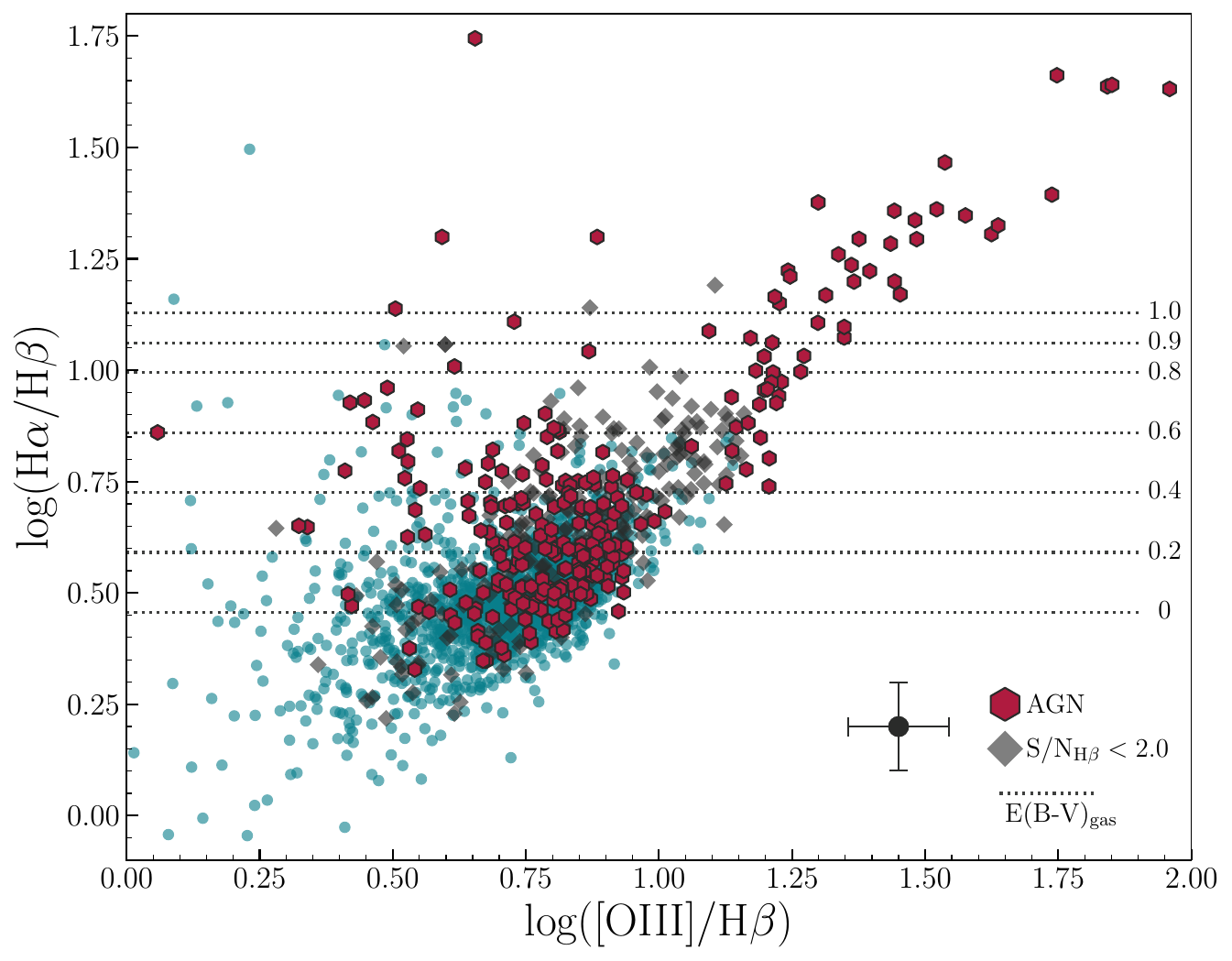}
    \caption{log(R3) against log($\mathrm{H\alpha/H\beta}$) for all objects in our sample. All AGN candidates selected in Section \ref{ssec_agn} are highlighted as red hexagons. Objects without a robust $> 2\sigma~\rm{H\beta}$ detection are shown as grey diamonds. Grey dashed lines represent the theoretical Balmer decrement at various $\rm E(B-V)$ steps, assuming the \cite{calzetti_2000} dust reddening law. $\rm E(B-V)$ steps are based on the case B value for star forming galaxies with $\rm E(B-V)$=0 corresponding to $\mathrm{log(H\alpha/H\beta)\simeq0.46}$}
    \label{fig:o3hb_bd}
\end{figure}

Before considering other line ratios to discern the ISM conditions of our sample, it is important to constrain the effect of dust on each source. Figure \ref{fig:o3hb_bd} presents the relation between R3 and the Balmer decrement H$\alpha$/H$\beta$. Overplotted are the Balmer decrement at various steps of $\rm E(B-V)$ derived from the \cite{calzetti_2000} dust extinction law. Across the complete sample, we find a distribution of $[68\%,~17\%,~6\%,~2\%,~1\%,<1\%]$ for bins of $\pm 0.1~\rm E(B-V)$ around mid points of $\rm E(B-V) = [0.1, 0.3, 0.5, 0.7, 0.9]$. The remaining 5\% have Balmer decrements associated with $\rm E(B-V) > 1.0$.

A subset of galaxies in the lowest $\rm E(B-V)$ bin have Balmer decrements below the case B value. In many cases this is simply due to the scattering introduced by the line flux errors, but in other cases the departure from case B conditions is real, as has been discussed in a variety of studies \citep[e.g.,][]{sandles_bd, Scarlata24}. Studying the conditions under which this can happen is beyond the scope of this work, and for our purposes we simply assume these galaxies are dust free. In total, $\sim$10\% of our sample have Balmer decrements at least 3$\sigma$ below the typical case B value, with a further $\sim$20\% being below but within 3$\sigma$ of the case B value.

Figure \ref{fig:o3hb_bd} plainly shows a positive correlation between Balmer decrement and R3. This is largely expected, owing to the presence of H$\beta$ in both ratios, but also physically motivated by the simultaneous build-up of dust and metals in the ISM \citep[e.g.,][]{Inoue03,Li19}. Indeed, we find that almost all galaxies with an \ohb $\geq 8$ ($\mathrm{log(R3) \geq 0.9}$) require a modest amount of dust, with the largest ratios requiring significant levels of reddening to explain their Balmer decrement. However, this region of parameter space with high H$\alpha$/H$\beta$ and R3 ratios is dominated by sources classified as AGN (See Section \ref{ssec_agn}) where metallicity and dust measurements are less reliable.

\subsection{Redshift Evolution of Main Line Ratios}\label{ssec_line_evo}

\begin{figure*}[t]
    \centering
    \includegraphics[width=1\linewidth]{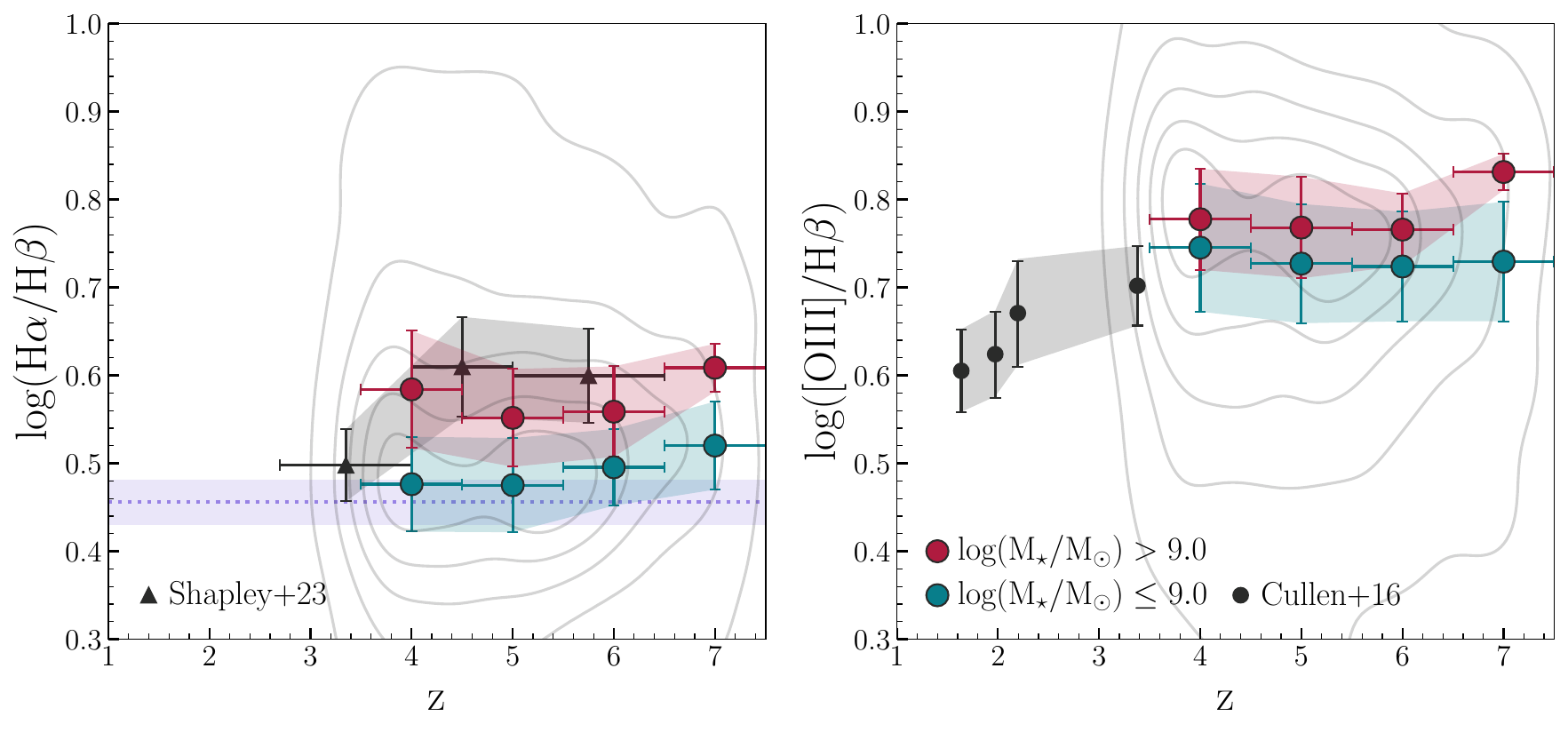}
    \caption{The evolution of Balmer decrement (H$\alpha$/H$\beta$, left) and R3 (right) with redshift for low- ($\mathrm{log(M_\star/M_\odot) \leq 9.0}$, blue) and high-mass ($\mathrm{log(M_\star/M_\odot) > 9.0}$, red) galaxies in our sample.  The total (non-binned) line ratio distributions are plotted as grey contours and the case B Balmer decrement value for a range of physical parameters is shaded purple. The dashed line represents a Balmer decrement of 2.86. Ratios obtained from \cite{cullen_2016, shapely_2023} are shown as grey circles and triangles respectively. These medians are calculated by merging the results across the parameters studied in these works.}
    \label{fig:lr_z_evol}
\end{figure*}

As the most accessible emission lines in our sample, we first consider the redshift evolution of the Balmer decrement (H$\alpha$/H$\beta$) and R3. These line ratios probe different ISM conditions, with the Balmer decrement mainly tracing dust content (see Section \ref{ssec_linefluxcal}) and R3 primarily reflecting gas-phase metallicity, with also some dependence on ionisation parameter and gas density. The evolution of these line ratios with redshift and various physical parameters can therefore provide insight into the typical ISM conditions of galaxies across cosmic time.

In Figure \ref{fig:lr_z_evol}, we present the median line ratios for all sources in redshift bins of width $z\pm0.5$ centred around $z=[4,5,6,7]$ with a total of [766, 560, 373, 156] galaxies in each bin. We show two medians for each line ratio separated between low-mass ($\mathrm{log(M_\star/M_\odot) \leq 9.0}$) and higher mass ($\mathrm{log(M_\star/M_\odot) > 9.0}$) sources. At first glance,  we see flat trends in both cases. We find Pearson rank coefficients of $\rm r_{BD}=0.024,~r_{R3}=-0.030$ and p-values of $\rm p_{BD}=0.37,~p_{R3}=0.27$  for the low-mass sample; and $\rm r_{BD}=0.02,~r_{R3}=0.078$ and $\rm p_{BD}=0.24,~p_{R3}=0.74$ for the high-mass sample. These values imply no correlation between both Balmer decrement and R3 with redshift, suggesting that ISM conditions are slow to evolve from the Epoch of Reionization towards Cosmic Noon. Meanwhile, we find an increase in both Balmer decrement and R3 between the low- and high-mass samples at all redshifts. This mass dependence in part explains the broad distribution of both Balmer decrements and R3 values and is explored in significant detail throughout the rest of this work.

The lack of evolution in Balmer decrement also suggests minimal dust build up in nebular regions across the spanned redshift range, although spectroscopic samples are naturally biased against the most heavily obscured sources, so the latter, if they exist, would probably be under-represented in our sample. At the same time, we note that our median points all lie above the case B value, implying at least some dust presence, even at $z\sim7$. Such early dust build-up has previously been observed with JWST \citep[e.g][]{witstok_2023} and ALMA \citep{hanae_2022}. For the high-mass sample, our median Balmer decrement values are broadly consistent with the results of \cite{shapely_2023}, who also report a lack of evolution beyond $z > 4$. Similar results have been reported by \cite{sandles_bd} and \cite{woodrum_2025}. Moreover, no R3 evolution implies minimal growth in the typical metallicity of the ISM, consistent with a weakly evolving mass-metallicity relation (MZR) at high-$z$ as seen in recent works \citep[e.g.,][]{nakajima_2023, curti_2024, marszewski_2025}.

We do however see a significant rise in R3 compared to \cite{cullen_2016}, suggesting that the galaxy median R3 value has peaked at $z\sim4$. Whilst typical relations between R3 and metallicity can provide bivariate metallicity solutions, we can safely assume that on average the R3 rise from $z=0$ to 4 is due to a combination of decreasing gas-phase metallicities and increasing ionisation parameter (e.g. \citealt{trump_2023, curti_2024}). This evolution flattens out at higher redshifts. The median values of our R3 measurements are log(R3)$=[0.78,0.74,0.73,0.74]$ for bins of $\pm0.5$ around mid points $z=[4,5,6,7]$. Such values are close to the peak of typical R3 metallicity diagnostics \citep[e.g., ][]{sanders23, cataldi25} which at those values indicate a moderately sub-solar metallicity. Meanwhile the lower redshift results from \cite{cullen_2016} indicate a rising  metallicity to $z=0$, as extremely metal-poor solutions for R3 can in general be disregarded at low redshifts.  We note that the stellar mass range of \cite{cullen_2016} coincides with the high-mass galaxies in our sample, but an evolution from $z\sim1.5$ to $z\sim4$ is still present.

\subsection{Main Line Ratios versus Stellar Mass}\label{ssec_line_evo_mass}

\begin{figure}
    \centering
    \includegraphics[width=0.9\linewidth]{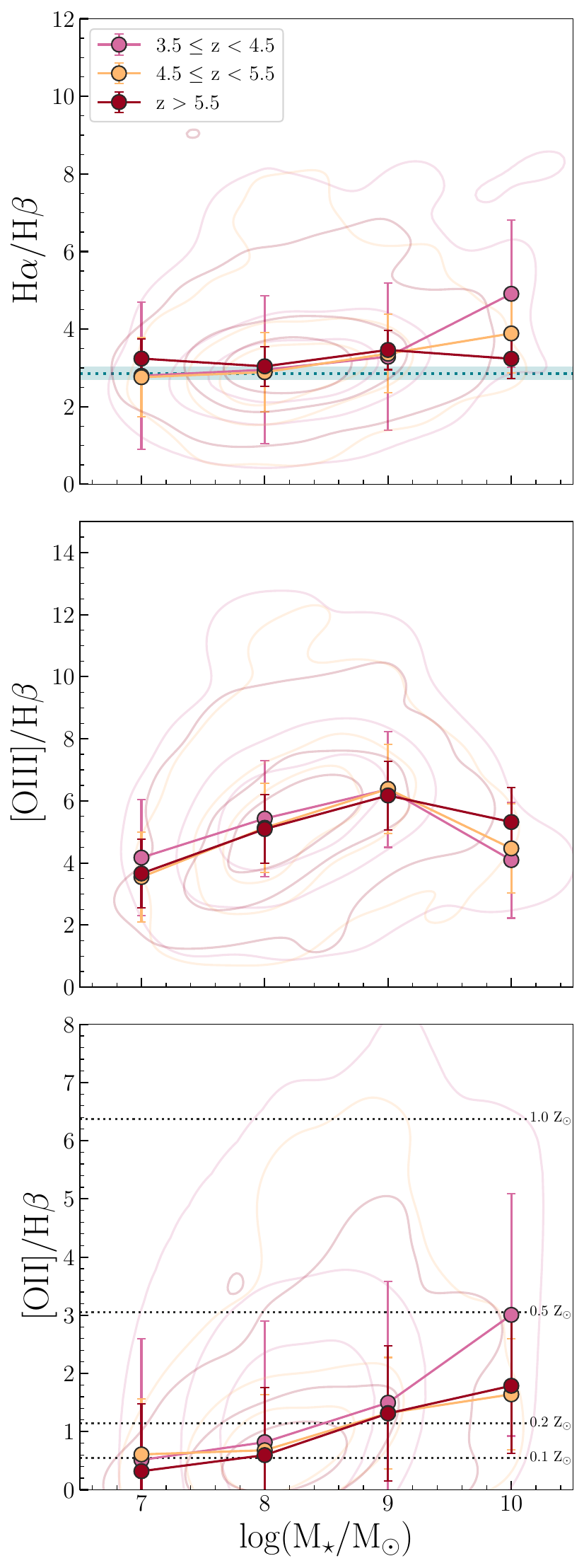}
    \caption{The evolution of Balmer decrement (Upper), R3 (Middle), and dust-attenuation corrected R2 (Lower) with galaxy stellar mass. Objects are split into three redshift bins and four stellar mass bins, with marked points determined as the median of these bins. The redshift binned distribution for each line ratio is shown with contours of matching colours to the median values. Metallicity estimations for R2 are calculated from the linear relation derived in \cite{Sanders_2024}.
    }
    \label{fig:lr_mass_evol}
\end{figure}

We also wish to consider the evolution of these line ratios as galaxies experience stellar mass growth. Figure \ref{fig:lr_mass_evol} shows the distribution and evolution of both Balmer decrement, R3, and R2 with stellar mass. R2 is introduced here to break any multiplicity in metallicity that can be derived from \ohb. We see minimal evolution in Balmer decrement from $\mathrm{M}_{\star}\sim10^7-10^9~\mathrm{M_\odot}$, but a notable rise at $\mathrm{M}_{\star}\sim10^{10}~\mathrm{M_\odot}$ in the two lower redshift bins, consistent with a rising dust content in more massive galaxies observed towards cosmic noon \citep[e.g.][]{shapely_2022, maheson_2024}.

For R3, we observe a steadily rising ratio from $\mathrm{M}_{\star}\sim10^7-10^9~\mathrm{M_\odot}$, with a decline at the highest masses. This is indicative of rising gas-phase metallicity with stellar mass, with the decline at the highest masses a result of the bimodal relationship between R3 and metallicity. A priori, we expect that these most massive sources lie on the high metallicity branch of this relationship, whilst sources at similar R3 but lower stellar mass are mostly expected to exist on the lower metallicity branch. This is verified by a similarly increasing R2, which sees a drastic rise from $\mathrm{M}_{\star}\sim10^9-10^{10}~\mathrm{M_\odot}$. Whilst the vast majority of the lowest mass galaxies have low R2 values indicating that they are metal poor, the average galaxy with $\mathrm{M}_{\star}\sim10^{10}~\mathrm{M_\odot}$ has already enriched to about a half the solar value by redshift $z\approx 4.5$. An important caveat to this is the effect of ionisation parameter, which can similarly raise R3. We consider this scenario in detail in Section \ref{ssec_ionised_gas}. 

We note that the behaviour we observe for R3 with stellar mass is at odds with the conclusions obtained by \citet{Backhaus24}, even if the measured R3 values are broadly consistent. However, the negative correlation they describe factors in a sample of high-mass galaxies from $z\sim0-1.5$. Combining their results with the results of this work suggests a turnover in the evolution of R3 with respect to stellar mass beyond $10^9~\mathrm{M_\odot}$.

It is also worth noting that, whilst the average evolution of R3 is consistent across redshift ranges, the distribution of R3 shows a significant population of sources with extreme line ratios present at $z < 5.5$. Such extreme line ratios have been observed at lower redshifts \citep[e.g.,][]{amorin15, Yang17}, with their origin a source of considerable work \citep[See e.g.][and sources therein]{scharre_2026}. In the local Universe, high R3 values are associated with AGN, but lower metallicities, younger stellar populations and harder ionisation fields can produce similarly extreme values \citep[e.g.,][]{Feltre16,Gutkin16,Yang17,Plat19,Lecroq24}. More recently, results utilising a combination of \texttt{CLOUDY} models have similarly been shown to produce these ratios \citep{Moreschini26}. Even if such hard-ionisation fields are also expected to be present in galaxies at $z>5.5$, their metallicities are still too low for R3 to reach a maximum value.

From Figure \ref{fig:lr_mass_evol} we also see that, for low-mass galaxies, the mean Balmer decrement H$\alpha$/H$\beta$ is consistent with the case B value at all redshifts, indicating basically no dust attenuation.  The mean H$\alpha$/H$\beta$ ratios only have a significant departure from case B at the highest masses, and increasingly towards lower redshifts. Nonetheless, across the full distribution of sources we see a fraction with substantial Balmer decrements indicative of a modest to large dust content. This includes $\sim25\%$ of low-mass sources with a Balmer decrement corresponding to an E(B$-$V) value $\geq 0.2$. 

The percentage of sources with E(B$-$V) $\geq 0.2$ in the lower stellar-mass bins of $\mathrm{\log(M_\star/M_\odot)} \leq 7.0$ and $7.0 <\mathrm{\log(M_\star/M_\odot)} \leq 8.0$ is 22\% and 19\% respectively, with the fraction rising for sources with $8.0 < \mathrm{\log(M_\star/M_\odot)} \leq 9.0$ to 28\%. This is consistent with previous works that found a substantial level of dust attenuation in a significant fraction of massive and intermediate-mass galaxies at $z=3-5$ \citep[e.g.,][]{Caputi2015, Deshmukh2018}. For some low-mass galaxies, in particular, a high Balmer decrement can be produced by high gas densities in place or in addition to nebular dust attenuation \citep[e.g.,][]{PLRD}.  However, the investigation of these differences is beyond the scope of this work, so in the following we will assume that all Balmer decrements above the case B value are exclusively produced by dust attenuation, which is the dominant factor for this effect.

Whether low-mass galaxies are capable of retaining their dust and metals in the presence of active star formation is uncertain, with stellar feedback potentially clearing new star forming regions over time. We further discuss the evolution of metal content with star-formation history in Section \ref{sec:burst}, while we will investigate the evolution of dust content in a future work.

\subsection{General Evolution of the Gas and Ionisation Conditions Inferred from Main Line Ratios}\label{ssec_ionised_gas}

\begin{figure*}
    \centering
    \includegraphics[width=0.8\linewidth]{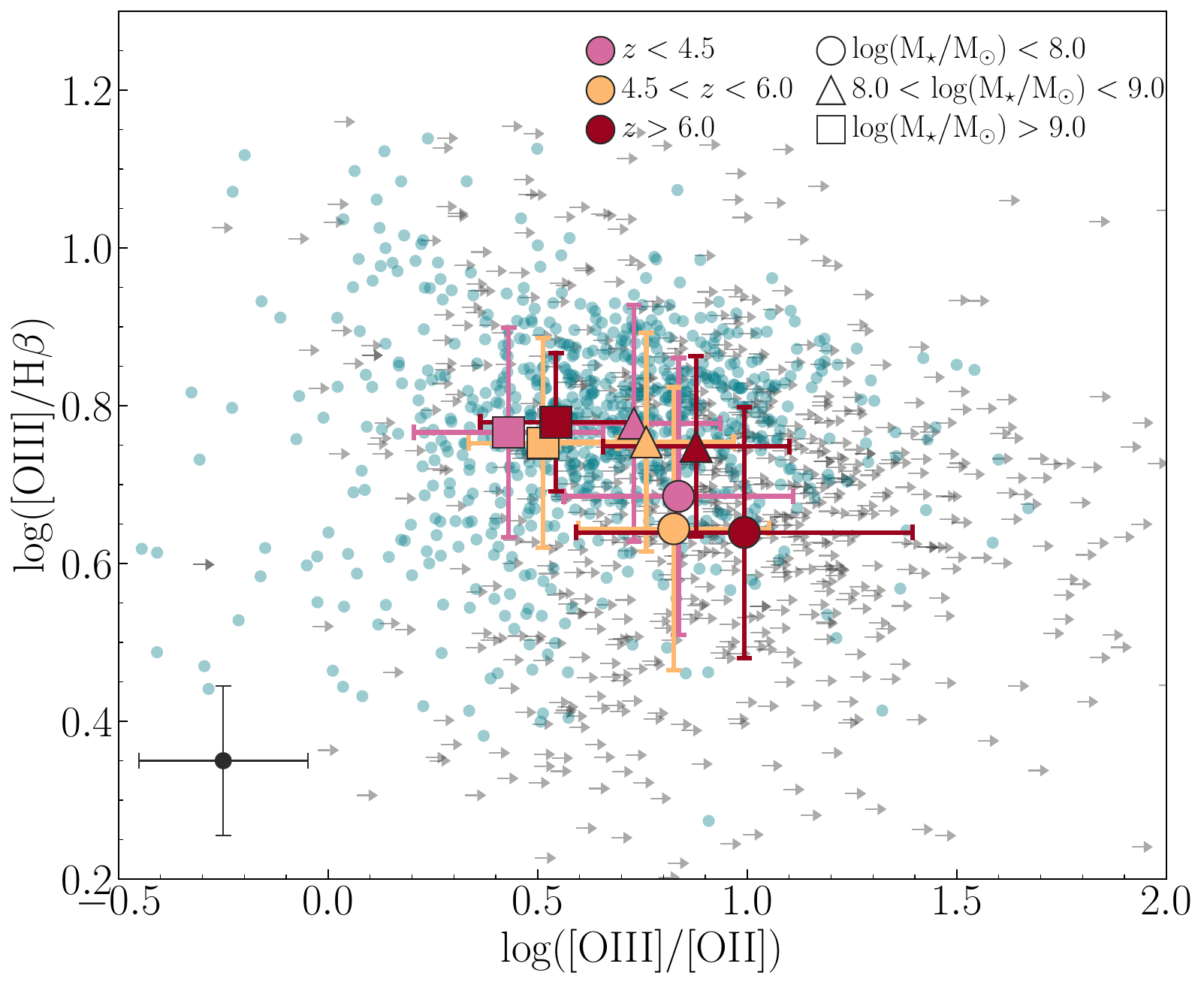}
    \caption{Dust corrected \otot~against \ohb~for all galaxies within our sample. Objects with a minimum 2$\sigma$ [O{\small II}] detection are marked as blue points, whilst lower limits in \otot~due to a non detection of [O{\small II}] are plotted as grey arrows. Median points are calculated in redshift bins of $z < 4.5$ (pink), $4.5 < z < 6.0$ (yellow), $z > 6.0$ (red) and mass bins of $\mathrm{\log(M_\star/M_\odot) < 8.0}$ (circles), $8.0 < \mathrm{\log(M_\star/M_\odot) < 9.0}$ (triangles) and $\mathrm{\log(M_\star/M_\odot) > 9.0}$ (squares). Error bars correspond to the upper 84th and lower 16th percentiles. }
    \label{fig:o3o2_all_evo}
\end{figure*}

Whilst R3 is known to trace gas-phase metallicity, it is similarly sensitive to ionisation parameter. As previously discussed, in the local Universe galaxies house older stellar populations that produce lower ionisation parameters, allowing R3 to predominantly trace metallicity. At high redshift this is not the case, so some consideration must be placed on a potentially evolving $\mathrm{log(U)}$. The most accessible probe of $\mathrm{log(U)}$ available to this study is the ratio of O32  \citep[e.g.,][]{Papovich22}, where a larger ratio represents a larger ionisation parameter (if one considers a constant ionising source). In Figure \ref{fig:o3o2_all_evo}, we present the relation between R3 and O32, where all line fluxes have been corrected for dust extinction. Also shown are the median ratios determined in redshift bins of $z < 4.5$, $4.5 < z < 6.0$, $z > 6.0$ and mass bins of $\mathrm{\log(M_\star/M_\odot) < 8.0}$, $8.0 < \mathrm{\log(M_\star/M_\odot) < 9.0}$, and $\mathrm{\log(M_\star/M_\odot) > 9.0}$.

From the median points in Figure \ref{fig:o3o2_all_evo}, we see an increasing R3 with stellar mass, but also crucially a falling O32. This evolution is somewhat consistent across redshift bins, but lower redshift bins do produce lower O32 values, with this effect being most notable between the $z > 6.0$ and $4.5 < z < 6.0$ bins. This broadly implies a lower ionisation parameter and is consistent with ageing, more metal rich stellar populations at higher stellar masses.

\begin{figure*}
    \centering
    \includegraphics[width=1\linewidth]{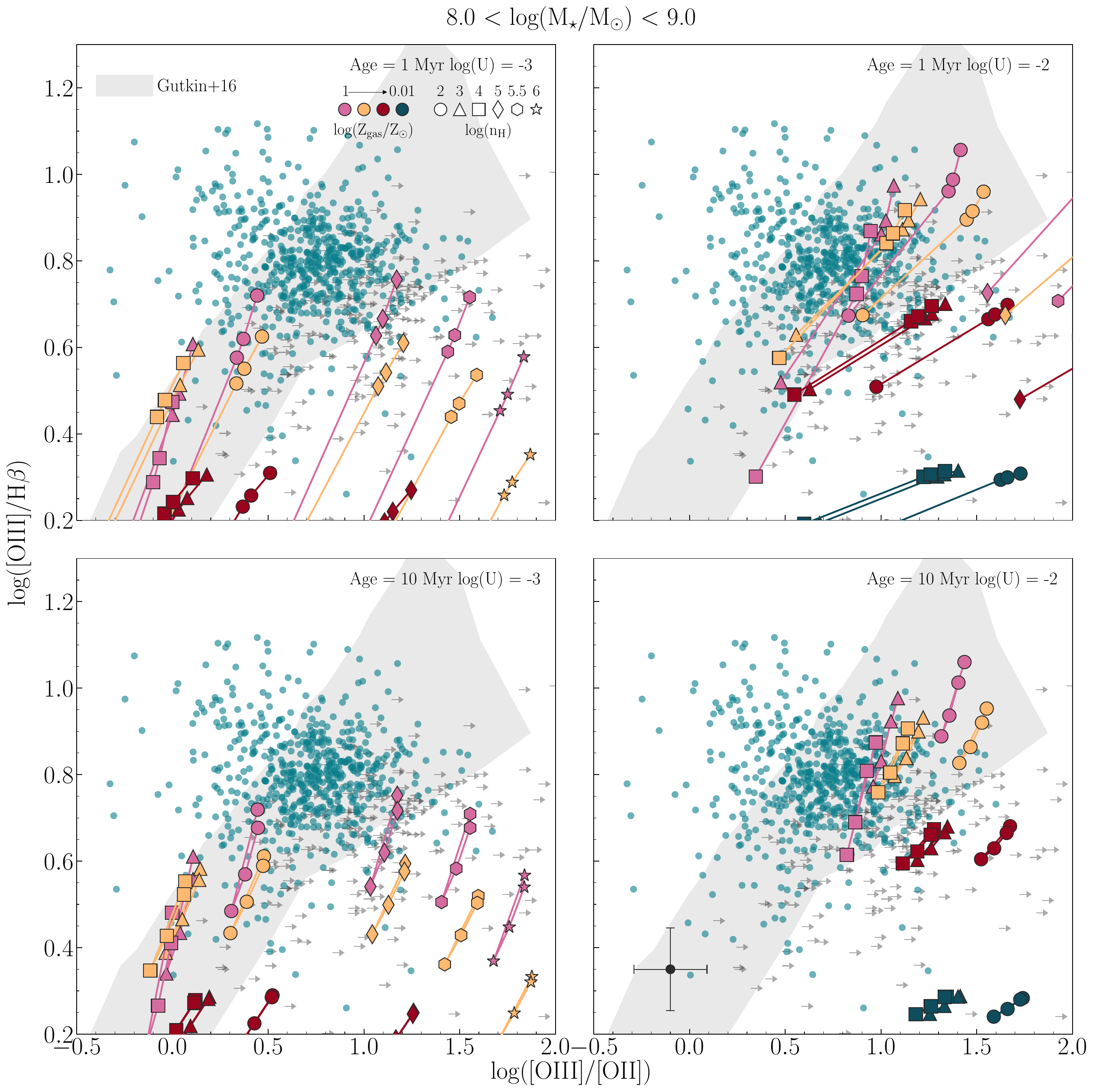}
    \caption{Dust corrected log(O32) vs log(R3) for normal (non-AGN) galaxies with stellar mass $8.0 < \mathrm{\log(M_\star/M_\odot)} < 9.0$. Over-plotted are a selection of \texttt{CLOUDY} models described in Section \ref{ssec_photmod}. All figures include gas-phase metallicities of [1, 0.2, 0.1, 0.01] $\mathrm{Z_\odot}$ and log gas densities of [2-5,5.5,6] cm$^{-3}$. Each combination has stellar metallicity vary between [1, 0.2, 0.1, 0.01] Z$_\odot$ whilst each individual plot has a fixed combination of stellar age and ionisation parameter shown in the upper right corner of each figure. The grey shaded region represents a set of models from \cite{Gutkin16} encompassing gas-phase metallicities of Z=[0.014, 0.004, 0.002], hydrogen densities of [100, 1000] cm$^{-3}$ and a continuous star formation history with age 100~Myr.}
    \label{fig:o3o2_cloudy_intmass}
\end{figure*}

\begin{figure*}
    \centering
    \includegraphics[width=1\linewidth]{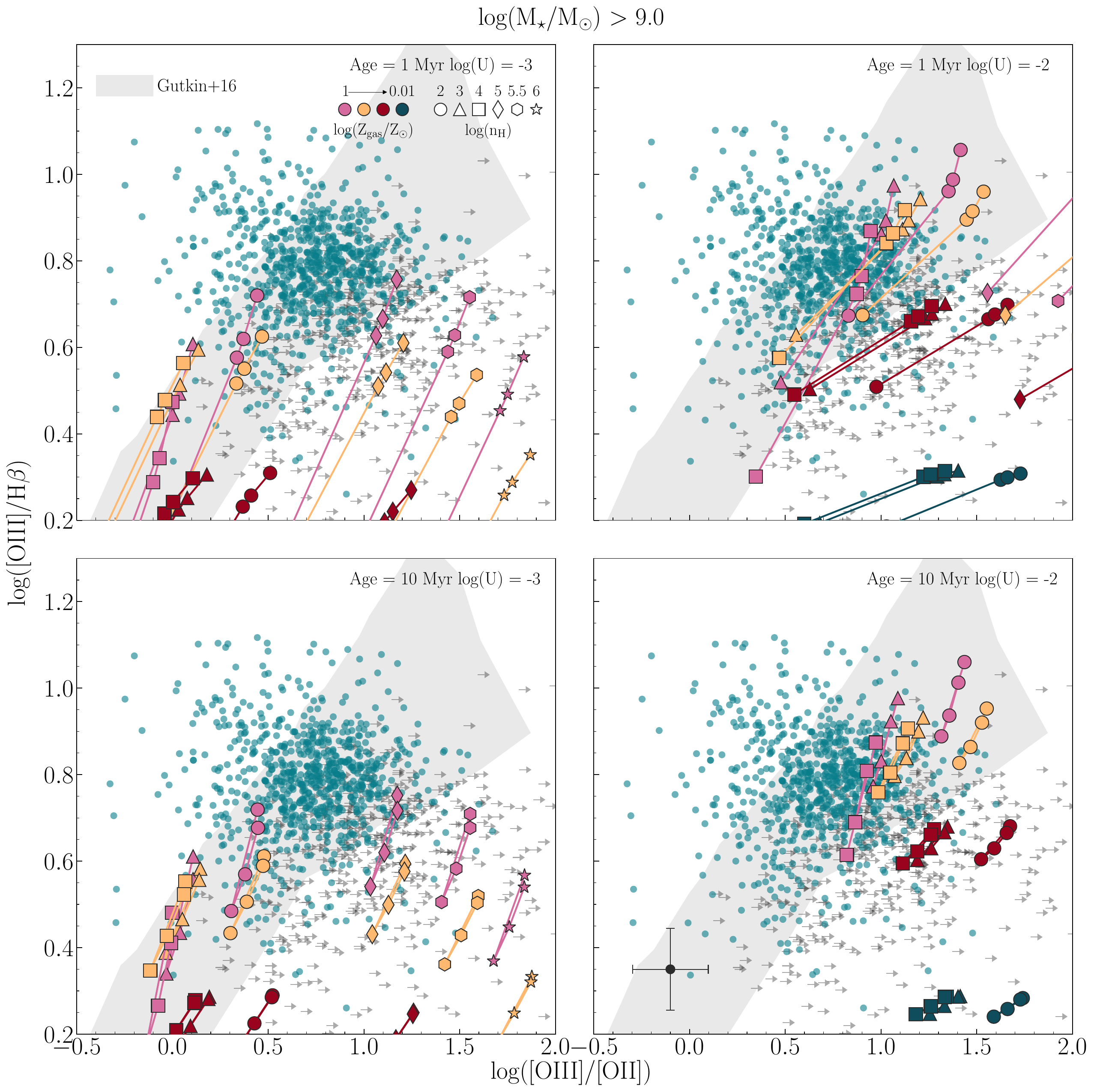}
    \caption{As Figure \ref{fig:o3o2_cloudy_intmass}, but with stellar masses $\mathrm{\log(M_\star/M_\odot)} > 9.0$}
    \label{fig:o3o2_cloudy_himass}
\end{figure*}

\begin{figure*}[t]
    \centering
    \includegraphics[width=1\linewidth]{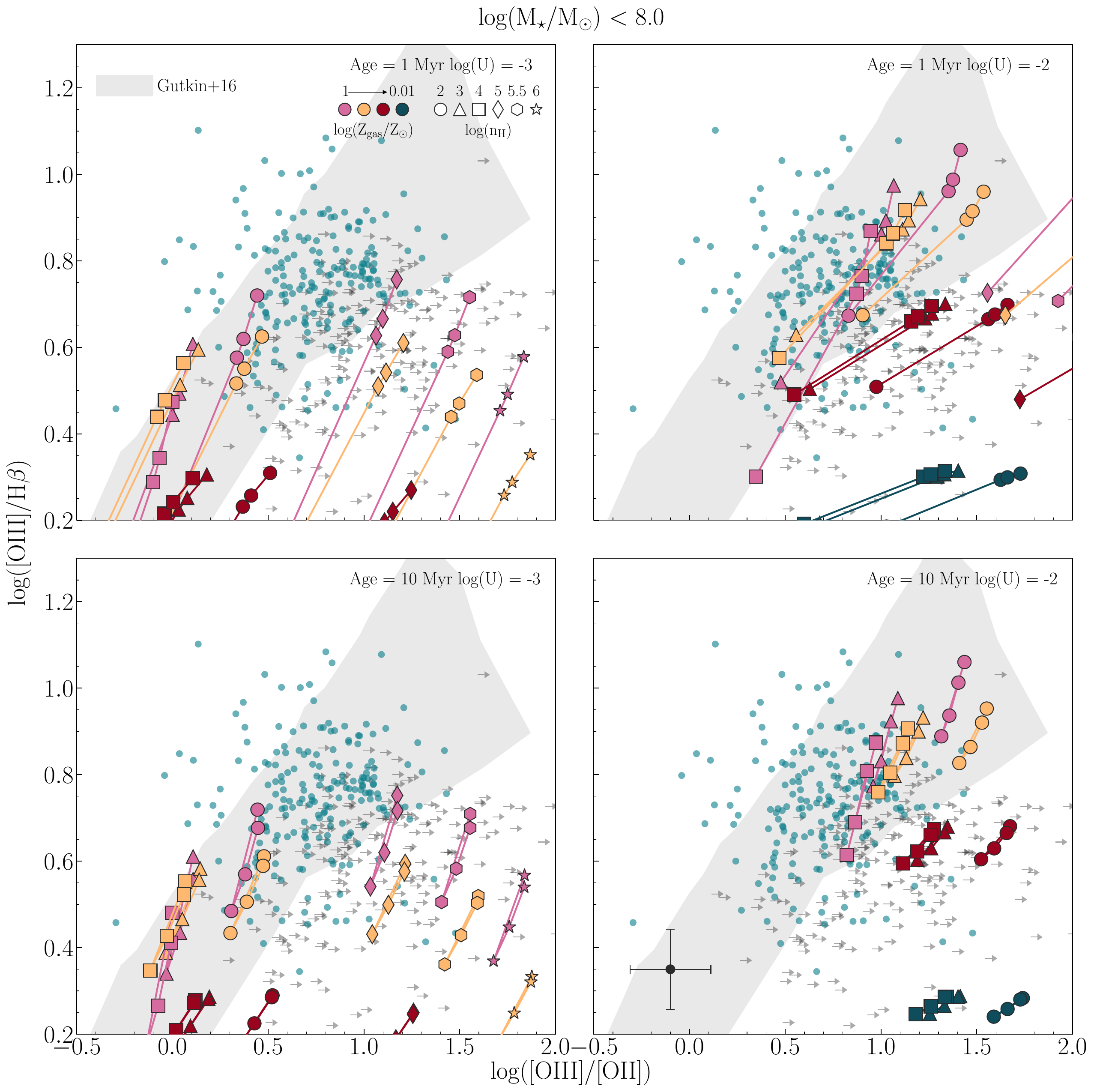}
    \caption{As Figure \ref{fig:o3o2_cloudy_intmass}, but with stellar masses $\mathrm{\log(M_\star/M_\odot)} < 8.0$}
    \label{fig:o3o2_cloudy_lowmass}
\end{figure*}

In an effort to more robustly constrain the ISM conditions of our sample, we use the \texttt{CLOUDY} modelling described in Section \ref{ssec_photmod}. In Figure \ref{fig:o3o2_cloudy_intmass}, we replot the line ratios from Figure \ref{fig:o3o2_all_evo} for galaxies with stellar masses $8.0 < \mathrm{\log(M_\star/M_\odot) < 9.0}$ and overplot a collection of \texttt{CLOUDY} models with various physical parameters. Identical figures are produced for higher masses ($\mathrm{\log(M_\star/M_\odot) > 9.0}$, Figure \ref{fig:o3o2_cloudy_himass}) and the lowest mass galaxies in our sample ($\mathrm{\log(M_\star/M_\odot) < 8.0}$, Figure \ref{fig:o3o2_cloudy_lowmass}).

Each figure contains a broad parameter space of stellar metallicities, gas-phase metallicities, gas densities, and combinations of stellar age and ionisation parameter. We also include observed ratios from a subsample of stellar ionising models produced by \cite{Gutkin16} (Hereafter G16). These models are produced with a continuous star formation over 100~Myr and the parameter space included on each figure spans gas-phase metallicities of Z=[0.014, 0.004, 0.002] and hydrogen densities of [100, 1000] cm$^{-3}$. The models from G16 also encompass two initial mass functions, a number of dust-to-metal ratios and varying C/O abundances.

A first comparison of these figures reveals the evolution in O32 described in Figure \ref{fig:o3o2_all_evo}, with 65\% of objects in the highest mass bin producing log ratios $< 0.5$, corresponding roughly to $\mathrm{Z_{gas}} >1/3~\mathrm{Z_\odot}$. This percentage drops rapidly to 23\% and 10\% in the intermediate and low-mass bins respectively, indicative of a substantial increase in metal enrichment at the highest masses. However, these results also show a non-negligible population of low-mass galaxies that have enriched to a significant fraction of solar metallicity.

We find that the lowest mass galaxies in our sample are broadly reproduced by a combination of gas-phase metallicities and gas densities, with the majority of points described well by young stellar ages (1~Myr) and a $\log(\mathrm{U})=-2$. We find no need for high gas densities or extremely low gas-phase metallicities, with the combination of high O32 and low  R3 being reproduced by gas of 1/10$~\mathrm{Z_\odot}$ at the youngest ages. Invoking an age of 10~Myr introduces the need for higher gas densities and higher gas-phase metallicities.

For a significant subset of the intermediate mass and most massive galaxies ($\mathrm{\log(M_\star/M_\odot) > 8.0}$), the line ratios shown in Fig.\ref{fig:o3o2_cloudy_intmass} and Figure \ref{fig:o3o2_cloudy_lowmass} are not reproduced by any of the \texttt{CLOUDY} models produced for this work. However, many of these points are reproduced by the models from G16. A key difference between these models and the models produced for this work is the star formation history of the stellar population. The G16 models follow 100~Myr of continuous star formation whilst the models of this work are produced with a burst of star formation at much younger times. This difference in SFH may be responsible for the different line ratio space covered by the two model sets, with the G16 models capable of producing lower O32 ratios whilst maintaining high R3 values. This can be attributed to the presence of older stellar populations with lower ionising potential that preferentially excite [O{\small II}] over [O{\small III}].

Even with these older stellar populations, a sample of objects with $\mathrm{\log(M_\star/M_\odot) > 8.0}$ are still not reproduced by either set of models. These objects have the unusual combination of high R3, but low O32 values. At face value, such a combination would be representative of a metal-rich ISM illuminated by older stellar populations. As these objects consist of, primarily, the most massive galaxies in our sample, such a combination is perhaps feasible and notably missed by all the models used in this work.

\section{Low-mass Galaxies}\label{sec_lowmass}

From the results presented so far throughout this work, it is clear that there is significant evolution between the low ($\mathrm{\log(M_\star/M_\odot)} < 9.0$) and high ($\mathrm{\log(M_\star/M_\odot)} > 9.0$) stellar mass regimes. The high-mass galaxies in our sample  are, on average, richer in both dust and metals. Their significant mass at these redshifts also suggests more rapid growth at early times. Irrespective of their formation history, it is clear that these galaxies are at a more advanced evolutionary stage that makes evaluating their ISM conditions more complex. The spectral analysis of massive and intermediate-mass galaxies at high-$z$ has already been the focus of a number of studies \citep[e.g.,][]{RB24,Chakraborty25,Tang25}. To this end, we restrict the remainder of our analysis to the low-mass regime, where we can anticipate the bulk of our sample to evolve in similar ways.

\begin{figure*}[t]
    \centering
    \includegraphics[width=1.0\linewidth]{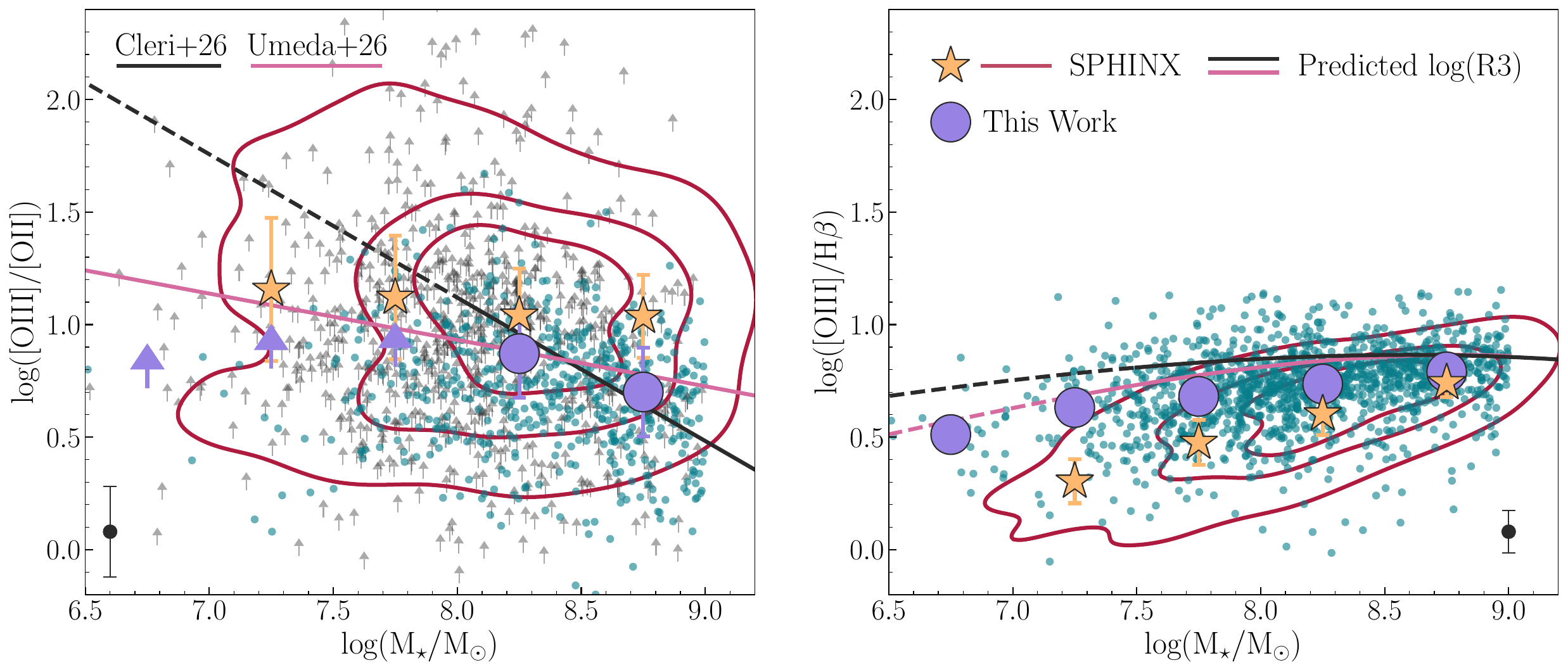}
    \caption{log(O32) (left) and log(R3) (right) against stellar mass for galaxies with $\mathrm{\log(M_\star/M_\odot)} < 9.0$. Median points are shown as purple circles (with lower limits as purple arrows) and are determined in bins of $\pm0.25~ \mathrm{\log(M_\star/M_\odot)}$ centred around $ \mathrm{\log(M_\star/M_\odot)=[6.75,7.25,7.75,8.25,8.75]}$. Overplotted in red contours is the distribution of mock galaxies from the SPHINX data release \citep{katz_2023}. The median points from SPHINX are shown as yellow stars and follow the same bins as this work with the exception of the lowest mass bin which is excluded. The mass evolution of O32 from \cite{cleri_2026} and \cite{umeda_2026} are shown as grey and pink lines respectively, with the dashed portion of the lines representing an extrapolation beyond the mass range in their works. Predicted log(R3) curves using the mass-metallicity relations from \cite{nakajima_2023} (grey) and \cite{Isobe26} (pink) with the log(R3) metallicity calibration from \cite{cataldi25} are shown in the right panel.}
    \label{fig:o_both_mass_evol}
\end{figure*}

\subsection{Line Ratio Evolution in Low-mass Systems}\label{ssec_lm_lr_evo}

Whilst it is important to understand the mass evolution of line ratios across the full mass range of our sample (as in Figure \ref{fig:lr_mass_evol}), it is particularly crucial to further constrain this evolution at the low-mass end. In Figure \ref{fig:o_both_mass_evol}, we show the ratios of log(O32) and log(R3) against stellar mass, with median points re-binned to more finely show the evolution down to $\mathrm{log(M_\star/M_\odot)<7.0}$.

For comparison, we consider also a sub-sample of mock galaxies produced within the SPHINX simulations \citep{rosdahl_2018,rosdahl_2022} available in the SPHINX public data release \citep{katz_2023}. We select galaxies following a proportional distribution to the redshift and mass ranges of our low-mass sample and over-plot their distribution as red contours as well as show median points as yellow stars. Furthermore, we show the observed (and extrapolated) log(O32)-stellar mass relation from \cite{cleri_2026}, as well as predicted log(R3) ratios determined using the mass-metallicity relations from \cite{nakajima_2023} and \cite{Isobe26} with the log(R3) calibration calculated by \cite{cataldi25}.

O32 shows a significant rise with decreasing stellar mass between the two highest mass bins, indicating stronger ionisation conditions at lower masses. Such evolution is expected and follows the results from \cite{cleri_2026}. However, at $\mathrm{log(M_\star/M_\odot) < 8.25}$, this relation appears to flatten and our results appear to be more in line with that of \cite{umeda_2026} (although when considering upper limits both relations are strictly still feasible). Such flattening is similarly seen in the SPHINX galaxies, where the overall distribution of O32 values also becomes much wider at the lowest masses. Physically, this lack of evolution with stellar mass can be interpreted as a homogeneity of the ionisation parameter, possibly owing to galaxies at these low stellar masses housing similarly young, hot stellar populations. However, this must be heavily cautioned by the fact that the three median points at $\mathrm{log(M_\star/M_\odot) \leq 7.75}$ are dominated by [OII] upper limits which, in reality, may raise these medians to show a more expected rising trend. A similar (but equally tentative) result is presented in \cite{shen_2025}.

The evolution of R3 meanwhile follows the typically expected increase with stellar mass, even at the lowest masses. This rise is similarly shown by SPHINX, but with a far steeper slope. In both cases, the low R3 values at $\mathrm{log(M_\star/M_\odot) \sim 7.25}$ suggest, at first glance, a combination of extremely low gas-phase metallicity and significant recent star formation. For example, the R3-metallicity calibration in \cite{cataldi25} implies that a log(R3) value of $\sim$0.5 (the median value of the lowest stellar mass bin in this work) corresponds to a metallicity $\mathrm{Z_{gas} \sim 0.03~Z_\odot}$. Whilst any R3 calibration is bivariate at this value, the high metallicity solution (approximately 65\% solar from \citealt{cataldi25}) is unlikely to be the case for such low mass systems. To confirm this, we make use of the \texttt{genesis-metallicity} code \citep{langeroodi_2026} to estimate gas-phase metallicities from a combination of H$\beta$ equivalent width and [O{\small II}], [O{\small III}] and H$\beta$ line fluxes. From this we find all low-mass sources have gas-phase metallicities $< 65\%~\rm Z_\odot$, thus reliably laying on the low-metallicity branch of most R3 calibrations.

These low metallicities have, until very recently, not been predicted by mass-metallicity relations, with our median values consistently being placed under the inferred values of log(R3) at $\mathrm{log(M_\star/M_\odot) < 8.25}$ from the \cite{nakajima_2023} relation. This is predominantly due to a lack of statistics at the lowest stellar masses in this latter work.  Indeed, by analysing a larger selection of low-mass galaxies, \cite{Isobe26} produced an updated MZR to which our results are in excellent agreement. Based on these predictions, $\sim25\%$ of our low-mass sample would be considered extremely metal-poor (following the 4\% Z$_\odot$ upper limit defined in \citealt{Isobe26}), with this fraction rising to $\sim55\%$ at stellar masses $\mathrm{log(M_\star/M_\odot) < 7.5}$. We caution however that these fractions correspond to specific line ratio diagnostics and calibrations used to determine metallicities, and the significance of any findings at low masses should be confirmed by further observations.

\begin{figure*}[]
    \centering
    \includegraphics[width=1\linewidth]{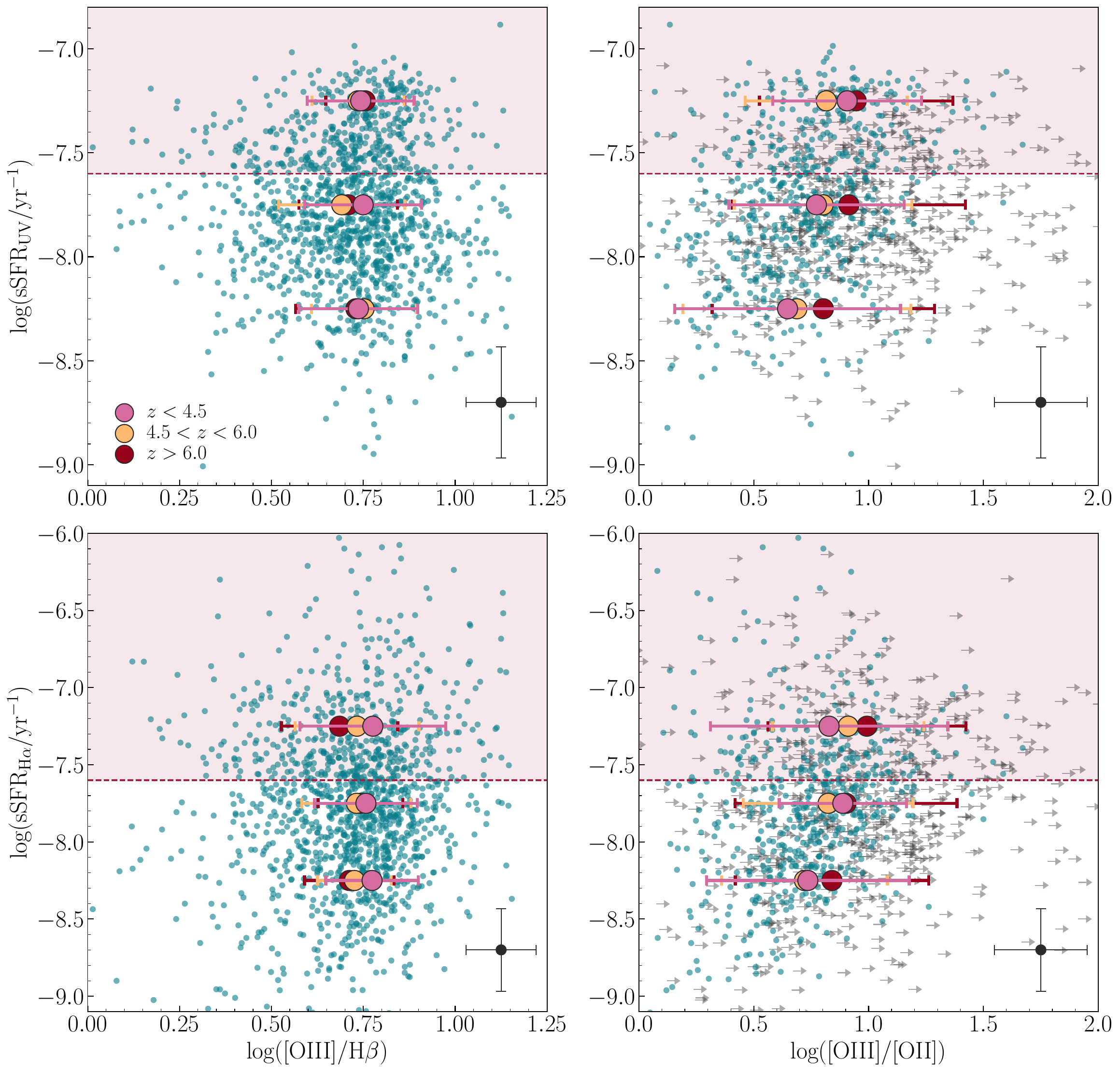}
    \caption{Dust-attenuation corrected \ohb~(left) and \otot~against sSFR for our low-mass galaxy sample, with SFRs estimated from rest-frame UV photometry (upper panels) and H$\alpha$ line flux (lower panels). Median points are estimated in redshift bins of $z < 4.5$ (pink), $4.5 < z < 6.0$ (yellow), and $z > 6.0$ (red), with sSFR bins of $\pm0.25$ around mid-points of [-8.25,-7.75,-7.25]/yr$^{-1}$. Errors on each median point are the upper 84th and lower 16th percentiles. The starburst regime as defined in \cite{caputi_2017,caputi_2021} is shaded red.}
    \label{fig:line_rat_ssfr}
\end{figure*}

\subsection{Main Line Ratios versus sSFR}\label{ssec_lr_ssfr}

So far, we have considered the evolution of various line ratios with stellar mass, with R3 showing the strongest trend in the low-mass regime. However, we similarly wish to investigate the effects of star formation over different timescales on these line ratios. In this Section we investigate the line ratio behaviour with specific star formation rate (sSFR), while in Section~\ref{sec:burst} we investigate the effects of burstiness. The latter implies variations on a galaxy's SFR and can occur to galaxies at any position of the SFR-M$^\ast$ plane, i.e., with any value of sSFR \citep{mcclymont25,navarro-carrera_2026}.

Figure \ref{fig:line_rat_ssfr} shows R3 and O32 plotted against sSFR measures derived from both UV and Balmer line indicators (see Section \ref{ssec_physparam}). We present median values in sSFR bins of width $\pm0.25$ centred around $\mathrm{\log (sSFR)=[-8.25,-7.75,-7.25]/yr}^{-1}$ and redshift bins of $z < 4.5$, $4.5 < z < 6.0$, and $z > 6.0$. For all combinations, we find a broad distribution of values, with any instance of an evolution with sSFR being very minor and subject to large variance. We similarly see no evolution across all three redshift bins, although we note that, for a given sSFR$\mathrm{_{H\alpha}}$, higher redshift bins show higher log(R3) on average, with this discrepancy becoming most notable at higher sSFRs. Furthermore, we find no discernable change between galaxies on the main sequence and those in the starburst regime defined by \cite{caputi_2017, caputi_2021}. Of the two line ratios presented, log(O32) shows a tentative positive trend with both sSFR measures, but again with large variance.

The tentative correlation found for log(O32) and lack of correlation of R3 with sSFR suggests that the ionisation parameter (rather than the galaxy metallicity) could be dependent of the sSFR. In any case, the broad distribution of line ratios at fixed sSFR, and lack of evolution with redshift, are likely indicative of the volatility of low-mass galaxies. The star formation in high redshift galaxies may be highly stochastic, with frequent burst and lull phases \citep{Endsley25, Looser25, navarro-carrera_2026}. These fluctuations, in combination with the shallow potential wells of low-mass systems likely produce a rapidly changing ISM. For example, a galaxy with high sSFR may well be producing more metals, but these metals may then be rapidly ejected from the galaxy through stellar feedback. Meanwhile a lower sSFR galaxy will produce heavier elements at a slower rate, but metals from a previous burst may have had time to fall back into the galaxy and enrich the ISM \citep[e.g][who argue the majority of outflows return to the host galaxy in low-mass systems]{outflows_1,xu_2025}. The wide range of ratios observed in this case may therefore be an effect of observing galaxies at different stages in their star formation cycles, which we investigate further in Section~\ref{sec:burst}.

In the local Universe, and up to Cosmic Noon, R3 has been shown to increase with sSFR \citep{dickey_2016,holden_2016,strom_2017}. However, the samples in these works contain almost exclusively galaxies with stellar mass $\mathrm{\log(M_\star/M_\odot)} > 9.0$, so any comparison must be treated with caution. In any case, it is worth noting that, for the sSFR range probed in our work, this relation appears to flatten off in line with the scatter reported here. Similar trends have been found for O32 out to redshift $z\sim3$ \citep{nakajima_2014, sanders_2016}, with both works reporting increasing O32 with sSFR, albeit again outside of the stellar mass and sSFR regimes studied here. 

\begin{figure}
    \centering
    \includegraphics[width=1\linewidth]{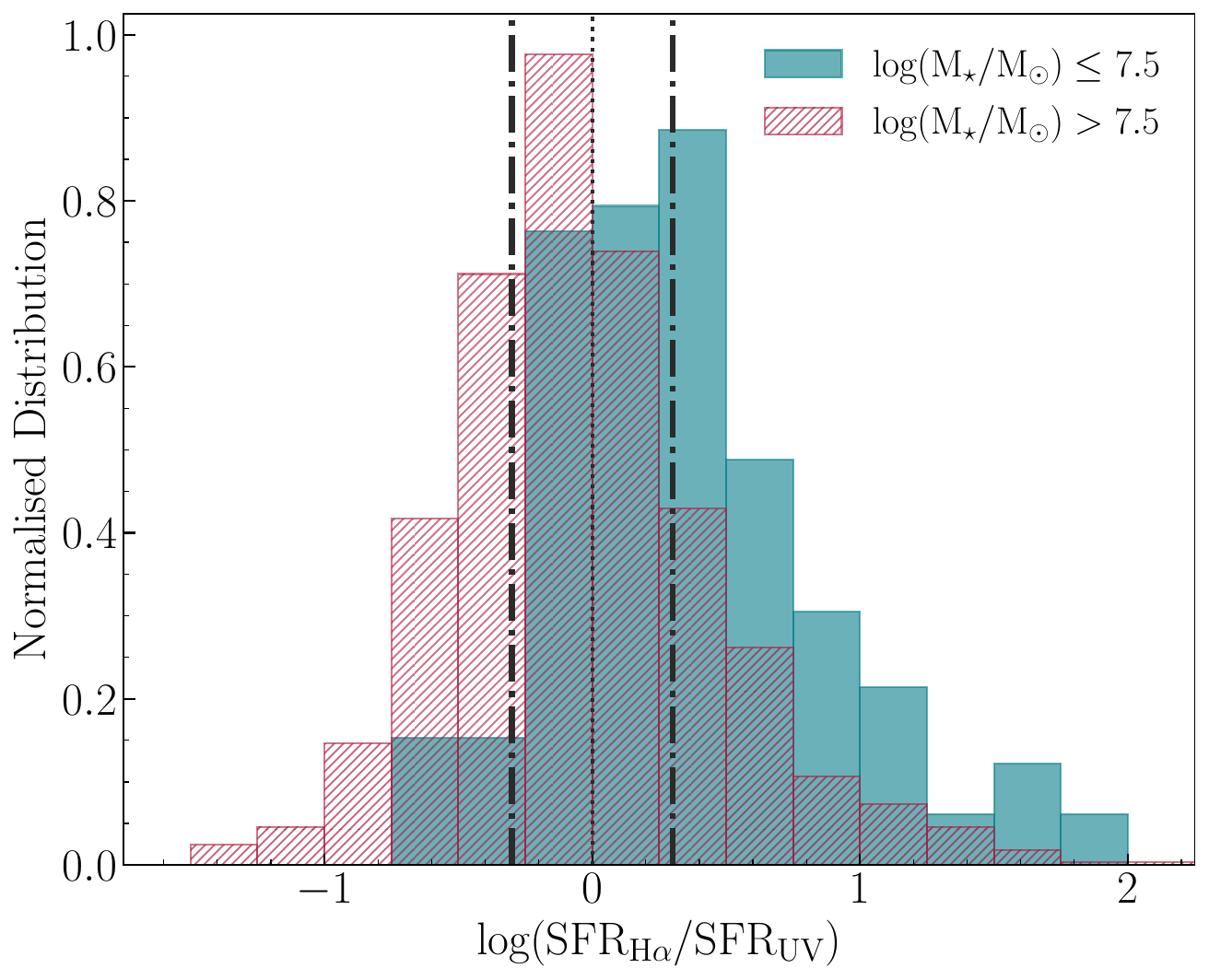}
    \caption{Normalised distribution of burstiness, defined as SFR$_{\mathrm{H\alpha}}/$SFR$_{\mathrm{UV}}$ for all galaxies with stellar mass $\mathrm{\log(M_\star/M_\odot})\leq9.0$. Galaxies are split into two bins around $\mathrm{\log(M_\star/M_\odot})=7.5$ with the number of galaxies in the lower and higher mass bins being 131 and 1326 respectively. Candidate AGN are omitted. Overplotted are demarcations at 0 and $\pm0.3$ which represent different timescales for bursts in star formation (e.g. \citealt{faisst_2019, navarro-carrera_2026}).}
    \label{fig:burst_dist}
\end{figure}

\begin{figure*}[t]
    \centering
    \includegraphics[width=1\linewidth]{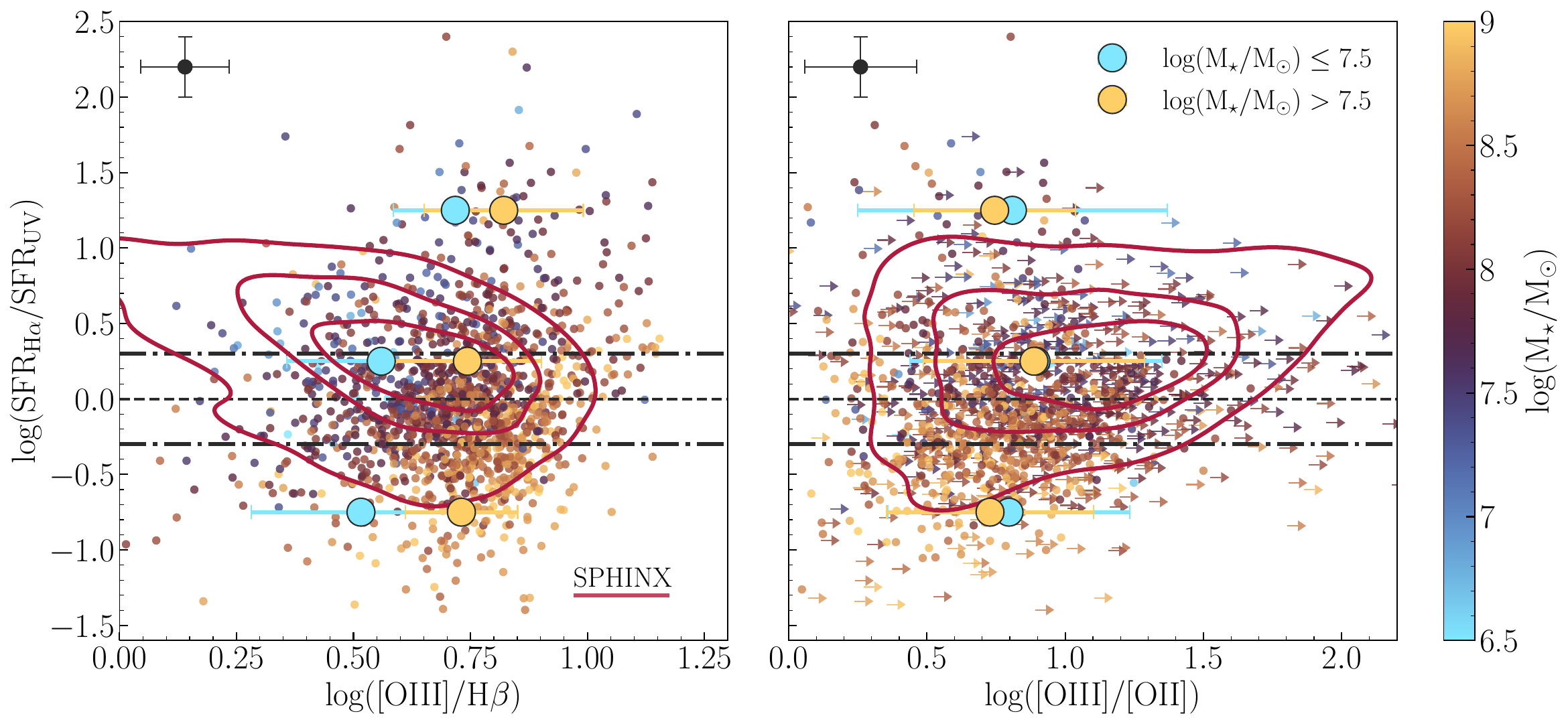}
    \caption{Dust-attenuation corrected \ohb~(left) and \otot~(right) against burstiness derived from the ratio between SFR$_{\mathrm{H\alpha}}$ and SFR$_{\mathrm{UV}}$. Galaxies are colour coded by stellar mass. Over-plotted are lines at $\log(\mathrm{SFR_{H\alpha}/SFR_{UV}}) = 0,$ and $\pm0.3$. The distribution of low-mass galaxies from the SPHINX data release \citep{katz_2023} are shown as red contours. Median points are determined in burstiness bins of $\pm0.1$ around mid points [-0.75, 0.25, 1.25] and are split by mass around $\mathrm{log(M_\star/M_\odot)=7.5}$. } 
    \label{fig:burst_line_rat}
\end{figure*}

\begin{figure*}
    \centering
    \includegraphics[width=1\linewidth]{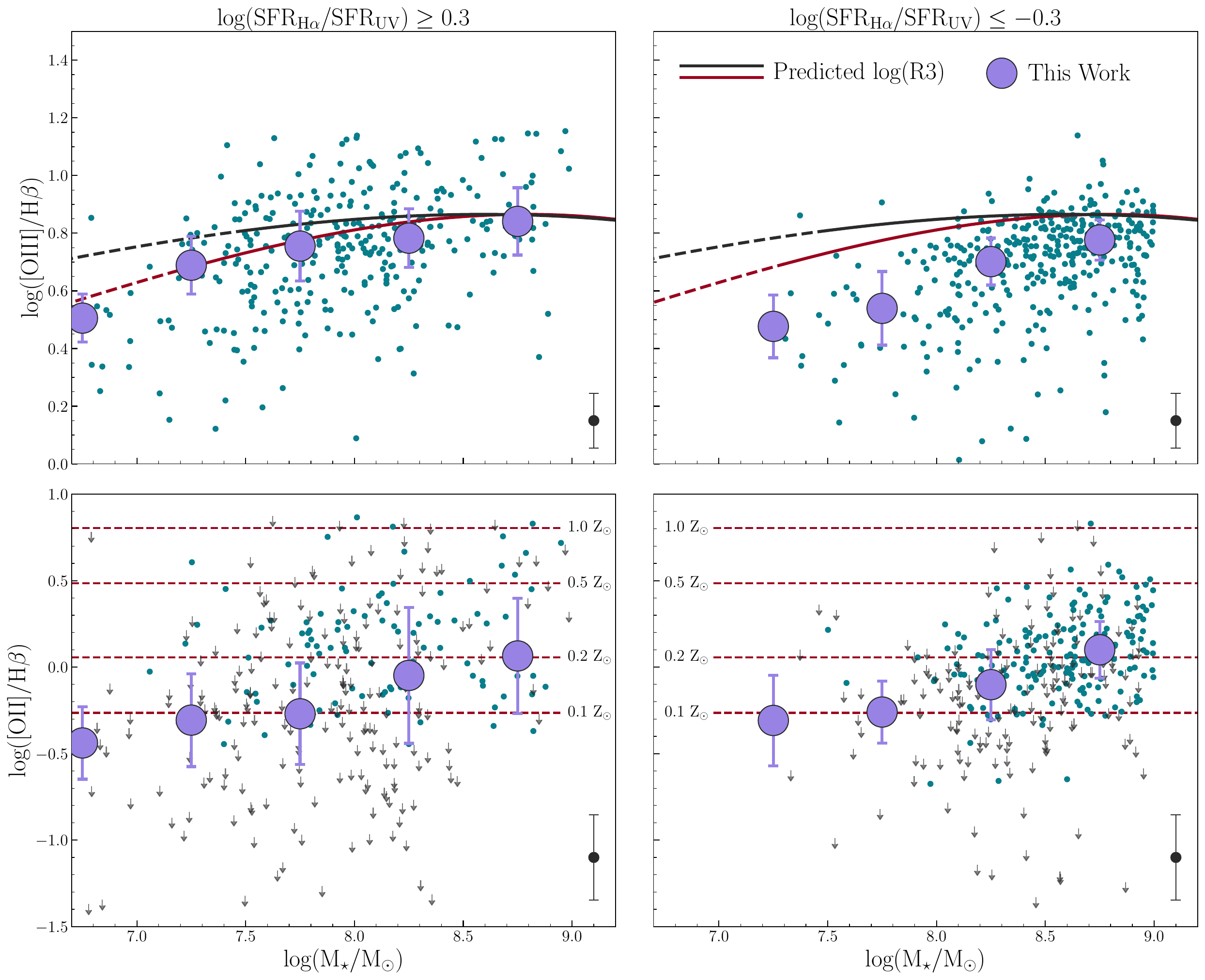}
    \caption{\textbf{Upper:} R3 against stellar mass for galaxies with $\log(\mathrm{SFR_{H\alpha}/SFR_{UV}}) \geq 0.3$ (left) and $\log(\mathrm{SFR_{H\alpha}/SFR_{UV}}) \leq -0.3$ (right). Sources with both $\mathrm{S/N_{H\beta } < 2.0}$ and $\mathrm{S/N_{[OII]} < 2.0}$ are excluded from both panels owing to no reliable R2 measurement. The same predicted log(R3) curves from Figure \ref{fig:o_both_mass_evol} are also shown. \textbf{Lower:} R2 against stellar mass for the same burstiness cuts. Presented gas-phase metallicity values are determined from the R2 calibration in \cite{Sanders_2024}. Median points in all panels are shown as purple circles and follow the bins defined in Figure \ref{fig:o_both_mass_evol}. The much higher degree of scattering in the line ratios under high-burstiness conditions (left panels) is evident.}
    \label{fig:o3hb_burst_split}
\end{figure*}

\subsection{Burstiness}\label{ssec_burstiness}
\label{sec:burst}

The ratio of star formation rate measures from Balmer lines and UV luminosity has historically been used to quantify the so-called `burstiness' of a galaxy's star formation history \citep{glazebrook_1999, weisz_2012, dominguez_2015, faisst_2019, navarro-carrera_2026}. This is driven by the idea that these measures trace star formation over 10~Myr and 100~Myr timescales respectively (although see \citealt{emami_2019} and sources therein discussing other factors that influence this ratio). Above, we briefly mentioned the potential consequences of a bursty (or stochastic) star formation history which we endeavour to analyse here. 

Figure \ref{fig:burst_dist} shows the distribution of burstiness, measured as $\mathrm{\log(SFR_{H\alpha}/SFR_{UV})}$, for the low-mass galaxies in our sample. We split this sample further in two around $\mathrm{\log(M_\star/M_\odot)=7.5}$ with a total of 131 and 1326 galaxies in the lower mass and higher mass bins respectively. From here it is evident that the lowest mass galaxies in our sample are typically dominated by bursty star formation, with a median burstiness of 0.28 and 73\% of sources having a burstiness value $> 0$. Higher mass galaxies meanwhile exhibit a median of -0.08, with a slight preference to burstiness values $< 0$ of 58\% to 42\%. These two populations are confirmed through a Kolmogorov–Smirnov test (KS test) to be independent of each other through the rejection of the null hypothesis with p-value $\ll 0.05$. 

Given that burstiness varies significantly with stellar mass, it is important to determine any effect that it may have on a galaxy's ISM conditions. In Figure~\ref{fig:burst_line_rat}, we show log(R3) and log(O32) against the ratio of $\mathrm{\log(SFR_{H\alpha}/SFR_{UV})}$ for the low-mass galaxies in our sample (see Section \ref{ssec_physparam} for details on SFR calculations). Points in Figure~\ref{fig:burst_line_rat} are colour-coded by stellar mass. As in Figure \ref{fig:o_both_mass_evol}, we present mock galaxies from SPHINX as red contours. Median values are calculated for the same stellar mass bins as Figure \ref{fig:burst_dist}, with burstiness bins of $\pm0.1$ around mid points [-0.75, 0.25, 1.25].

As is the case in Figure \ref{fig:line_rat_ssfr}, we find a broad distribution of points, but note a more apparent (albeit still tentative) positive correlation between R3 and burstiness. We report Pearson rank coefficients of $\rm r=[0.30, 0.15]$ for the low- and high-mass splits respectively, with both having $\rm p \ll 0.05$, so any correlation is stronger for the lowest stellar masses. We find no correlation present between O32 and burstiness. Overall, the scatter in both ratios is broadly reproduced by SPHINX, although we remark some differences with respect to our results: on the one hand, in the case of log(R3), we do not see the low line ratio tail at high burstiness exhibited by the mock galaxies. These systems are the most metal-poor in our SPHINX subsample and the discrepancy to our results is likely a sampling effect. On the other hand, we do see a significant subset of sources with high burstiness and high R3 values that do not appear in SPHINX. SPHINX notably scales galaxies following a solar chemical abundance which has the effect of reducing R3 in metal-rich systems which may account for the lack of mock galaxies at the high R3 values observed in this work. 

Across both Figure \ref{fig:burst_dist} and Figure~\ref{fig:burst_line_rat}, we include demarcations representing different timescales for starburst activity, with lines at $\mathrm{log(SFR_{H\alpha}/SFR_{UV}) = \pm0.3}$ representing a burst within the last 50~Myr in the positive case, and after 200~Myr in the negative case \citep{faisst_2019, navarro-carrera_2026}. In this way we can infer (from Figure \ref{fig:burst_dist}) that the majority of the lowest mass galaxies have experienced a burst within the last 50~Myr, with very few sources having their most recent burst on longer timescales. Higher mass galaxies meanwhile have a much greater variation in their star formation histories, with preference to either a continuous star forming process, or a burst at much earlier times. Whilst the preference for low-mass galaxies to be bursty is anticipated, the observed difference between the two samples could be due to a selection effect: very low mass galaxies with low burstiness may have Balmer lines too faint to be present in the spectroscopic sample analysed here.

For the majority of objects with a more continuous star formation history ($\mathrm{-0.3 \leq log(SFR_{H\alpha}/SFR_{UV}) \leq 0.3}$), we reproduce the observed mass dependence in R3, and the weaker dependence in O32, as initially presented in Figure \ref{fig:o_both_mass_evol}. The situation at low/high burstiness values is, however, more complex. To try and disentangle these situations, in Figure \ref{fig:o3hb_burst_split} we show R3 and R2 as a function of stellar mass only for sources with either high ($\log(\mathrm{SFR_{H\alpha}/SFR_{UV}}) \geq 0.3$) or low ($\log(\mathrm{SFR_{H\alpha}/SFR_{UV}}) \leq -0.3$) burstiness. From this figure, we find two notable results. One concerning the scatter of both line ratios and another concerning the offset that the observed values have with respect to the predicted R3 curves. 

We first note a significant scatter in the high-burstiness sample compared to a tighter correlation in the low-burstiness regime. We calculate the median absolute deviation (MAD) for mass bins of $\pm0.25\mathrm{~log(M_\star/M_\odot)}$ around central values $\mathrm{log(M_\star/M_\odot)}=[7.75, 8.25, 8.75]$, finding MAD values for R3 of $[0.12, 0.10, 0.12]$ and
$[0.13, 0.08, 0.07]$ for the high and low burstiness samples respectively. We similarly find MAD values for R2 of $[0.29, 0.39, 0.33]$ and $[0.18, 0.20, 0.16]$ for high and low burstiness. For both line ratios, the difference in scatter is most prominent at higher stellar masses, whilst at lower stellar masses the low-burstiness sample exhibits a similar or larger scatter (although we caution a low-sample size in the low-mass, low-burstiness regime may affect this result). 

As suggested previously, this scatter is indicative of the disruptive nature of a recent starburst on a galaxies ISM conditions, where both chemical enrichment and material loss through stellar winds occur over short timescales. These processes can dramatically change line ratios, with the present scatter an effect of galaxies being at different stages of this cycle. This scatter being reproduced by R2 further supports this and highlights the need for caution when determining ISM conditions from line ratios in highly bursty systems.

Furthermore, the low scatter of low-burstiness galaxies suggests more settled ISM conditions where chemical enrichment is occurring in line with galaxy growth. Crucially, the majority of the low-burstiness sample also resides below predicted R3 values, whilst galaxies with high burstiness are, on average, consistent with the predictions of \cite{Isobe26}. The low burstiness sample is constrained to a most recent burst occurring $\sim200$ Myr ago, with any proceeding star formation being slower (potentially in a lulling phase) and more continuous. These galaxies being below the predicted R3 values suggests this reduction in star formation has reduced the rate of chemical enrichment. This could be evidence of stellar outflows removing metals from the galaxy ISM as the burst has terminated. Comparing with the observed R2 values, such stripping of metals would be most prevalent in the lowest mass systems which have typical predicted metallicities $\mathrm{< 0.2Z_\odot}$.

From Figures~\ref{fig:burst_line_rat} and \ref{fig:o3hb_burst_split} we can also see that high R3 values ($\mathrm{log(R3) \geq 0.8}$) are almost exclusively found amongst low-mass galaxies with high burstiness. (We remind the reader that all identified AGN are being left out of the analysis in this paper). New star-formation bursts can lead to quick chemical enrichment and/or, at least, a temporary enhancement of some line ratios, preventing reliable metallicity estimations. Even if the regular conversions of main line ratios into oxygen abundances still apply, the inferred metallicities would in most cases be only valid for a short period of time and not indicative of the final level of metal enrichment that those galaxies would have after the ISM conditions stabilise again after the burst. This is also supported by a significant number of objects with high-burstiness reaching high R2 values consistent with metallicities $> 0.2\mathrm{Z_\odot}$ and a smaller minority reaching metallicities $0.5\mathrm{Z_\odot}$. These R2 values are, conversely, almost unobserved in the low-burstiness sample except at the highest masses, i.e., $\mathrm{log(M_\star/M_\odot)}= 8.5-9.0$. 

At the same time, low R3 values ($\mathrm{log(R3) \leq 0.3}$) are virtually not found amongst high-burstiness galaxies, except at the lowest stellar masses $\mathrm{log(M_\star/M_\odot)}<7.5$ for sources that would make the very low end of the mass-metallicity relation \citep[e.g.,][]{Chemerynska24}.  Although this result may need to be confirmed with larger spectroscopic samples, it suggests that the search for the most metal-poor sources would be most efficiently conducted amongst low-burstiness galaxies (or galaxies with smoother star-formation histories). Of course, one has to consider the impact of ionisation conditions on R3, but for the lowest-mass galaxies we expect a naturally higher log(U) owing to the presence of very young stellar populations. In this sense, a low R3 value is likely to be a reflection of truely metal-poor ISM conditions.

Furthermore, the fact that some galaxies show such low R3 values after 200~Myr of a burst, but not in the first 50~Myr, illustrates how a very recent burst may drastically affect observed line ratios, rendering any inference on a galaxy's level of chemical enrichment very uncertain (or at least unstable) in such a non-equilibrium situation. It is also in line with the previously described scenario in which a star-formation burst enriches the galaxy with a significant amount of metals, but these metals can be quickly lost via stellar winds.

\begin{figure*}
    \centering
    \includegraphics[width=1\linewidth]{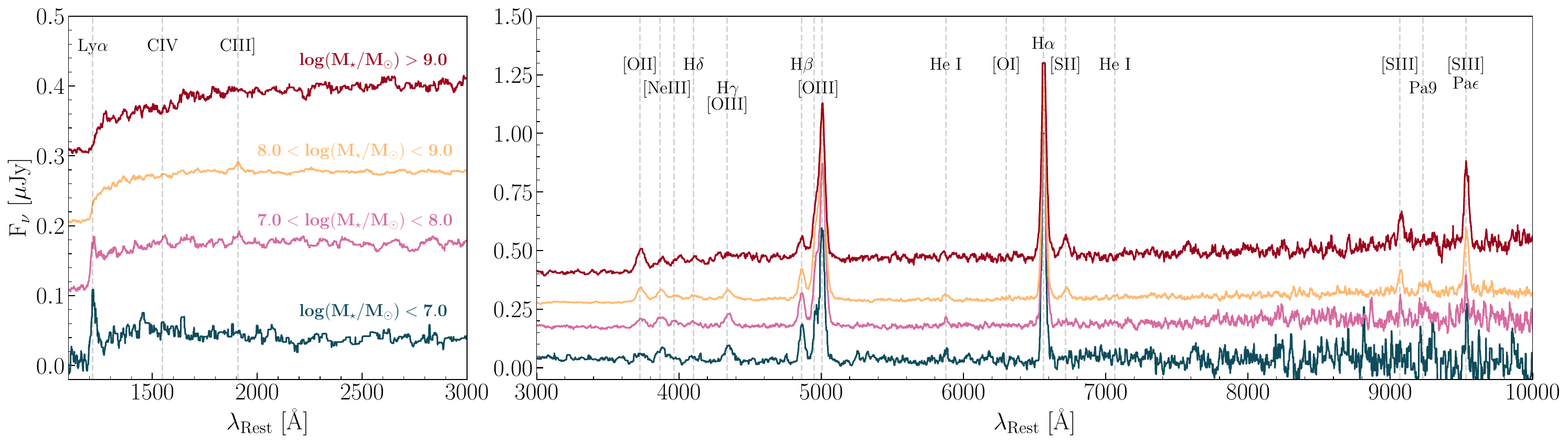}
    \caption{Median stacked spectra of star forming galaxies in bins of stellar mass, separated to show individually the Ly$\alpha$ profile and other UV lines (left) and optical emission lines (right). Stacks are formed from bins of $\mathrm{log(M_\star/M_\odot) = [< 7.0, 7.0-8.0,8.0-9.0, >9.0]}$ with N = [27, 496, 863, 232]. Each spectrum has been shifted in flux density by an arbitrary value for clarity. Notable emission lines are labelled.}
    \label{fig:spec_stack}
\end{figure*}

\subsection{Other common spectral lines present in low-mass galaxy spectra}\label{ssec_spec_stack}

To determine the presence of other emission lines across our sample, we perform a median spectral stack of star-forming galaxies in different low mass bins, and include a mass bin at $\mathrm{log(M_\star/M_\odot) > 9.0}$ for comparison. This produces a sample of N$=[27,496,863,232]$ per mass bin. To perform the stacking, each individual spectra is shifted to rest-frame and resampled to a common wavelength grid of fixed wavelength separation. Individual spectra are normalised by their H$\alpha$ flux, this choice is made over a continuum wavelength as not all sources in our sample have a secure continuum detection. The resulting median spectra are presented in Figure \ref{fig:spec_stack} which has been split to highlight the Ly$\alpha$ region and the bulk of the most prominent optical emission lines.

Our low-mass sample is generally consistent with the high-mass sample in terms of emission lines present, although they vary in strength. The most significant line evolution occurs, as expected, following the build up of metals with stellar mass, across the oxygen and sulphur emission lines. Indeed, the high-mass stack exhibits considerably stronger [OII] emission as well as significantly stronger sulphur lines. Aside from this, the high-mass stack also shows a rising optical slope, as well as a non-negligible Balmer break not seen in any of the low-mass medians. As well as greater dust content, these features suggest the presence of older stellar populations, meaning the most massive galaxies in our sample likely experienced a burst in star formation some time ago, and have since seen a decline (but not turn-off) in SFR.

Whilst these features vary between the low-mass and high-mass regimes, the three low-mass medians are markedly consistent with one another (although some lines fall below the detection level of the lowest-mass stack). However, a notable discrepancy is that of the Balmer lines. Visually, it is clear that the detected Balmer lines H$\alpha$, H$\beta$ and H$\gamma$ (if we assume that the contribution of [O{\small III}$\lambda4363$ is small) remain roughly consistent from $\mathrm{log(M_\star/M_\odot)\sim7-9}$. At stellar masses $\mathrm{log(M_\star/M_\odot) < 7.0}$ however, all three lines are substantially more prominent, i.e., they have higher equivalent widths (EW), even exceeding that of the high-mass regime. 

Indeed, estimating the EW of the H$\alpha$ line for each stack gives values of [965.9, 432.2, 367.5, 202.0]$\,\mathrm{\AA}$ in order of increasing stellar mass according to the bins defined above. We see a falling EW with increasing stellar mass and more extreme EWs for the lowest mass galaxies in our sample.  This is likely indicative of a rise in bursty star formation at low-masses \citep[e.g.][]{atek_2022,perry_2025,rinaldi_2025,navarro-carrera_2026}, with such heightened EWs suggesting particularly rapid star formation and very young stellar populations. 

The other marked difference between stellar mass bins is the presence/absence of Ly$\alpha$ emission. For stellar masses $\mathrm{log(M_\star/M_\odot) < 8.0}$, we see evidence for Ly$\alpha$ emission, with the strongest line present in the lowest stellar mass bin (consistent with the findings of e.g., \citealt{Iani24}). Above $10^8\mathrm{M_\odot}$, this Ly$\alpha$ emission disappears and is instead replaced by a smooth Lyman break. Given the stellar masses of our sample are well distributed across the redshift range studied in this work, this evolution is unlikely to be a consequence of a changing intergalactic medium. Instead, we associate this reduction with an increase in dust reddening. This is then further evidence for an increase in dust content in higher mass galaxies, even out to redshift $z\sim7$. 

Interestingly, we further see a tentative CIII] detection from stellar masses $\mathrm{log(M_\star/M_\odot)\sim7-9}$. This line appears absent from the lowest and highest stellar mass bins, but we caution that, at least for the lowest mass bin, this may be an effect of low S/N. The presence of this line in star-forming galaxies has been noted in \cite{feltre_2020}, where they report a stronger detection in galaxies with stellar mass $\mathrm{log(M_\star/M_\odot) < 8.4}$. As a high ionisation line, CIII] emission suggests a particularly hard ionising field that would more typically be associated with AGN, but here it is likely to imply extremely young and hot stellar populations. This is further supported by the presence of helium emission in the lower mass bins, with this line being most prominent at the lowest stellar masses.

\section{Discussion and Summary}\label{sec_summary}

Throughout this work, we have exploited a large data set of NIRSpec spectroscopy in an effort to describe the ISM evolution of high-redshift galaxies across cosmic time. Deep spectroscopy, in combination with the magnification of lensing clusters, permits us to somewhat constrain these properties down to previously unexplored stellar masses. By analysing various line ratios and star formation diagnostics, we have investigated the build up of dust and metals in low mass galaxies and some of the mechanisms that may regulate these quantities.

We have shown throughout Section \ref{sec_fullsample} that the ISM conditions of galaxies at low stellar masses $\mathrm{\log(M_\star/M_\odot) \leq 9.0}$ are significantly different to those with higher stellar masses, with low-mass systems exhibiting lower metallicities and dust content (Figure \ref{fig:lr_mass_evol}). We find that the typical ISM conditions of low-mass galaxies remain consistent across cosmic time (Figures \ref{fig:lr_z_evol} and \ref{fig:lr_mass_evol}), with any evolution being driven primarily by galaxy growth. This is consistent with recent findings of a weakly evolving mass-metallicity relation \citep[e.g.][]{nakajima_2023, sarkar_2025, lam_2026} and has been similarly suggested in simulations \citep{ma_2016} 

Another interesting, although perhaps not surprising, finding is that the models of a young (1-10~Myr) burst of star formation can explain the main line ratios observed for most galaxies with $\mathrm{\log(M_\star/M_\odot) \leq 8.0}$ under average conditions (gas density, ionisation parameter). Meanwhile, for a significant fraction of the more massive galaxies, invoking continuous star-formation histories over at least 100~Myr are necessary to explain their simultaneous combinations of relatively high R3 and low O32 values (Figure \ref{fig:o3o2_cloudy_himass}).

In Section \ref{sec_lowmass}, we focus specifically on the low-mass galaxies in our sample and discuss the evolution of various line ratios in more detail as a function of global galaxy parameters. In particular, Figure \ref{fig:o_both_mass_evol} shows the evolution of O32 and R3, with R3 following the expected trend of an increasing line ratio with stellar mass. We find these ratios to be consistent with the mass-metallicity relation of \cite{Isobe26}, whilst our lowest mass points are systematically below the MZR from \cite{nakajima_2023}. Meanwhile, we report a much greater diversity in O32 values, with only a modest evolution with stellar mass that flattens at the lowest stellar masses. This general evolution is in broad agreement with results from the SPHINX models, but is in significant disagreement with a direct extrapolation of the trend presented by \cite{cleri_2026}. However, care should be taken to draw strong conclusions as, at the lowest masses, the majority of [O{\small{II}}] lines are undetected, producing only lower limits on O32. 

Whilst the determination of many line ratios (and the ISM conditions associated with them) are at present limited for the lowest-mass galaxies at high $z$, they are regularly observed in the local Universe. Specifically, the so-called `blueberry' \citep{Yang17} and `green-pea' \citep{cardamone_2009} galaxies exhibit extremely prominent emission lines, enhanced levels of star formation and generally compact sizes \citep[e.g.][]{kouro_2024, yang_2026} reminiscent of higher redshift sources. Blueberry galaxies in particular allow for a useful comparison, as they frequently reach $\mathrm{log(M_\star/M_\odot) <7.5}$, a regime which remains sparsely populated at high-redshift. These sources typically present log(O32) values $> 1.0$ but, consistent with our results, also exhibit significant scatter and remain consistently below the extrapolation from \cite{cleri_2026}.  Understanding to which extent this remains true for the high redshift population, where ionising conditions are expected to be more extreme, requires deep spectroscopy capable of recovering faint [O{\small II}] emission in the lowest-mass systems.

Another consistent result from these works is the intrinsic scatter of main line ratios and inferred metallicities in low-mass galaxies. Such scatter is observed from Cosmic Dawn \citep{li_2023}, to Cosmic Noon \citep{he_2026} and the more local Universe \citep{lian_2018} and is frequently attributed to stochastic star formation, stellar feedback and inflows of pristine gas \citep{forbes_2014, vanloon_2021, laseter_2025, mcclymont_2026}. Whilst gas kinematics is beyond the scope of this work, we do identify a general increase in burstiness in the lowest mass galaxies in our sample (Figure \ref{fig:burst_dist}), as expected. This increase is more prominent for the lowest mass galaxies in our sample, but both mass bins show some scatter at any given burstiness.

In Sections \ref{ssec_lr_ssfr} and \ref{ssec_burstiness}, we investigated in detail this spectral-line ratio scatter and the two-fold effect that bursty star formation can have on ISM conditions.  A clear picture emerges in which galaxies undergoing a burst have sudden changes of chemical enrichment which are part of the unstable conditions created by the burst phenomenon.  The formation of new stars can produce harder ionising spectra that boost ionisation sensitive ratios, whilst the expulsion of gas through stellar feedback can strip a galaxy of any newly formed dust and metals, reducing metallicity sensitive ratios. We posit that the scatter in line ratios is representative of observing galaxies at different stages of the baryon cycle, with the most extreme ratios corresponding to a more metal rich ISM (perhaps yet to be expelled or returning to the galaxy through inflows) illuminated by newly formed stellar populations. 

These results highlight the challenge of trying to infer gas metallicities in galaxies at the initial phases of a bursty episode, in line with findings from galaxy formation models \citep[e.g.,][]{Pallottini25, Menon26}.  Meanwhile, for low-mass sources, the lowest R3 values are produced for those objects with negative burstiness. These sources have likely undergone a single burst of star formation at an early time and have since evolved more slowly, allowing the ISM to remain particularly metal poor after the expulsion of most metals via outflows. Alternatively, such sources may reside in isolated environments, where the inflow of pristine gas maintains metal-poor ISM conditions.

In Section \ref{ssec_spec_stack}, we showed stacked spectra across four stellar mass bins and compare the presence/strengths of various emission lines. At the highest masses we see significantly brighter oxygen and sulphur lines consistent with a more metal rich ISM, as well as a lack of Ly$\alpha$ emission caused by dust attenuation. The lowest mass galaxies meanwhile exhibit prominent Balmer lines, with EWs significantly higher than that of more massive sources. This is likely related to the prevalence of very young ages amongst them. At the same time, they lack many metal indicators ([O{\small II}], [S{\small II}]) present at higher masses, as expected for being at an early stage of chemical enrichment.

To summarize, we have discussed the physical conditions of low-mass galaxies at high redshift. Whilst galaxies at stellar masses $\mathrm{log(M_\star/M_\odot) < 9.0}$ are now routinely observed with JWST's spectroscopic instruments, the lowest mass sources at these redshifts remain scarce. It is in this regime that galaxy ISM properties remain poorly constrained, with bursty star formation producing volatile conditions that make predicting ISM properties non-trivial. Conversely, a defining conclusion in this work is that the lowest metallicity sources are those at low mass with low burstiness. The mechanisms and perhaps environmental conditions that allow such sources to remain metal-poor are not currently understood and will require significant observational efforts to constrain. These sources are naturally challenging to observe, but the deep spectroscopy required is essential in helping understand the build-up (or lack thereof) of dust and metals in nascent galaxies.

\begin{acknowledgments}

This work is based on observations made with the NASA/ESA/CSA JWST. The data were obtained from the Mikulski Archive for Space Telescopes at the Space Telescope Science Institute, which is operated by the Association of Universities for Research in Astronomy, Inc., under NASA contract NAS 5-03127 for JWST. These observations are associated with the JWST programs GTO \#1208, GO \#2561, GO \#4111, GO \#6386, GO \#1837, GO \#1345, GTO \#1180/1181 GO \#1210, GO \#1286, GO \#3215, GO \#1176, GO \#1895, GO \#1963, GO \#2079, GO \#2514, GO \#3990, GO \#6434. The authors acknowledge the teams led by coPIs I. Labbe and R. Bezanson, D. Eisenstein and N. Luetzgendorf, C. Williams and P. Oesch, J. Dunlop, S. Finkelstein and M. Dickinson for developing their respective observing
programs with a zero-exclusive-access period. Some of the observations analyzed here can be accessed via \dataset[doi: 10.17909/8tdj-8n28]{\doi{10.17909/8tdj-8n28}} and \dataset[doi: 10.17909/18nv-np70]{\doi{10.17909/18nv-np70}}. Some of the analyzed data products were retrieved from the Dawn JWST Archive (DJA). DJA is an initiative of the Cosmic Dawn Center (DAWN), which is funded by the Danish National Research Foundation under grant DNRF140. Also based on observations made with the NASA/ESA Hubble Space Telescope obtained from the Space Telescope Science Institute, which is operated by the Association of Universities for Research in Astronomy, Inc., under NASA contract NAS 526555.

RAC, KIC and GD acknowledge funding from the Dutch Research Council
(NWO) through the award of the Vici Grant VI.C.212.036 and Aspasia Grant 015.017.006.

\end{acknowledgments}

%

\vspace{5mm}
\facilities{{\sl JWST, HST}}.

\software{\texttt{ASTROPY} \citep{bradley_astropyphotutils_2022}, 
\texttt{CIGALE} \citep{cigale_2019},
\texttt{CLOUDY} \citep{cloudy_13, cloudy_2025},
\texttt{GENESIS-METALLICITY} \citep{langeroodi_2026}
\texttt{MSAEXP} \citep{msaexp_zenodo},
\texttt{NUMPY} \citep{harris_array_2020},
\texttt{TOPCAT} \citep{taylor_topcat_2022},
\texttt{UNITE} \citep{hviding_unite}.
}


\appendix

\section{Differing dust-attenuation laws}\label{ap:diff_dust}

\begin{figure}
    \centering
    \includegraphics[width=1.0\linewidth]{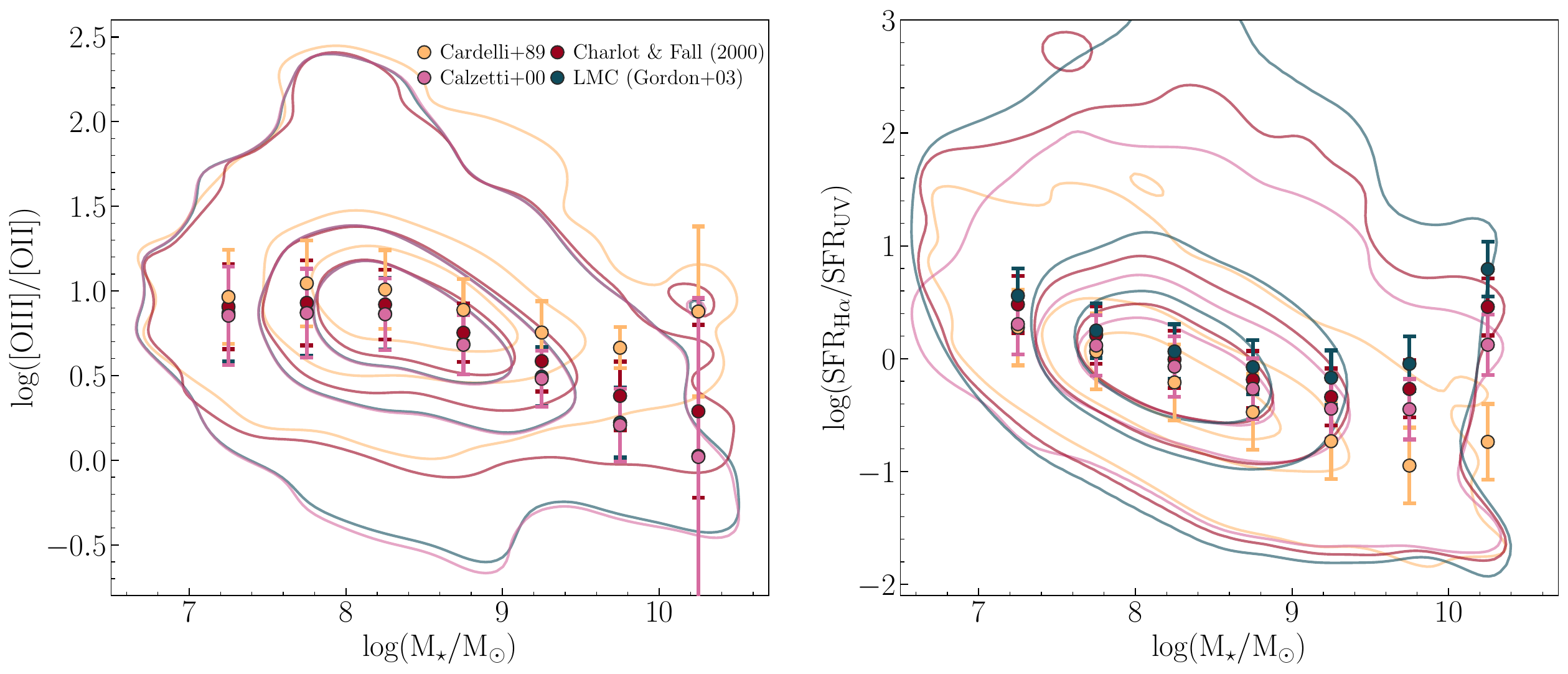}
    \caption{Dust corrected O32 (left) and burstiness (right) against stellar mass for a variety of dust extinction laws. Median points are represented by coloured circles whilst the overall distributions are shown as contours of matching colour. The dust extinction laws used are from \cite{cardelli_1989} (yellow) \cite{calzetti_2000} (pink, this work), \cite{charlot_2000} (red) and LMC curve from \cite{gordon_2003} (blue).}
    \label{fig:diff_dust}
\end{figure}

In Figure \ref{fig:diff_dust} we present the O32 and burstiness values produced under different dust extinction laws. These quantities are chosen as they are among the most sensitive to dust extinction in this work. We present values for four different extinction laws, \cite{cardelli_1989}, \cite{calzetti_2000} (as used in this work), \cite{charlot_2000} and \cite{gordon_2003}. We find excellent agreement in both quantities up to $\mathrm{log(M_\star/M_\odot) \sim 9.5}$, where results begin to diverge. Also of note is that the extinction law defined in \cite{cardelli_1989} produces systematically higher and lower results for O32 and burstiness respectively, although these results again remain with errors up to $\mathrm{log(M_\star/M_\odot) \sim 9.5}$.

\bibliography{high_oiii}{}

@ARTICLE{kewley_2019,
       author = {{Kewley}, Lisa J. and {Nicholls}, David C. and {Sutherland}, Ralph S.},
        title = "{Understanding Galaxy Evolution Through Emission Lines}",
      journal = {\araa},
         year = 2019,
        month = aug,
       volume = {57},
        pages = {511-570},
          doi = {10.1146/annurev-astro-081817-051832},
archivePrefix = {arXiv},
       eprint = {1910.09730},
 primaryClass = {astro-ph.GA},
       adsurl = {https://ui.adsabs.harvard.edu/abs/2019ARA&A..57..511K}
}

@ARTICLE{kaasinen_2018,
       author = {{Kaasinen}, Melanie and {Kewley}, Lisa and {Bian}, Fuyan and {Groves}, Brent and {Kashino}, Daichi and {Silverman}, John and {Kartaltepe}, Jeyhan},
        title = "{The ionization parameter of star-forming galaxies evolves with the specific star formation rate}",
      journal = {\mnras},
         year = 2018,
        month = jul,
       volume = {477},
       number = {4},
        pages = {5568-5589},
          doi = {10.1093/mnras/sty1012},
archivePrefix = {arXiv},
       eprint = {1804.10621},
 primaryClass = {astro-ph.GA},
       adsurl = {https://ui.adsabs.harvard.edu/abs/2018MNRAS.477.5568K}
}

@ARTICLE{cleri_2026,
       author = {{Cleri}, Nikko J. and {Lewis}, Zach J. and {Leja}, Joel and {Helton}, Jakob M. and {Burnham}, Emilie and {Curtis}, Olivia and {de Graaff}, Anna and {Hirschmann}, Michaela and {Katz}, Harley and {Maseda}, Michael V. and {McConachie}, Ian and {Plat}, Adele and {Scharre}, Lucie},
        title = "{RUBIES: The Evolution of the Ionization Parameter from 0 < z < 9}",
      journal = {arXiv e-prints},
         year = 2026,
        month = may,
          eid = {arXiv:2605.30410},
        pages = {arXiv:2605.30410},
          doi = {10.48550/arXiv.2605.30410},
archivePrefix = {arXiv},
       eprint = {2605.30410},
 primaryClass = {astro-ph.GA},
       adsurl = {https://ui.adsabs.harvard.edu/abs/2026arXiv260530410C}
}

@ARTICLE{kaasinen_2017,
       author = {{Kaasinen}, Melanie and {Bian}, Fuyan and {Groves}, Brent and {Kewley}, Lisa J. and {Gupta}, Anshu},
        title = "{The COSMOS-[O II] survey: evolution of electron density with star formation rate}",
      journal = {\mnras},
         year = 2017,
        month = mar,
       volume = {465},
       number = {3},
        pages = {3220-3234},
          doi = {10.1093/mnras/stw2827},
archivePrefix = {arXiv},
       eprint = {1611.01166},
 primaryClass = {astro-ph.GA},
       adsurl = {https://ui.adsabs.harvard.edu/abs/2017MNRAS.465.3220K}
}

@ARTICLE{topping_2025,
       author = {{Topping}, Michael W. and {Sanders}, Ryan L. and {Shapley}, Alice E. and {Pahl}, Anthony J. and {Reddy}, Naveen A. and {Stark}, Daniel P. and {Berg}, Danielle A. and {Clarke}, Leonardo and {Cullen}, Fergus and {Dunlop}, James S. and {Ellis}, Richard S. and {Schreiber}, N.~M. F{\"o}rster and {Illingworth}, Garth D. and {Jones}, Tucker and {Narayanan}, Desika and {Pettini}, Max and {Schaerer}, Daniel},
        title = "{The AURORA survey: the evolution of multiphase electron densities at high redshift}",
      journal = {\mnras},
         year = 2025,
        month = aug,
       volume = {541},
       number = {2},
        pages = {1707-1721},
          doi = {10.1093/mnras/staf903},
archivePrefix = {arXiv},
       eprint = {2502.08712},
 primaryClass = {astro-ph.GA},
       adsurl = {https://ui.adsabs.harvard.edu/abs/2025MNRAS.541.1707T}
}

@ARTICLE{perez-montero_2017,
       author = {{P{\'e}rez-Montero}, Enrique},
        title = "{Ionized Gaseous Nebulae Abundance Determination from the Direct Method}",
      journal = {\pasp},
         year = 2017,
        month = apr,
       volume = {129},
       number = {974},
        pages = {043001},
          doi = {10.1088/1538-3873/aa5abb},
archivePrefix = {arXiv},
       eprint = {1702.04255},
 primaryClass = {astro-ph.GA},
       adsurl = {https://ui.adsabs.harvard.edu/abs/2017PASP..129d3001P}
}

@ARTICLE{Tremonti04,
       author = {{Tremonti}, Christy A. and {Heckman}, Timothy M. and {Kauffmann}, Guinevere and {Brinchmann}, Jarle and {Charlot}, St{\'e}phane and {White}, Simon D.~M. and {Seibert}, Mark and {Peng}, Eric W. and {Schlegel}, David J. and {Uomoto}, Alan and {Fukugita}, Masataka and {Brinkmann}, Jon},
        title = "{The Origin of the Mass-Metallicity Relation: Insights from 53,000 Star-forming Galaxies in the Sloan Digital Sky Survey}",
      journal = {\apj},
         year = 2004,
        month = oct,
       volume = {613},
       number = {2},
        pages = {898-913},
          doi = {10.1086/423264},
archivePrefix = {arXiv},
       eprint = {astro-ph/0405537},
 primaryClass = {astro-ph},
       adsurl = {https://ui.adsabs.harvard.edu/abs/2004ApJ...613..898T}
}

@ARTICLE{andrews_2013,
       author = {{Andrews}, Brett H. and {Martini}, Paul},
        title = "{The Mass-Metallicity Relation with the Direct Method on Stacked Spectra of SDSS Galaxies}",
      journal = {\apj},
         year = 2013,
        month = mar,
       volume = {765},
       number = {2},
          eid = {140},
        pages = {140},
          doi = {10.1088/0004-637X/765/2/140},
archivePrefix = {arXiv},
       eprint = {1211.3418},
 primaryClass = {astro-ph.CO},
       adsurl = {https://ui.adsabs.harvard.edu/abs/2013ApJ...765..140A}
}

@ARTICLE{moustakas_2011,
       author = {{Moustakas}, John and {Zaritsky}, Dennis and {Brown}, Michael and {Cool}, Richard and {Dey}, Arjun and {Eisenstein}, Daniel J. and {Gonzalez}, Anthony H. and {Jannuzi}, Buell and {Jones}, Christine and {Kochanek}, Chris S. and {Murray}, Stephen S. and {Wild}, Vivienne},
        title = "{Evolution of the Stellar Mass-Metallicity Relation Since z=0.75}",
      journal = {arXiv e-prints},
         year = 2011,
        month = dec,
          eid = {arXiv:1112.3300},
        pages = {arXiv:1112.3300},
          doi = {10.48550/arXiv.1112.3300},
archivePrefix = {arXiv},
       eprint = {1112.3300},
 primaryClass = {astro-ph.CO},
       adsurl = {https://ui.adsabs.harvard.edu/abs/2011arXiv1112.3300M}
}

@ARTICLE{pagel_1979,
       author = {{Pagel}, B.~E.~J. and {Edmunds}, M.~G. and {Blackwell}, D.~E. and {Chun}, M.~S. and {Smith}, G.},
        title = "{On the composition of H II regions in southern galaxies - I. NGC 300 and 1365.}",
      journal = {\mnras},
         year = 1979,
        month = oct,
       volume = {189},
        pages = {95-113},
          doi = {10.1093/mnras/189.1.95},
       adsurl = {https://ui.adsabs.harvard.edu/abs/1979MNRAS.189...95P}
}

@ARTICLE{dopita_1986,
       author = {{Dopita}, M.~A. and {Evans}, I.~N.},
        title = "{Theoretical Models for H II Regions. II. The Extragalactic H II Region Abundance Sequence}",
      journal = {\apj},
         year = 1986,
        month = aug,
       volume = {307},
        pages = {431},
          doi = {10.1086/164432},
       adsurl = {https://ui.adsabs.harvard.edu/abs/1986ApJ...307..431D}
}

@ARTICLE{marino_2013,
       author = {{Marino}, R.~A. and {Rosales-Ortega}, F.~F. and {S{\'a}nchez}, S.~F. and {Gil de Paz}, A. and {V{\'\i}lchez}, J. and {Miralles-Caballero}, D. and {Kehrig}, C. and {P{\'e}rez-Montero}, E. and {Stanishev}, V. and {Iglesias-P{\'a}ramo}, J. and {D{\'\i}az}, A.~I. and {Castillo-Morales}, A. and {Kennicutt}, R. and {L{\'o}pez-S{\'a}nchez}, A.~R. and {Galbany}, L. and {Garc{\'\i}a-Benito}, R. and {Mast}, D. and {Mendez-Abreu}, J. and {Monreal-Ibero}, A. and {Husemann}, B. and {Walcher}, C.~J. and {Garc{\'\i}a-Lorenzo}, B. and {Masegosa}, J. and {Del Olmo Orozco}, A. and {Mour{\~a}o}, A.~M. and {Ziegler}, B. and {Moll{\'a}}, M. and {Papaderos}, P. and {S{\'a}nchez-Bl{\'a}zquez}, P. and {Gonz{\'a}lez Delgado}, R.~M. and {Falc{\'o}n-Barroso}, J. and {Roth}, M.~M. and {van de Ven}, G. and {CALIFA Team}},
        title = "{The O3N2 and N2 abundance indicators revisited: improved calibrations based on CALIFA and T$_{e}$-based literature data}",
      journal = {\aap},
         year = 2013,
        month = nov,
       volume = {559},
          eid = {A114},
        pages = {A114},
          doi = {10.1051/0004-6361/201321956},
archivePrefix = {arXiv},
       eprint = {1307.5316},
 primaryClass = {astro-ph.CO},
       adsurl = {https://ui.adsabs.harvard.edu/abs/2013A&A...559A.114M}
}

@ARTICLE{maiolino_2008,
       author = {{Maiolino}, R. and {Nagao}, T. and {Grazian}, A. and {Cocchia}, F. and {Marconi}, A. and {Mannucci}, F. and {Cimatti}, A. and {Pipino}, A. and {Ballero}, S. and {Calura}, F. and {Chiappini}, C. and {Fontana}, A. and {Granato}, G.~L. and {Matteucci}, F. and {Pastorini}, G. and {Pentericci}, L. and {Risaliti}, G. and {Salvati}, M. and {Silva}, L.},
        title = "{AMAZE. I. The evolution of the mass-metallicity relation at z > 3}",
      journal = {\aap},
         year = 2008,
        month = sep,
       volume = {488},
       number = {2},
        pages = {463-479},
          doi = {10.1051/0004-6361:200809678},
archivePrefix = {arXiv},
       eprint = {0806.2410},
 primaryClass = {astro-ph},
       adsurl = {https://ui.adsabs.harvard.edu/abs/2008A&A...488..463M}
}

@ARTICLE{curti_2020,
       author = {{Curti}, Mirko and {Maiolino}, Roberto and {Cirasuolo}, Michele and {Mannucci}, Filippo and {Williams}, Rebecca J. and {Auger}, Matt and {Mercurio}, Amata and {Hayden-Pawson}, Connor and {Cresci}, Giovanni and {Marconi}, Alessandro and {Belfiore}, Francesco and {Cappellari}, Michele and {Cicone}, Claudia and {Cullen}, Fergus and {Meneghetti}, Massimo and {Ota}, Kazuaki and {Peng}, Yingjie and {Pettini}, Max and {Swinbank}, Mark and {Troncoso}, Paulina},
        title = "{The KLEVER Survey: spatially resolved metallicity maps and gradients in a sample of 1.2 < z < 2.5 lensed galaxies}",
      journal = {\mnras},
         year = 2020,
        month = feb,
       volume = {492},
       number = {1},
        pages = {821-842},
          doi = {10.1093/mnras/stz3379},
archivePrefix = {arXiv},
       eprint = {1910.13451},
 primaryClass = {astro-ph.GA},
       adsurl = {https://ui.adsabs.harvard.edu/abs/2020MNRAS.492..821C}
}

@ARTICLE{berg_2019,
       author = {{Berg}, Danielle A. and {Erb}, Dawn K. and {Henry}, Richard B.~C. and {Skillman}, Evan D. and {McQuinn}, Kristen B.~W.},
        title = "{The Chemical Evolution of Carbon, Nitrogen, and Oxygen in Metal-poor Dwarf Galaxies}",
      journal = {\apj},
         year = 2019,
        month = mar,
       volume = {874},
       number = {1},
          eid = {93},
        pages = {93},
          doi = {10.3847/1538-4357/ab020a},
archivePrefix = {arXiv},
       eprint = {1901.08160},
 primaryClass = {astro-ph.GA},
       adsurl = {https://ui.adsabs.harvard.edu/abs/2019ApJ...874...93B}
}

@ARTICLE{peeples_2008,
       author = {{Peeples}, Molly S. and {Pogge}, Richard W. and {Stanek}, K.~Z.},
        title = "{Outliers from the Mass-Metallicity Relation. I. A Sample of Metal-Rich Dwarf Galaxies from SDSS}",
      journal = {\apj},
         year = 2008,
        month = oct,
       volume = {685},
       number = {2},
        pages = {904-914},
          doi = {10.1086/591492},
archivePrefix = {arXiv},
       eprint = {0804.2671},
 primaryClass = {astro-ph},
       adsurl = {https://ui.adsabs.harvard.edu/abs/2008ApJ...685..904P}
}

@ARTICLE{calabro_2017,
       author = {{Calabr{\`o}}, A. and {Amor{\'\i}n}, R. and {Fontana}, A. and {P{\'e}rez-Montero}, E. and {Lemaux}, B.~C. and {Ribeiro}, B. and {Bardelli}, S. and {Castellano}, M. and {Contini}, T. and {De Barros}, S. and {Garilli}, B. and {Grazian}, A. and {Guaita}, L. and {Hathi}, N.~P. and {Koekemoer}, A.~M. and {Le F{\`e}vre}, O. and {Maccagni}, D. and {Pentericci}, L. and {Schaerer}, D. and {Talia}, M. and {Tasca}, L.~A.~M. and {Zucca}, E.},
        title = "{Characterization of star-forming dwarf galaxies at 0.1 {\ensuremath{\lesssim}}z {\ensuremath{\lesssim}} 0.9 in VUDS: probing the low-mass end of the mass-metallicity relation}",
      journal = {\aap},
         year = 2017,
        month = may,
       volume = {601},
          eid = {A95},
        pages = {A95},
          doi = {10.1051/0004-6361/201629762},
archivePrefix = {arXiv},
       eprint = {1701.04418},
 primaryClass = {astro-ph.GA},
       adsurl = {https://ui.adsabs.harvard.edu/abs/2017A&A...601A..95C}
}

@ARTICLE{Erb06,
       author = {{Erb}, Dawn K. and {Shapley}, Alice E. and {Pettini}, Max and {Steidel}, Charles C. and {Reddy}, Naveen A. and {Adelberger}, Kurt L.},
        title = "{The Mass-Metallicity Relation at z>\raisebox{-0.5ex}\textasciitilde2}",
      journal = {\apj},
         year = 2006,
        month = jun,
       volume = {644},
       number = {2},
        pages = {813-828},
          doi = {10.1086/503623},
archivePrefix = {arXiv},
       eprint = {astro-ph/0602473},
 primaryClass = {astro-ph},
       adsurl = {https://ui.adsabs.harvard.edu/abs/2006ApJ...644..813E}
}

@ARTICLE{laseter24,
       author = {{Laseter}, Isaac H. and {Maseda}, Michael V. and {Curti}, Mirko and {Maiolino}, Roberto and {D'Eugenio}, Francesco and {Cameron}, Alex J. and {Looser}, Tobias J. and {Arribas}, Santiago and {Baker}, William M. and {Bhatawdekar}, Rachana and {Boyett}, Kristan and {Bunker}, Andrew J. and {Carniani}, Stefano and {Charlot}, Stephane and {Chevallard}, Jacopo and {Curtis-lake}, Emma and {Egami}, Eiichi and {Eisenstein}, Daniel J. and {Hainline}, Kevin and {Hausen}, Ryan and {Ji}, Zhiyuan and {Kumari}, Nimisha and {Perna}, Michele and {Rawle}, Tim and {Rix}, Hans-Walter and {Robertson}, Brant and {Rodr{\'\i}guez Del Pino}, Bruno and {Sandles}, Lester and {Scholtz}, Jan and {Smit}, Renske and {Tacchella}, Sandro and {{\"U}bler}, Hannah and {Williams}, Christina C. and {Willott}, Chris and {Witstok}, Joris},
        title = "{JADES: Detecting [OIII]{\ensuremath{\lambda}}4363 emitters and testing strong line calibrations in the high-z Universe with ultra-deep JWST/NIRSpec spectroscopy up to z {\ensuremath{\sim}} 9.5}",
      journal = {\aap},
         year = 2024,
        month = jan,
       volume = {681},
          eid = {A70},
        pages = {A70},
          doi = {10.1051/0004-6361/202347133},
archivePrefix = {arXiv},
       eprint = {2306.03120},
 primaryClass = {astro-ph.GA},
       adsurl = {https://ui.adsabs.harvard.edu/abs/2024A&A...681A..70L}
}

@ARTICLE{Maier14,
       author = {{Maier}, C. and {Lilly}, S.~J. and {Ziegler}, B.~L. and {Contini}, T. and {P{\'e}rez Montero}, E. and {Peng}, Y. and {Balestra}, I.},
        title = "{The Mass-Metallicity and Fundamental Metallicity Relations at z > 2 Using Very Large Telescope and Subaru Near-infrared Spectroscopy of zCOSMOS Galaxies}",
      journal = {\apj},
         year = 2014,
        month = sep,
       volume = {792},
       number = {1},
          eid = {3},
        pages = {3},
          doi = {10.1088/0004-637X/792/1/3},
archivePrefix = {arXiv},
       eprint = {1406.6069},
 primaryClass = {astro-ph.GA},
       adsurl = {https://ui.adsabs.harvard.edu/abs/2014ApJ...792....3M}
}

@ARTICLE{Endsley25,
       author = {{Endsley}, Ryan and {Chisholm}, John and {Stark}, Daniel P. and {Topping}, Michael W. and {Whitler}, Lily},
        title = "{The Burstiness of Star Formation at z {\ensuremath{\sim}} 6: A Huge Diversity in the Recent Star Formation Histories of Very UV-faint Galaxies}",
      journal = {\apj},
         year = 2025,
        month = jul,
       volume = {987},
       number = {2},
          eid = {189},
        pages = {189},
          doi = {10.3847/1538-4357/addc74},
archivePrefix = {arXiv},
       eprint = {2410.01905},
 primaryClass = {astro-ph.GA},
       adsurl = {https://ui.adsabs.harvard.edu/abs/2025ApJ...987..189E}
}

@ARTICLE{mcclymont25,
       author = {{McClymont}, William and {Tacchella}, Sandro and {Smith}, Aaron and {Kannan}, Rahul and {Puchwein}, Ewald and {Borrow}, Josh and {Garaldi}, Enrico and {Keating}, Laura and {Vogelsberger}, Mark and {Zier}, Oliver and {Shen}, Xuejian and {Popovic}, Filip and {Simmonds}, Charlotte},
        title = "{The THESAN-ZOOM project: burst, quench, repeat ─ unveiling the evolution of high-redshift galaxies along the star-forming main sequence}",
      journal = {\mnras},
         year = 2025,
        month = nov,
       volume = {544},
       number = {1},
        pages = {513-534},
          doi = {10.1093/mnras/staf1660},
archivePrefix = {arXiv},
       eprint = {2503.00106},
 primaryClass = {astro-ph.GA},
       adsurl = {https://ui.adsabs.harvard.edu/abs/2025MNRAS.544..513M}
}

@ARTICLE{planck_2016,
       author = {{Planck Collaboration} and {Ade}, P.~A.~R. and {Aghanim}, N. and {Arnaud}, M. and {Ashdown}, M. and {Aumont}, J. and {Baccigalupi}, C. and {Banday}, A.~J. and {Barreiro}, R.~B. and {Bartlett}, J.~G. and {Bartolo}, N. and {Battaner}, E. and {Battye}, R. and {Benabed}, K. and {Beno{\^\i}t}, A. and {Benoit-L{\'e}vy}, A. and {Bernard}, J.-P. and {Bersanelli}, M. and {Bielewicz}, P. and {Bock}, J.~J. and {Bonaldi}, A. and {Bonavera}, L. and {Bond}, J.~R. and {Borrill}, J. and {Bouchet}, F.~R. and {Boulanger}, F. and {Bucher}, M. and {Burigana}, C. and {Butler}, R.~C. and {Calabrese}, E. and {Cardoso}, J.-F. and {Catalano}, A. and {Challinor}, A. and {Chamballu}, A. and {Chary}, R.-R. and {Chiang}, H.~C. and {Chluba}, J. and {Christensen}, P.~R. and {Church}, S. and {Clements}, D.~L. and {Colombi}, S. and {Colombo}, L.~P.~L. and {Combet}, C. and {Coulais}, A. and {Crill}, B.~P. and {Curto}, A. and {Cuttaia}, F. and {Danese}, L. and {Davies}, R.~D. and {Davis}, R.~J. and {de Bernardis}, P. and {de Rosa}, A. and {de Zotti}, G. and {Delabrouille}, J. and {D{\'e}sert}, F.-X. and {Di Valentino}, E. and {Dickinson}, C. and {Diego}, J.~M. and {Dolag}, K. and {Dole}, H. and {Donzelli}, S. and {Dor{\'e}}, O. and {Douspis}, M. and {Ducout}, A. and {Dunkley}, J. and {Dupac}, X. and {Efstathiou}, G. and {Elsner}, F. and {En{\ss}lin}, T.~A. and {Eriksen}, H.~K. and {Farhang}, M. and {Fergusson}, J. and {Finelli}, F. and {Forni}, O. and {Frailis}, M. and {Fraisse}, A.~A. and {Franceschi}, E. and {Frejsel}, A. and {Galeotta}, S. and {Galli}, S. and {Ganga}, K. and {Gauthier}, C. and {Gerbino}, M. and {Ghosh}, T. and {Giard}, M. and {Giraud-H{\'e}raud}, Y. and {Giusarma}, E. and {Gjerl{\o}w}, E. and {Gonz{\'a}lez-Nuevo}, J. and {G{\'o}rski}, K.~M. and {Gratton}, S. and {Gregorio}, A. and {Gruppuso}, A. and {Gudmundsson}, J.~E. and {Hamann}, J. and {Hansen}, F.~K. and {Hanson}, D. and {Harrison}, D.~L. and {Helou}, G. and {Henrot-Versill{\'e}}, S. and {Hern{\'a}ndez-Monteagudo}, C. and {Herranz}, D. and {Hildebrandt}, S.~R. and {Hivon}, E. and {Hobson}, M. and {Holmes}, W.~A. and {Hornstrup}, A. and {Hovest}, W. and {Huang}, Z. and {Huffenberger}, K.~M. and {Hurier}, G. and {Jaffe}, A.~H. and {Jaffe}, T.~R. and {Jones}, W.~C. and {Juvela}, M. and {Keih{\"a}nen}, E. and {Keskitalo}, R. and {Kisner}, T.~S. and {Kneissl}, R. and {Knoche}, J. and {Knox}, L. and {Kunz}, M. and {Kurki-Suonio}, H. and {Lagache}, G. and {L{\"a}hteenm{\"a}ki}, A. and {Lamarre}, J.-M. and {Lasenby}, A. and {Lattanzi}, M. and {Lawrence}, C.~R. and {Leahy}, J.~P. and {Leonardi}, R. and {Lesgourgues}, J. and {Levrier}, F. and {Lewis}, A. and {Liguori}, M. and {Lilje}, P.~B. and {Linden-V{\o}rnle}, M. and {L{\'o}pez-Caniego}, M. and {Lubin}, P.~M. and {Mac{\'\i}as-P{\'e}rez}, J.~F. and {Maggio}, G. and {Maino}, D. and {Mandolesi}, N. and {Mangilli}, A. and {Marchini}, A. and {Maris}, M. and {Martin}, P.~G. and {Martinelli}, M. and {Mart{\'\i}nez-Gonz{\'a}lez}, E. and {Masi}, S. and {Matarrese}, S. and {McGehee}, P. and {Meinhold}, P.~R. and {Melchiorri}, A. and {Melin}, J.-B. and {Mendes}, L. and {Mennella}, A. and {Migliaccio}, M. and {Millea}, M. and {Mitra}, S. and {Miville-Desch{\^e}nes}, M.-A. and {Moneti}, A. and {Montier}, L. and {Morgante}, G. and {Mortlock}, D. and {Moss}, A. and {Munshi}, D. and {Murphy}, J.~A. and {Naselsky}, P. and {Nati}, F. and {Natoli}, P. and {Netterfield}, C.~B. and {N{\o}rgaard-Nielsen}, H.~U. and {Noviello}, F. and {Novikov}, D. and {Novikov}, I. and {Oxborrow}, C.~A. and {Paci}, F. and {Pagano}, L. and {Pajot}, F. and {Paladini}, R. and {Paoletti}, D. and {Partridge}, B. and {Pasian}, F. and {Patanchon}, G. and {Pearson}, T.~J. and {Perdereau}, O. and {Perotto}, L. and {Perrotta}, F. and {Pettorino}, V. and {Piacentini}, F. and {Piat}, M. and {Pierpaoli}, E. and {Pietrobon}, D. and {Plaszczynski}, S. and {Pointecouteau}, E. and {Polenta}, G. and {Popa}, L. and {Pratt}, G.~W. and {Pr{\'e}zeau}, G.},
        title = "{Planck 2015 results. XIII. Cosmological parameters}",
      journal = {\aap},
         year = 2016,
        month = sep,
       volume = {594},
          eid = {A13},
        pages = {A13},
          doi = {10.1051/0004-6361/201525830},
archivePrefix = {arXiv},
       eprint = {1502.01589},
 primaryClass = {astro-ph.CO},
       adsurl = {https://ui.adsabs.harvard.edu/abs/2016A&A...594A..13P}
}

@ARTICLE{chabrier_2003,
       author = {{Chabrier}, Gilles},
        title = "{Galactic Stellar and Substellar Initial Mass Function}",
      journal = {\pasp},
         year = 2003,
        month = jul,
       volume = {115},
       number = {809},
        pages = {763-795},
          doi = {10.1086/376392},
archivePrefix = {arXiv},
       eprint = {astro-ph/0304382},
 primaryClass = {astro-ph},
       adsurl = {https://ui.adsabs.harvard.edu/abs/2003PASP..115..763C}
}

@software{msaexp_zenodo,
  author       = {Brammer, Gabriel},
  title        = {msaexp: NIRSpec analyis tools},
  month        = sep,
  year         = 2023,
  publisher    = {Zenodo},
  version      = {0.6.17},
  doi          = {10.5281/zenodo.8319596},
  url          = {https://doi.org/10.5281/zenodo.8319596},
}

@ARTICLE{canucs_phot,
       author = {{Sarrouh}, Ghassan T.~E. and {Asada}, Yoshihisa and {Martis}, Nicholas S. and {Willott}, Chris J. and {Iyer}, Kartheik G. and {Noirot}, Ga{\"e}l and {Muzzin}, Adam and {Sawicki}, Marcin and {Brammer}, Gabriel and {Desprez}, Guillaume and {Rihtar{\v{s}}i{\v{c}}}, Gregor and {Zabl}, Johannes and {Abraham}, Roberto and {Brada{\v{c}}}, Maru{\v{s}}a and {Doyon}, Ren{\'e} and {Antwi-Danso}, Jacqueline and {Berek}, Samantha and {Brown}, Westley and {Estrada-Carpenter}, Vince and {Favaro}, Jeremy and {Felicioni}, Giordano and {Forrest}, Ben and {Gaspar}, Gaia and {Gould}, Katriona M.~L. and {Gledhill}, Rachel and {Harshan}, Anishya and {Jahan}, Nusrath and {Jagga}, Naadiyah and {Jude{\v{z}}}, Jon and {Marchesini}, Danilo and {Markov}, Vladan and {Matharu}, Jasleen and {MacFarland}, Shannon and {Merchant}, Maya and {M{\'e}rida}, Rosa M. and {Mowla}, Lamiya and {Myers}, Katherine and {Omori}, Kiyoaki C. and {Pacifici}, Camilla and {Ravindranath}, Swara and {Robbins}, Luke and {Strait}, Victoria and {Sok}, Visal and {Tan}, Vivian Yun Yan and {Tripodi}, Roberta and {Wilson}, Gillian and {Withers}, Sunna},
        title = "{CANUCS/Technicolor Data Release 1: Imaging, Photometry, Slit Spectroscopy, and Stellar Population Parameters}",
      journal = {\apjs},
         year = 2026,
        month = jan,
       volume = {282},
       number = {1},
          eid = {3},
        pages = {3},
          doi = {10.3847/1538-4365/ae1611},
archivePrefix = {arXiv},
       eprint = {2506.21685},
 primaryClass = {astro-ph.GA},
       adsurl = {https://ui.adsabs.harvard.edu/abs/2026ApJS..282....3S}
}

@article{canucs_spec,
    author = {Desprez, Guillaume and Martis, Nicholas S and Asada, Yoshihisa and Sawicki, Marcin and Willott, Chris J and Muzzin, Adam and Abraham, Roberto G and Bradač, Maruša and Brammer, Gabe and Estrada-Carpenter, Vicente and Iyer, Kartheik G and Matharu, Jasleen and Mowla, Lamiya and Noirot, Gaël and Sarrouh, Ghassan T E and Strait, Victoria and Gledhill, Rachel and Rihtaršič, Gregor},
    title = {ΛCDM not dead yet: massive high-z Balmer break galaxies are less common than previously reported},
    journal = {Monthly Notices of the Royal Astronomical Society},
    volume = {530},
    number = {3},
    pages = {2935-2952},
    year = {2024},
    month = {05},
    issn = {0035-8711},
    doi = {10.1093/mnras/stae1084},
    url = {https://doi.org/10.1093/mnras/stae1084},
    eprint = {https://academic.oup.com/mnras/article-pdf/530/3/2935/57361748/stae1084.pdf},
}

@ARTICLE{CLASH,
       author = {{Postman}, Marc and {Coe}, Dan and {Ben{\'\i}tez}, Narciso and {Bradley}, Larry and {Broadhurst}, Tom and {Donahue}, Megan and {Ford}, Holland and {Graur}, Or and {Graves}, Genevieve and {Jouvel}, Stephanie and {Koekemoer}, Anton and {Lemze}, Doron and {Medezinski}, Elinor and {Molino}, Alberto and {Moustakas}, Leonidas and {Ogaz}, Sara and {Riess}, Adam and {Rodney}, Steve and {Rosati}, Piero and {Umetsu}, Keiichi and {Zheng}, Wei and {Zitrin}, Adi and {Bartelmann}, Matthias and {Bouwens}, Rychard and {Czakon}, Nicole and {Golwala}, Sunil and {Host}, Ole and {Infante}, Leopoldo and {Jha}, Saurabh and {Jimenez-Teja}, Yolanda and {Kelson}, Daniel and {Lahav}, Ofer and {Lazkoz}, Ruth and {Maoz}, Dani and {McCully}, Curtis and {Melchior}, Peter and {Meneghetti}, Massimo and {Merten}, Julian and {Moustakas}, John and {Nonino}, Mario and {Patel}, Brandon and {Reg{\"o}s}, Enik{\"o} and {Sayers}, Jack and {Seitz}, Stella and {Van der Wel}, Arjen},
        title = "{The Cluster Lensing and Supernova Survey with Hubble: An Overview}",
      journal = {\apjs},
         year = 2012,
        month = apr,
       volume = {199},
       number = {2},
          eid = {25},
        pages = {25},
          doi = {10.1088/0067-0049/199/2/25},
archivePrefix = {arXiv},
       eprint = {1106.3328},
 primaryClass = {astro-ph.CO},
       adsurl = {https://ui.adsabs.harvard.edu/abs/2012ApJS..199...25P}
}

@ARTICLE{F_Fields,
       author = {{Lotz}, J.~M. and {Koekemoer}, A. and {Coe}, D. and {Grogin}, N. and {Capak}, P. and {Mack}, J. and {Anderson}, J. and {Avila}, R. and {Barker}, E.~A. and {Borncamp}, D. and {Brammer}, G. and {Durbin}, M. and {Gunning}, H. and {Hilbert}, B. and {Jenkner}, H. and {Khandrika}, H. and {Levay}, Z. and {Lucas}, R.~A. and {MacKenty}, J. and {Ogaz}, S. and {Porterfield}, B. and {Reid}, N. and {Robberto}, M. and {Royle}, P. and {Smith}, L.~J. and {Storrie-Lombardi}, L.~J. and {Sunnquist}, B. and {Surace}, J. and {Taylor}, D.~C. and {Williams}, R. and {Bullock}, J. and {Dickinson}, M. and {Finkelstein}, S. and {Natarajan}, P. and {Richard}, J. and {Robertson}, B. and {Tumlinson}, J. and {Zitrin}, A. and {Flanagan}, K. and {Sembach}, K. and {Soifer}, B.~T. and {Mountain}, M.},
        title = "{The Frontier Fields: Survey Design and Initial Results}",
      journal = {\apj},
         year = 2017,
        month = mar,
       volume = {837},
       number = {1},
          eid = {97},
        pages = {97},
          doi = {10.3847/1538-4357/837/1/97},
archivePrefix = {arXiv},
       eprint = {1605.06567},
 primaryClass = {astro-ph.GA},
       adsurl = {https://ui.adsabs.harvard.edu/abs/2017ApJ...837...97L}
}

@ARTICLE{UNCOVER,
       author = {{Bezanson}, Rachel and {Labbe}, Ivo and {Whitaker}, Katherine E. and {Leja}, Joel and {Price}, Sedona H. and {Franx}, Marijn and {Brammer}, Gabriel and {Marchesini}, Danilo and {Zitrin}, Adi and {Wang}, Bingjie and {Weaver}, John R. and {Furtak}, Lukas J. and {Atek}, Hakim and {Coe}, Dan and {Cutler}, Sam E. and {Dayal}, Pratika and {van Dokkum}, Pieter and {Feldmann}, Robert and {F{\"o}rster Schreiber}, Natascha M. and {Fujimoto}, Seiji and {Geha}, Marla and {Glazebrook}, Karl and {de Graaff}, Anna and {Greene}, Jenny E. and {Juneau}, St{\'e}phanie and {Kassin}, Susan and {Kriek}, Mariska and {Khullar}, Gourav and {Maseda}, Michael and {Mowla}, Lamiya A. and {Muzzin}, Adam and {Nanayakkara}, Themiya and {Nelson}, Erica J. and {Oesch}, Pascal A. and {Pacifici}, Camilla and {Pan}, Richard and {Papovich}, Casey and {Setton}, David J. and {Shapley}, Alice E. and {Smit}, Renske and {Stefanon}, Mauro and {Taylor}, Edward N. and {Williams}, Christina C.},
        title = "{The JWST UNCOVER Treasury Survey: Ultradeep NIRSpec and NIRCam Observations before the Epoch of Reionization}",
      journal = {\apj},
         year = 2024,
        month = oct,
       volume = {974},
       number = {1},
          eid = {92},
        pages = {92},
          doi = {10.3847/1538-4357/ad66cf},
archivePrefix = {arXiv},
       eprint = {2212.04026},
 primaryClass = {astro-ph.GA},
       adsurl = {https://ui.adsabs.harvard.edu/abs/2024ApJ...974...92B}
}

@ARTICLE{MegaScience,
       author = {{Suess}, Katherine A. and {Weaver}, John R. and {Price}, Sedona H. and {Pan}, Richard and {Wang}, Bingjie and {Bezanson}, Rachel and {Brammer}, Gabriel and {Cutler}, Sam E. and {Labb{\'e}}, Ivo and {Leja}, Joel and {Williams}, Christina C. and {Whitaker}, Katherine E. and {Atek}, Hakim and {Dayal}, Pratika and {de Graaff}, Anna and {Feldmann}, Robert and {Franx}, Marijn and {Fudamoto}, Yoshinobu and {Fujimoto}, Seiji and {Furtak}, Lukas J. and {Goulding}, Andy D. and {Greene}, Jenny E. and {Khullar}, Gourav and {Kokorev}, Vasily and {Kriek}, Mariska and {Lorenz}, Brian and {Marchesini}, Danilo and {Maseda}, Michael V. and {Matthee}, Jorryt and {Miller}, Tim B. and {Mitsuhashi}, Ikki and {Mowla}, Lamiya A. and {Muzzin}, Adam and {Naidu}, Rohan P. and {Nanayakkara}, Themiya and {Nelson}, Erica J. and {Oesch}, Pascal A. and {Setton}, David J. and {Shipley}, Heath and {Smit}, Renske and {Spilker}, Justin S. and {van Dokkum}, Pieter and {Zitrin}, Adi},
        title = "{Medium Bands, Mega Science: A JWST/NIRCam Medium-band Imaging Survey of A2744}",
      journal = {\apj},
         year = 2024,
        month = nov,
       volume = {976},
       number = {1},
          eid = {101},
        pages = {101},
          doi = {10.3847/1538-4357/ad75fe},
archivePrefix = {arXiv},
       eprint = {2404.13132},
 primaryClass = {astro-ph.GA},
       adsurl = {https://ui.adsabs.harvard.edu/abs/2024ApJ...976..101S}
}

@ARTICLE{UNCOVER_spec,
       author = {{Price}, Sedona H. and {Bezanson}, Rachel and {Labbe}, Ivo and {Furtak}, Lukas J. and {de Graaff}, Anna and {Greene}, Jenny E. and {Kokorev}, Vasily and {Setton}, David J. and {Suess}, Katherine A. and {Brammer}, Gabriel and {Cutler}, Sam E. and {Leja}, Joel and {Pan}, Richard and {Wang}, Bingjie and {Weaver}, John R. and {Whitaker}, Katherine E. and {Atek}, Hakim and {Burgasser}, Adam J. and {Chemerynska}, Iryna and {Dayal}, Pratika and {Feldmann}, Robert and {F{\"o}rster Schreiber}, Natascha M. and {Fudamoto}, Yoshinobu and {Fujimoto}, Seiji and {Glazebrook}, Karl and {Goulding}, Andy D. and {Khullar}, Gourav and {Kriek}, Mariska and {Marchesini}, Danilo and {Maseda}, Michael V. and {Miller}, Tim B. and {Muzzin}, Adam and {Nanayakkara}, Themiya and {Nelson}, Erica and {Oesch}, Pascal A. and {Shipley}, Heath and {Smit}, Renske and {Taylor}, Edward N. and {Dokkum}, Pieter van and {Williams}, Christina C. and {Zitrin}, Adi},
        title = "{The UNCOVER Survey: First Release of Ultradeep JWST/NIRSpec PRISM Spectra for {\ensuremath{\sim}}700 Galaxies from z {\ensuremath{\sim}} 0.3─13 in A2744}",
      journal = {\apj},
         year = 2025,
        month = mar,
       volume = {982},
       number = {1},
          eid = {51},
        pages = {51},
          doi = {10.3847/1538-4357/adaec1},
archivePrefix = {arXiv},
       eprint = {2408.03920},
 primaryClass = {astro-ph.GA},
       adsurl = {https://ui.adsabs.harvard.edu/abs/2025ApJ...982...51P}
}

@ARTICLE{UNCOVER_magnif,
       author = {{Furtak}, Lukas J. and {Zitrin}, Adi and {Weaver}, John R. and {Atek}, Hakim and {Bezanson}, Rachel and {Labb{\'e}}, Ivo and {Whitaker}, Katherine E. and {Leja}, Joel and {Price}, Sedona H. and {Brammer}, Gabriel B. and {Wang}, Bingjie and {Marchesini}, Danilo and {Pan}, Richard and {Dayal}, Pratika and {van Dokkum}, Pieter and {Feldmann}, Robert and {Fujimoto}, Seiji and {Franx}, Marijn and {Khullar}, Gourav and {Nelson}, Erica J. and {Mowla}, Lamiya A.},
        title = "{UNCOVERing the extended strong lensing structures of Abell 2744 with the deepest JWST imaging}",
      journal = {\mnras},
         year = 2023,
        month = aug,
       volume = {523},
       number = {3},
        pages = {4568-4582},
          doi = {10.1093/mnras/stad1627},
archivePrefix = {arXiv},
       eprint = {2212.04381},
 primaryClass = {astro-ph.GA},
       adsurl = {https://ui.adsabs.harvard.edu/abs/2023MNRAS.523.4568F}
}

@ARTICLE{UNCOVER_magnif_2,
       author = {{Weaver}, John R. and {Cutler}, Sam E. and {Pan}, Richard and {Whitaker}, Katherine E. and {Labb{\'e}}, Ivo and {Price}, Sedona H. and {Bezanson}, Rachel and {Brammer}, Gabriel and {Marchesini}, Danilo and {Leja}, Joel and {Wang}, Bingjie and {Furtak}, Lukas J. and {Zitrin}, Adi and {Atek}, Hakim and {Chemerynska}, Iryna and {Coe}, Dan and {Dayal}, Pratika and {van Dokkum}, Pieter and {Feldmann}, Robert and {F{\"o}rster Schreiber}, Natascha M. and {Franx}, Marijn and {Fujimoto}, Seiji and {Fudamoto}, Yoshinobu and {Glazebrook}, Karl and {de Graaff}, Anna and {Greene}, Jenny E. and {Juneau}, St{\'e}phanie and {Kassin}, Susan and {Kriek}, Mariska and {Khullar}, Gourav and {Maseda}, Michael V. and {Mowla}, Lamiya A. and {Muzzin}, Adam and {Nanayakkara}, Themiya and {Nelson}, Erica J. and {Oesch}, Pascal A. and {Pacifici}, Camilla and {Papovich}, Casey and {Setton}, David J. and {Shapley}, Alice E. and {Shipley}, Heath V. and {Smit}, Renske and {Stefanon}, Mauro and {Taylor}, Edward N. and {Weibel}, Andrea and {Williams}, Christina C.},
        title = "{The UNCOVER Survey: A First-look HST + JWST Catalog of 60,000 Galaxies near A2744 and beyond}",
      journal = {\apjs},
         year = 2024,
        month = jan,
       volume = {270},
       number = {1},
          eid = {7},
        pages = {7},
          doi = {10.3847/1538-4365/ad07e0},
archivePrefix = {arXiv},
       eprint = {2301.02671},
 primaryClass = {astro-ph.GA},
       adsurl = {https://ui.adsabs.harvard.edu/abs/2024ApJS..270....7W}
}

@INPROCEEDINGS{GOODS_legacy,
       author = {{Dickinson}, Mark and {Giavalisco}, Mauro and {GOODS Team}},
        title = "{The Great Observatories Origins Deep Survey}",
    booktitle = {The Mass of Galaxies at Low and High Redshift},
         year = 2003,
       editor = {{Bender}, Ralf and {Renzini}, Alvio},
        month = jan,
        pages = {324},
          doi = {10.1007/10899892_78},
archivePrefix = {arXiv},
       eprint = {astro-ph/0204213},
 primaryClass = {astro-ph},
       adsurl = {https://ui.adsabs.harvard.edu/abs/2003mglh.conf..324D}
}

@ARTICLE{GOODS_legacy_2,
       author = {{Giavalisco}, M. and {Ferguson}, H.~C. and {Koekemoer}, A.~M. and {Dickinson}, M. and {Alexander}, D.~M. and {Bauer}, F.~E. and {Bergeron}, J. and {Biagetti}, C. and {Brandt}, W.~N. and {Casertano}, S. and {Cesarsky}, C. and {Chatzichristou}, E. and {Conselice}, C. and {Cristiani}, S. and {Da Costa}, L. and {Dahlen}, T. and {de Mello}, D. and {Eisenhardt}, P. and {Erben}, T. and {Fall}, S.~M. and {Fassnacht}, C. and {Fosbury}, R. and {Fruchter}, A. and {Gardner}, J.~P. and {Grogin}, N. and {Hook}, R.~N. and {Hornschemeier}, A.~E. and {Idzi}, R. and {Jogee}, S. and {Kretchmer}, C. and {Laidler}, V. and {Lee}, K.~S. and {Livio}, M. and {Lucas}, R. and {Madau}, P. and {Mobasher}, B. and {Moustakas}, L.~A. and {Nonino}, M. and {Padovani}, P. and {Papovich}, C. and {Park}, Y. and {Ravindranath}, S. and {Renzini}, A. and {Richardson}, M. and {Riess}, A. and {Rosati}, P. and {Schirmer}, M. and {Schreier}, E. and {Somerville}, R.~S. and {Spinrad}, H. and {Stern}, D. and {Stiavelli}, M. and {Strolger}, L. and {Urry}, C.~M. and {Vandame}, B. and {Williams}, R. and {Wolf}, C.},
        title = "{The Great Observatories Origins Deep Survey: Initial Results from Optical and Near-Infrared Imaging}",
      journal = {\apjl},
         year = 2004,
        month = jan,
       volume = {600},
       number = {2},
        pages = {L93-L98},
          doi = {10.1086/379232},
archivePrefix = {arXiv},
       eprint = {astro-ph/0309105},
 primaryClass = {astro-ph},
       adsurl = {https://ui.adsabs.harvard.edu/abs/2004ApJ...600L..93G}
}

@ARTICLE{JADES_DR5_1,
       author = {{Robertson}, Brant E. and {Johnson}, Benjamin D. and {Tacchella}, Sandro and {Eisenstein}, Daniel J. and {Hainline}, Kevin and {Alberts}, Stacey and {Arribas}, Santiago and {Baker}, William M. and {Bunker}, Andrew J. and {Cameron}, Alex J. and {Carniani}, Stefano and {Carreira}, Courtney and {Chevallard}, Jacopo and {Circosta}, Chiara and {Curtis-Lake}, Emma and {Danhaive}, A. Lola and {Duan}, Qiao and {Egami}, Eiichi and {Hausen}, Ryan and {Helton}, Jakob M. and {Ji}, Zhiyuan and {Maiolino}, Roberto and {P{\'e}rez-Gonz{\'a}lez}, Pablo G. and {Pusk{\'a}s}, D{\'a}vid and {Rieke}, Marcia and {Rinaldi}, Pierluigi and {Sun}, Fengwu and {Sun}, Yang and {{\"U}bler}, Hannah and {Trussler}, James A.~A. and {Villanueva}, Natalia C. and {Whitler}, Lily and {Williams}, Christina C. and {Willmer}, Christopher N.~A. and {Willott}, Chris and {Wu}, Zihao and {Zhu}, Yongda},
        title = "{JWST Advanced Deep Extragalactic Survey (JADES) Data Release 5: Photometric Catalog}",
      journal = {arXiv e-prints},
         year = 2026,
        month = jan,
          eid = {arXiv:2601.15956},
        pages = {arXiv:2601.15956},
          doi = {10.48550/arXiv.2601.15956},
archivePrefix = {arXiv},
       eprint = {2601.15956},
 primaryClass = {astro-ph.GA},
       adsurl = {https://ui.adsabs.harvard.edu/abs/2026arXiv260115956R}
}

@ARTICLE{JADES_overview,
       author = {{Eisenstein}, Daniel J. and {Willott}, Chris and {Alberts}, Stacey and {Arribas}, Santiago and {Bonaventura}, Nina and {Bunker}, Andrew J. and {Cameron}, Alex J. and {Carniani}, Stefano and {Charlot}, Stephane and {Curtis-Lake}, Emma and {D'Eugenio}, Francesco and {Ferruit}, Pierre and {Giardino}, Giovanna and {Hainline}, Kevin and {Hausen}, Ryan and {Jakobsen}, Peter and {Johnson}, Benjamin D. and {Maiolino}, Roberto and {Rauscher}, Bernard J. and {Rieke}, Marcia and {Rieke}, George and {Rix}, Hans-Walter and {Robertson}, Brant and {Stark}, Daniel P. and {Tacchella}, Sandro and {Williams}, Christina C. and {Willmer}, Christopher N.~A. and {Baker}, William M. and {Baum}, Stefi and {Bhatawdekar}, Rachana and {Boyett}, Kristan and {Chen}, Zuyi and {Chevallard}, Jacopo and {Circosta}, Chiara and {Curti}, Mirko and {Danhaive}, A. Lola and {DeCoursey}, Christa and {Endsley}, Ryan and {de Graaff}, Anna and {Dressler}, Alan and {Egami}, Eiichi and {Helton}, Jakob M. and {Hviding}, Raphael E. and {Ji}, Zhiyuan and {Jones}, Gareth C. and {Kumari}, Nimisha and {L{\"u}tzgendorf}, Nora and {Laseter}, Isaac and {Looser}, Tobias J. and {Lyu}, Jianwei and {Maseda}, Michael V. and {Nelson}, Erica and {Parlanti}, Eleonora and {Perna}, Michele and {Pusk{\'a}s}, D{\'a}vid and {Rawle}, Tim and {Rodr{\'\i}guez Del Pino}, Bruno and {Rujopakarn}, Wiphu and {Sandles}, Lester and {Saxena}, Aayush and {Scholtz}, Jan and {Sharpe}, Katherine and {Shivaei}, Irene and {Silcock}, Maddie S. and {Simmonds}, Charlotte and {Skarbinski}, Maya and {Smit}, Renske and {Stone}, Meredith and {Suess}, Katherine A. and {Sun}, Fengwu and {Tang}, Mengtao and {Topping}, Michael W. and {{\"U}bler}, Hannah and {Villanueva}, Natalia C. and {Wallace}, Imaan E.~B. and {Whitler}, Lily and {Witstok}, Joris and {Woodrum}, Charity},
        title = "{Overview of the JWST Advanced Deep Extragalactic Survey (JADES)}",
      journal = {\apjs},
         year = 2026,
        month = mar,
       volume = {283},
       number = {1},
          eid = {6},
        pages = {6},
          doi = {10.3847/1538-4365/ae3163},
archivePrefix = {arXiv},
       eprint = {2306.02465},
 primaryClass = {astro-ph.GA},
       adsurl = {https://ui.adsabs.harvard.edu/abs/2026ApJS..283....6E}
}

@ARTICLE{JADES_DR5_2,
       author = {{Johnson}, Benjamin D. and {Robertson}, Brant E. and {Eisenstein}, Daniel J. and {Tacchella}, Sandro and {Pusk{\'a}s}, D{\'a}vid and {Duan}, Qiao and {Wu}, Zihao and {Hainline}, Kevin and {Rieke}, Marcia and {Willott}, Chris and {Willmer}, Christopher N.~A. and {Trussler}, James A.~A. and {Alberts}, Stacey and {Arribas}, Santiago and {Baker}, William M. and {Bunker}, Andrew J. and {Cameron}, Alex J. and {Carniani}, Stefano and {Carreira}, Courtney and {Cargile}, Phillip A. and {Curtis-Lake}, Emma and {Egami}, Eiichi and {Hausen}, Ryan and {Helton}, Jakob M. and {Ji}, Zhiyuan and {Maiolino}, Roberto and {P{\'e}rez-Gonz{\'a}lez}, Pablo G. and {Rinaldi}, Pierluigi and {Sun}, Fengwu and {Sun}, Yang and {Villanueva}, Natalia C. and {Williams}, Christina C. and {Zhu}, Yongda},
        title = "{JWST Advanced Deep Extragalactic Survey (JADES) Data Release 5: NIRCam Imaging in GOODS-S and GOODS-N}",
      journal = {arXiv e-prints},
         year = 2026,
        month = jan,
          eid = {arXiv:2601.15954},
        pages = {arXiv:2601.15954},
          doi = {10.48550/arXiv.2601.15954},
archivePrefix = {arXiv},
       eprint = {2601.15954},
 primaryClass = {astro-ph.GA},
       adsurl = {https://ui.adsabs.harvard.edu/abs/2026arXiv260115954J}
}

@ARTICLE{GOODS_HST_1,
       author = {{Illingworth}, Garth and {Magee}, Daniel and {Bouwens}, Rychard and {Oesch}, Pascal and {Labbe}, Ivo and {van Dokkum}, Pieter and {Whitaker}, Katherine and {Holden}, Bradford and {Franx}, Marijn and {Gonzalez}, Valentino},
        title = "{The Hubble Legacy Fields (HLF-GOODS-S) v1.5 Data Products: Combining 2442 Orbits of GOODS-S/CDF-S Region ACS and WFC3/IR Images}",
      journal = {arXiv e-prints},
         year = 2016,
        month = jun,
          eid = {arXiv:1606.00841},
        pages = {arXiv:1606.00841},
          doi = {10.48550/arXiv.1606.00841},
archivePrefix = {arXiv},
       eprint = {1606.00841},
 primaryClass = {astro-ph.GA},
       adsurl = {https://ui.adsabs.harvard.edu/abs/2016arXiv160600841I}
}

@ARTICLE{GOODS_HST_2,
       author = {{Whitaker}, Katherine E. and {Ashas}, Mohammad and {Illingworth}, Garth and {Magee}, Daniel and {Leja}, Joel and {Oesch}, Pascal and {van Dokkum}, Pieter and {Mowla}, Lamiya and {Bouwens}, Rychard and {Franx}, Marijn and {Holden}, Bradford and {Labb{\'e}}, Ivo and {Rafelski}, Marc and {Teplitz}, Harry and {Gonzalez}, Valentino},
        title = "{The Hubble Legacy Field GOODS-S Photometric Catalog}",
      journal = {\apjs},
         year = 2019,
        month = sep,
       volume = {244},
       number = {1},
          eid = {16},
        pages = {16},
          doi = {10.3847/1538-4365/ab3853},
archivePrefix = {arXiv},
       eprint = {1908.05682},
 primaryClass = {astro-ph.GA},
       adsurl = {https://ui.adsabs.harvard.edu/abs/2019ApJS..244...16W}
}

@ARTICLE{JADES_DR4_1,
       author = {{Curtis-Lake}, Emma and {Cameron}, Alex J. and {Bunker}, Andrew J. and {Scholtz}, Jan and {Carniani}, Stefano and {Parlanti}, Eleonora and {D'Eugenio}, Francesco and {Jakobsen}, Peter and {Willmer}, Christopher N.~A. and {Arribas}, Santiago and {Baker}, William M. and {Charlot}, St{\'e}phane and {Chevallard}, Jacopo and {Circosta}, Chiara and {Curti}, Mirko and {Eisenstein}, Daniel J. and {Hainline}, Kevin and {Ji}, Zhiyuan and {Johnson}, Benjamin D. and {Jones}, Gareth C. and {Maiolino}, Roberto and {Maseda}, Michael V. and {P{\'e}rez-Gonz{\'a}lez}, Pablo G. and {Rawle}, Tim and {Rieke}, Marcia and {Rinaldi}, Pierluigi and {Robertson}, Brant and {Rodr{\'\i}gez Del Pino}, Bruno and {Saxena}, Aayush and {Shivaei}, Irene and {Smit}, Renske and {Tacchella}, Sandro and {{\"U}bler}, Hannah and {Venturi}, Giacomo and {Williams}, Christina C. and {Willott}, Chris and {Duan}, Qiao},
        title = "{JADES Data Release 4 Paper I: Sample Selection, Observing Strategy and Redshifts of the complete spectroscopic sample}",
      journal = {arXiv e-prints},
         year = 2025,
        month = oct,
          eid = {arXiv:2510.01033},
        pages = {arXiv:2510.01033},
          doi = {10.48550/arXiv.2510.01033},
archivePrefix = {arXiv},
       eprint = {2510.01033},
 primaryClass = {astro-ph.GA},
       adsurl = {https://ui.adsabs.harvard.edu/abs/2025arXiv251001033C}
}

@ARTICLE{JADES_DR4_2,
       author = {{Scholtz}, J. and {Carniani}, S. and {Parlanti}, E. and {D'Eugenio}, F. and {Curtis-Lake}, E. and {Jakobsen}, P. and {Bunker}, A.~J. and {Cameron}, A.~J. and {Arribas}, S. and {Baker}, W.~M. and {Charlot}, S. and {Chevellard}, J. and {Circosta}, C. and {Curti}, M. and {Duan}, Q. and {Eisenstein}, D.~J. and {Hainline}, K. and {Ji}, Z. and {Johnson}, B.~D. and {Jones}, G.~C. and {Kumari}, N. and {Maiolino}, R. and {Maseda}, M.~V. and {Perna}, M. and {P{\'e}rez-Gonz{\'a}lez}, P.~G. and {Rawle}, T. and {Rieke}, M. and {Rinaldi}, P. and {Robertson}, B. and {Saxena}, A. and {Shivaei}, I. and {Silcock}, M.~S. and {Sun}, Y. and {Rodr{\'\i}guez Del Pino}, B. and {Tacchella}, S. and {{\"U}bler}, H. and {Venturi}, G. and {Williams}, C.~C. and {Willmer}, C.~N.~A. and {Willott}, C. and {Witstok}, J.},
        title = "{JADES Data Release 4 -- Paper II: Data reduction, analysis and emission-line fluxes of the complete spectroscopic sample}",
      journal = {arXiv e-prints},
         year = 2025,
        month = oct,
          eid = {arXiv:2510.01034},
        pages = {arXiv:2510.01034},
          doi = {10.48550/arXiv.2510.01034},
archivePrefix = {arXiv},
       eprint = {2510.01034},
 primaryClass = {astro-ph.GA},
       adsurl = {https://ui.adsabs.harvard.edu/abs/2025arXiv251001034S}
}

@ARTICLE{CANDELS,
       author = {{Grogin}, Norman A. and {Kocevski}, Dale D. and {Faber}, S.~M. and {Ferguson}, Henry C. and {Koekemoer}, Anton M. and {Riess}, Adam G. and {Acquaviva}, Viviana and {Alexander}, David M. and {Almaini}, Omar and {Ashby}, Matthew L.~N. and {Barden}, Marco and {Bell}, Eric F. and {Bournaud}, Fr{\'e}d{\'e}ric and {Brown}, Thomas M. and {Caputi}, Karina I. and {Casertano}, Stefano and {Cassata}, Paolo and {Castellano}, Marco and {Challis}, Peter and {Chary}, Ranga-Ram and {Cheung}, Edmond and {Cirasuolo}, Michele and {Conselice}, Christopher J. and {Roshan Cooray}, Asantha and {Croton}, Darren J. and {Daddi}, Emanuele and {Dahlen}, Tomas and {Dav{\'e}}, Romeel and {de Mello}, Du{\'\i}lia F. and {Dekel}, Avishai and {Dickinson}, Mark and {Dolch}, Timothy and {Donley}, Jennifer L. and {Dunlop}, James S. and {Dutton}, Aaron A. and {Elbaz}, David and {Fazio}, Giovanni G. and {Filippenko}, Alexei V. and {Finkelstein}, Steven L. and {Fontana}, Adriano and {Gardner}, Jonathan P. and {Garnavich}, Peter M. and {Gawiser}, Eric and {Giavalisco}, Mauro and {Grazian}, Andrea and {Guo}, Yicheng and {Hathi}, Nimish P. and {H{\"a}ussler}, Boris and {Hopkins}, Philip F. and {Huang}, Jia-Sheng and {Huang}, Kuang-Han and {Jha}, Saurabh W. and {Kartaltepe}, Jeyhan S. and {Kirshner}, Robert P. and {Koo}, David C. and {Lai}, Kamson and {Lee}, Kyoung-Soo and {Li}, Weidong and {Lotz}, Jennifer M. and {Lucas}, Ray A. and {Madau}, Piero and {McCarthy}, Patrick J. and {McGrath}, Elizabeth J. and {McIntosh}, Daniel H. and {McLure}, Ross J. and {Mobasher}, Bahram and {Moustakas}, Leonidas A. and {Mozena}, Mark and {Nandra}, Kirpal and {Newman}, Jeffrey A. and {Niemi}, Sami-Matias and {Noeske}, Kai G. and {Papovich}, Casey J. and {Pentericci}, Laura and {Pope}, Alexandra and {Primack}, Joel R. and {Rajan}, Abhijith and {Ravindranath}, Swara and {Reddy}, Naveen A. and {Renzini}, Alvio and {Rix}, Hans-Walter and {Robaina}, Aday R. and {Rodney}, Steven A. and {Rosario}, David J. and {Rosati}, Piero and {Salimbeni}, Sara and {Scarlata}, Claudia and {Siana}, Brian and {Simard}, Luc and {Smidt}, Joseph and {Somerville}, Rachel S. and {Spinrad}, Hyron and {Straughn}, Amber N. and {Strolger}, Louis-Gregory and {Telford}, Olivia and {Teplitz}, Harry I. and {Trump}, Jonathan R. and {van der Wel}, Arjen and {Villforth}, Carolin and {Wechsler}, Risa H. and {Weiner}, Benjamin J. and {Wiklind}, Tommy and {Wild}, Vivienne and {Wilson}, Grant and {Wuyts}, Stijn and {Yan}, Hao-Jing and {Yun}, Min S.},
        title = "{CANDELS: The Cosmic Assembly Near-infrared Deep Extragalactic Legacy Survey}",
      journal = {\apjs},
         year = 2011,
        month = dec,
       volume = {197},
       number = {2},
          eid = {35},
        pages = {35},
          doi = {10.1088/0067-0049/197/2/35},
archivePrefix = {arXiv},
       eprint = {1105.3753},
 primaryClass = {astro-ph.CO},
       adsurl = {https://ui.adsabs.harvard.edu/abs/2011ApJS..197...35G}
}

@ARTICLE{CANDELS_2,
       author = {{Koekemoer}, Anton M. and {Faber}, S.~M. and {Ferguson}, Henry C. and {Grogin}, Norman A. and {Kocevski}, Dale D. and {Koo}, David C. and {Lai}, Kamson and {Lotz}, Jennifer M. and {Lucas}, Ray A. and {McGrath}, Elizabeth J. and {Ogaz}, Sara and {Rajan}, Abhijith and {Riess}, Adam G. and {Rodney}, Steve A. and {Strolger}, Louis and {Casertano}, Stefano and {Castellano}, Marco and {Dahlen}, Tomas and {Dickinson}, Mark and {Dolch}, Timothy and {Fontana}, Adriano and {Giavalisco}, Mauro and {Grazian}, Andrea and {Guo}, Yicheng and {Hathi}, Nimish P. and {Huang}, Kuang-Han and {van der Wel}, Arjen and {Yan}, Hao-Jing and {Acquaviva}, Viviana and {Alexander}, David M. and {Almaini}, Omar and {Ashby}, Matthew L.~N. and {Barden}, Marco and {Bell}, Eric F. and {Bournaud}, Fr{\'e}d{\'e}ric and {Brown}, Thomas M. and {Caputi}, Karina I. and {Cassata}, Paolo and {Challis}, Peter J. and {Chary}, Ranga-Ram and {Cheung}, Edmond and {Cirasuolo}, Michele and {Conselice}, Christopher J. and {Roshan Cooray}, Asantha and {Croton}, Darren J. and {Daddi}, Emanuele and {Dav{\'e}}, Romeel and {de Mello}, Duilia F. and {de Ravel}, Loic and {Dekel}, Avishai and {Donley}, Jennifer L. and {Dunlop}, James S. and {Dutton}, Aaron A. and {Elbaz}, David and {Fazio}, Giovanni G. and {Filippenko}, Alexei V. and {Finkelstein}, Steven L. and {Frazer}, Chris and {Gardner}, Jonathan P. and {Garnavich}, Peter M. and {Gawiser}, Eric and {Gruetzbauch}, Ruth and {Hartley}, Will G. and {H{\"a}ussler}, Boris and {Herrington}, Jessica and {Hopkins}, Philip F. and {Huang}, Jia-Sheng and {Jha}, Saurabh W. and {Johnson}, Andrew and {Kartaltepe}, Jeyhan S. and {Khostovan}, Ali A. and {Kirshner}, Robert P. and {Lani}, Caterina and {Lee}, Kyoung-Soo and {Li}, Weidong and {Madau}, Piero and {McCarthy}, Patrick J. and {McIntosh}, Daniel H. and {McLure}, Ross J. and {McPartland}, Conor and {Mobasher}, Bahram and {Moreira}, Heidi and {Mortlock}, Alice and {Moustakas}, Leonidas A. and {Mozena}, Mark and {Nandra}, Kirpal and {Newman}, Jeffrey A. and {Nielsen}, Jennifer L. and {Niemi}, Sami and {Noeske}, Kai G. and {Papovich}, Casey J. and {Pentericci}, Laura and {Pope}, Alexandra and {Primack}, Joel R. and {Ravindranath}, Swara and {Reddy}, Naveen A. and {Renzini}, Alvio and {Rix}, Hans-Walter and {Robaina}, Aday R. and {Rosario}, David J. and {Rosati}, Piero and {Salimbeni}, Sara and {Scarlata}, Claudia and {Siana}, Brian and {Simard}, Luc and {Smidt}, Joseph and {Snyder}, Diana and {Somerville}, Rachel S. and {Spinrad}, Hyron and {Straughn}, Amber N. and {Telford}, Olivia and {Teplitz}, Harry I. and {Trump}, Jonathan R. and {Vargas}, Carlos and {Villforth}, Carolin and {Wagner}, Cory R. and {Wandro}, Pat and {Wechsler}, Risa H. and {Weiner}, Benjamin J. and {Wiklind}, Tommy and {Wild}, Vivienne and {Wilson}, Grant and {Wuyts}, Stijn and {Yun}, Min S.},
        title = "{CANDELS: The Cosmic Assembly Near-infrared Deep Extragalactic Legacy Survey{\textemdash}The Hubble Space Telescope Observations, Imaging Data Products, and Mosaics}",
      journal = {\apjs},
         year = 2011,
        month = dec,
       volume = {197},
       number = {2},
          eid = {36},
        pages = {36},
          doi = {10.1088/0067-0049/197/2/36},
archivePrefix = {arXiv},
       eprint = {1105.3754},
 primaryClass = {astro-ph.CO},
       adsurl = {https://ui.adsabs.harvard.edu/abs/2011ApJS..197...36K}
}

@ARTICLE{COSMOS,
       author = {{Scoville}, N. and {Aussel}, H. and {Brusa}, M. and {Capak}, P. and {Carollo}, C.~M. and {Elvis}, M. and {Giavalisco}, M. and {Guzzo}, L. and {Hasinger}, G. and {Impey}, C. and {Kneib}, J.-P. and {LeFevre}, O. and {Lilly}, S.~J. and {Mobasher}, B. and {Renzini}, A. and {Rich}, R.~M. and {Sanders}, D.~B. and {Schinnerer}, E. and {Schminovich}, D. and {Shopbell}, P. and {Taniguchi}, Y. and {Tyson}, N.~D.},
        title = "{The Cosmic Evolution Survey (COSMOS): Overview}",
      journal = {\apjs},
         year = 2007,
        month = sep,
       volume = {172},
       number = {1},
        pages = {1-8},
          doi = {10.1086/516585},
archivePrefix = {arXiv},
       eprint = {astro-ph/0612305},
 primaryClass = {astro-ph},
       adsurl = {https://ui.adsabs.harvard.edu/abs/2007ApJS..172....1S}
}

@ARTICLE{UDS,
       author = {{Lawrence}, A. and {Warren}, S.~J. and {Almaini}, O. and {Edge}, A.~C. and {Hambly}, N.~C. and {Jameson}, R.~F. and {Lucas}, P. and {Casali}, M. and {Adamson}, A. and {Dye}, S. and {Emerson}, J.~P. and {Foucaud}, S. and {Hewett}, P. and {Hirst}, P. and {Hodgkin}, S.~T. and {Irwin}, M.~J. and {Lodieu}, N. and {McMahon}, R.~G. and {Simpson}, C. and {Smail}, I. and {Mortlock}, D. and {Folger}, M.},
        title = "{The UKIRT Infrared Deep Sky Survey (UKIDSS)}",
      journal = {\mnras},
         year = 2007,
        month = aug,
       volume = {379},
       number = {4},
        pages = {1599-1617},
          doi = {10.1111/j.1365-2966.2007.12040.x},
archivePrefix = {arXiv},
       eprint = {astro-ph/0604426},
 primaryClass = {astro-ph},
       adsurl = {https://ui.adsabs.harvard.edu/abs/2007MNRAS.379.1599L}
}

@ARTICLE{EGS,
       author = {{Davis}, M. and {Guhathakurta}, P. and {Konidaris}, N.~P. and {Newman}, J.~A. and {Ashby}, M.~L.~N. and {Biggs}, A.~D. and {Barmby}, P. and {Bundy}, K. and {Chapman}, S.~C. and {Coil}, A.~L. and {Conselice}, C.~J. and {Cooper}, M.~C. and {Croton}, D.~J. and {Eisenhardt}, P.~R.~M. and {Ellis}, R.~S. and {Faber}, S.~M. and {Fang}, T. and {Fazio}, G.~G. and {Georgakakis}, A. and {Gerke}, B.~F. and {Goss}, W.~M. and {Gwyn}, S. and {Harker}, J. and {Hopkins}, A.~M. and {Huang}, J.-S. and {Ivison}, R.~J. and {Kassin}, S.~A. and {Kirby}, E.~N. and {Koekemoer}, A.~M. and {Koo}, D.~C. and {Laird}, E.~S. and {Le Floc'h}, E. and {Lin}, L. and {Lotz}, J.~M. and {Marshall}, P.~J. and {Martin}, D.~C. and {Metevier}, A.~J. and {Moustakas}, L.~A. and {Nandra}, K. and {Noeske}, K.~G. and {Papovich}, C. and {Phillips}, A.~C. and {Rich}, R.~M. and {Rieke}, G.~H. and {Rigopoulou}, D. and {Salim}, S. and {Schiminovich}, D. and {Simard}, L. and {Smail}, I. and {Small}, T.~A. and {Weiner}, B.~J. and {Willmer}, C.~N.~A. and {Willner}, S.~P. and {Wilson}, G. and {Wright}, E.~L. and {Yan}, R.},
        title = "{The All-Wavelength Extended Groth Strip International Survey (AEGIS) Data Sets}",
      journal = {\apjl},
         year = 2007,
        month = may,
       volume = {660},
       number = {1},
        pages = {L1-L6},
          doi = {10.1086/517931},
archivePrefix = {arXiv},
       eprint = {astro-ph/0607355},
 primaryClass = {astro-ph},
       adsurl = {https://ui.adsabs.harvard.edu/abs/2007ApJ...660L...1D}
}

@ARTICLE{CEERS,
       author = {{Finkelstein}, Steven L. and {Bagley}, Micaela B. and {Arrabal Haro}, Pablo and {Dickinson}, Mark and {Ferguson}, Henry C. and {Kartaltepe}, Jeyhan S. and {Kocevski}, Dale D. and {Koekemoer}, Anton M. and {Lotz}, Jennifer M. and {Papovich}, Casey and {P{\'e}rez-Gonz{\'a}lez}, Pablo G. and {Pirzkal}, Nor and {Somerville}, Rachel S. and {Trump}, Jonathan R. and {Yang}, Guang and {Yung}, L.~Y. Aaron and {Fontana}, Adriano and {Grazian}, Andrea and {Grogin}, Norman A. and {Kewley}, Lisa J. and {Kirkpatrick}, Allison and {Larson}, Rebecca L. and {Pentericci}, Laura and {Ravindranath}, Swara and {Wilkins}, Stephen M. and {Almaini}, Omar and {Amor{\'\i}n}, Ricardo O. and {Barro}, Guillermo and {Bhatawdekar}, Rachana and {Bisigello}, Laura and {Brooks}, Madisyn and {Buat}, V{\'e}ronique and {Buitrago}, Fernando and {Burgarella}, Denis and {Calabr{\`o}}, Antonello and {Castellano}, Marco and {Cheng}, Yingjie and {Cleri}, Nikko J. and {Cole}, Justin W. and {Cooper}, M.~C. and {Cooper}, Olivia R. and {Costantin}, Luca and {Cox}, Isa G. and {Croton}, Darren and {Daddi}, Emanuele and {Davis}, Kelcey and {Dekel}, Avishai and {Elbaz}, David and {Fern{\'a}ndez}, Vital and {Fujimoto}, Seiji and {Gandolfi}, Giovanni and {Gardner}, Jonathan P. and {Gawiser}, Eric and {Giavalisco}, Mauro and {G{\'o}mez-Guijarro}, Carlos and {Guo}, Yuchen and {Gupta}, Ansh R. and {Hathi}, Nimish P. and {Harish}, Santosh and {Henry}, Aur{\'e}lien and {Hirschmann}, Michaela and {Hu}, Weida and {Hutchison}, Taylor A. and {Iyer}, Kartheik G. and {Jaskot}, Anne E. and {Jha}, Saurabh W. and {Jung}, Intae and {Kassin}, Susan A. and {Kokorev}, Vasily and {Kurczynski}, Peter and {Leung}, Gene C.~K. and {Llerena}, Mario and {Long}, Arianna S. and {Lucas}, Ray A. and {Lu}, Shiying and {McGrath}, Elizabeth J. and {McIntosh}, Daniel H. and {Merlin}, Emiliano and {Mobasher}, Bahram and {Morales}, Alexa M. and {Napolitano}, Lorenzo and {Pacucci}, Fabio and {Pandya}, Viraj and {Rafelski}, Marc and {Rodighiero}, Giulia and {Rose}, Caitlin and {Santini}, Paola and {Seill{\'e}}, Lise-Marie and {Simons}, Raymond C. and {Shen}, Lu and {Straughn}, Amber N. and {Tacchella}, Sandro and {Taylor}, Anthony J. and {Vanderhoof}, Brittany N. and {Vega-Ferrero}, Jes{\'u}s and {Weiner}, Benjamin J. and {Willmer}, Christopher N.~A. and {Zhu}, Peixin and {Bell}, Eric F. and {Wuyts}, Stijn and {Holwerda}, Benne W. and {Wang}, Xin and {Wang}, Weichen and {Zavala}, Jorge A. and {CEERS Collaboration}},
        title = "{The Cosmic Evolution Early Release Science Survey (CEERS)}",
      journal = {\apjl},
         year = 2025,
        month = apr,
       volume = {983},
       number = {1},
          eid = {L4},
        pages = {L4},
          doi = {10.3847/2041-8213/adbbd3},
archivePrefix = {arXiv},
       eprint = {2501.04085},
 primaryClass = {astro-ph.GA},
       adsurl = {https://ui.adsabs.harvard.edu/abs/2025ApJ...983L...4F}
}

@ARTICLE{DJ_Spec_1,
       author = {{Heintz}, Kasper E. and {Watson}, Darach and {Brammer}, Gabriel and {Vejlgaard}, Simone and {Hutter}, Anne and {Strait}, Victoria B. and {Matthee}, Jorryt and {Oesch}, Pascal A. and {Jakobsson}, P{\'a}ll and {Tanvir}, Nial R. and {Laursen}, Peter and {Naidu}, Rohan P. and {Mason}, Charlotte A. and {Killi}, Meghana and {Jung}, Intae and {Hsiao}, Tiger Yu-Yang and {Abdurro'uf} and {Coe}, Dan and {Arrabal Haro}, Pablo and {Finkelstein}, Steven L. and {Toft}, Sune},
        title = "{Strong damped Lyman-{\ensuremath{\alpha}} absorption in young star-forming galaxies at redshifts 9 to 11}",
      journal = {Science},
         year = 2024,
        month = may,
       volume = {384},
       number = {6698},
        pages = {890-894},
          doi = {10.1126/science.adj0343},
archivePrefix = {arXiv},
       eprint = {2306.00647},
 primaryClass = {astro-ph.GA},
       adsurl = {https://ui.adsabs.harvard.edu/abs/2024Sci...384..890H}
}

@ARTICLE{DJ_Spec_2,
       author = {{de Graaff}, Anna and {Brammer}, Gabriel and {Weibel}, Andrea and {Lewis}, Zach and {Maseda}, Michael V. and {Oesch}, Pascal A. and {Bezanson}, Rachel and {Boogaard}, Leindert A. and {Cleri}, Nikko J. and {Cooper}, Olivia R. and {Gottumukkala}, Rashmi and {Greene}, Jenny E. and {Hirschmann}, Michaela and {Hviding}, Raphael E. and {Katz}, Harley and {Labb{\'e}}, Ivo and {Leja}, Joel and {Matthee}, Jorryt and {McConachie}, Ian and {Miller}, Tim B. and {Naidu}, Rohan P. and {Price}, Sedona H. and {Rix}, Hans-Walter and {Setton}, David J. and {Suess}, Katherine A. and {Wang}, Bingjie and {Whitaker}, Katherine E. and {Williams}, Christina C.},
        title = "{RUBIES: A complete census of the bright and red distant Universe with JWST/NIRSpec}",
      journal = {\aap},
         year = 2025,
        month = may,
       volume = {697},
          eid = {A189},
        pages = {A189},
          doi = {10.1051/0004-6361/202452186},
archivePrefix = {arXiv},
       eprint = {2409.05948},
 primaryClass = {astro-ph.GA},
       adsurl = {https://ui.adsabs.harvard.edu/abs/2025A&A...697A.189D}
}

@software{grizli,
  author       = {Brammer, Gabriel},
  title        = {grizli},
  month        = sep,
  year         = 2023,
  publisher    = {Zenodo},
  version      = {1.9.11},
  doi          = {10.5281/zenodo.8370018},
  url          = {https://doi.org/10.5281/zenodo.8370018},
}

@ARTICLE{DJ_Phot,
       author = {{Valentino}, Francesco and {Brammer}, Gabriel and {Gould}, Katriona M.~L. and {Kokorev}, Vasily and {Fujimoto}, Seiji and {Jespersen}, Christian Kragh and {Vijayan}, Aswin P. and {Weaver}, John R. and {Ito}, Kei and {Tanaka}, Masayuki and {Ilbert}, Olivier and {Magdis}, Georgios E. and {Whitaker}, Katherine E. and {Faisst}, Andreas L. and {Gallazzi}, Anna and {Gillman}, Steven and {Gim{\'e}nez-Arteaga}, Clara and {G{\'o}mez-Guijarro}, Carlos and {Kubo}, Mariko and {Heintz}, Kasper E. and {Hirschmann}, Michaela and {Oesch}, Pascal and {Onodera}, Masato and {Rizzo}, Francesca and {Lee}, Minju and {Strait}, Victoria and {Toft}, Sune},
        title = "{An Atlas of Color-selected Quiescent Galaxies at z > 3 in Public JWST Fields}",
      journal = {\apj},
         year = 2023,
        month = apr,
       volume = {947},
       number = {1},
          eid = {20},
        pages = {20},
          doi = {10.3847/1538-4357/acbefa},
archivePrefix = {arXiv},
       eprint = {2302.10936},
 primaryClass = {astro-ph.GA},
       adsurl = {https://ui.adsabs.harvard.edu/abs/2023ApJ...947...20V}
}

@ARTICLE{PLRD,
       author = {{Caputi}, Karina I. and {Cooper}, Ryan A. and {Rinaldi}, Pierluigi and {Navarro-Carrera}, Rafael and {Iani}, Edoardo and {Tumborang}, Abigail A.},
        title = "{Pseudo Little Red Dot: An Active Black Hole Embedded in a Dense and Dusty, Metal-poor Starburst Galaxy at z = 5.96}",
      journal = {\apj},
         year = 2026,
        month = aug,
       volume = {1007},
       number = {2},
          eid = {203},
        pages = {203},
          doi = {10.3847/1538-4357/ae80b9},
       adsurl = {https://ui.adsabs.harvard.edu/abs/2026ApJ..1007..203C}
}

@ARTICLE{cliff,
       author = {{de Graaff}, Anna and {Rix}, Hans-Walter and {Naidu}, Rohan P. and {Labb{\'e}}, Ivo and {Wang}, Bingjie and {Leja}, Joel and {Matthee}, Jorryt and {Katz}, Harley and {Greene}, Jenny E. and {Hviding}, Raphael E. and {Baggen}, Josephine and {Bezanson}, Rachel and {Boogaard}, Leindert A. and {Brammer}, Gabriel and {Dayal}, Pratika and {van Dokkum}, Pieter and {Goulding}, Andy D. and {Hirschmann}, Michaela and {Maseda}, Michael V. and {McConachie}, Ian and {Miller}, Tim B. and {Nelson}, Erica and {Oesch}, Pascal A. and {Setton}, David J. and {Shivaei}, Irene and {Weibel}, Andrea and {Whitaker}, Katherine E. and {Williams}, Christina C.},
        title = "{A remarkable ruby: Absorption in dense gas, rather than evolved stars, drives the extreme Balmer break of a little red dot at z = 3.5}",
      journal = {\aap},
         year = 2025,
        month = sep,
       volume = {701},
          eid = {A168},
        pages = {A168},
          doi = {10.1051/0004-6361/202554681},
archivePrefix = {arXiv},
       eprint = {2503.16600},
 primaryClass = {astro-ph.GA},
       adsurl = {https://ui.adsabs.harvard.edu/abs/2025A&A...701A.168D}
}

@ARTICLE{qso1,
       author = {{Maiolino}, Roberto and {{\"U}bler}, Hannah and {D'Eugenio}, Francesco and {Scholtz}, Jan and {Juod{\v{z}}balis}, Ignas and {Ji}, Xihan and {Perna}, Michele and {Bromm}, Volker and {Dayal}, Pratika and {Koudmani}, Sophie and {Liu}, Boyuan and {Schneider}, Raffaella and {Sijacki}, Debora and {Valiante}, Rosa and {Trinca}, Alessandro and {Zhang}, Saiyang and {Volonteri}, Marta and {Inayoshi}, Kohei and {Carniani}, Stefano and {Nakajima}, Kimihiko and {Isobe}, Yuki and {Witstok}, Joris and {Jones}, Gareth C. and {Tacchella}, Sandro and {Arribas}, Santiago and {Bunker}, Andrew and {Cataldi}, Elisa and {Charlot}, Stephane and {Curti}, Giovanni Cresci Mirko and {Fabian}, Andrew C. and {Katz}, Harley and {Kumari}, Nimisha and {Laporte}, Nicolas and {Mazzolari}, Giovanni and {Robertson}, Brant and {Sun}, Fengwu and {Rodriguez Del Pino}, Bruno and {Venturi}, Giacomo},
        title = "{A black hole in a near pristine galaxy 700 Myr after the big bang}",
      journal = {\mnras},
         year = 2026,
        month = may,
       volume = {548},
       number = {1},
          eid = {staf2109},
        pages = {staf2109},
          doi = {10.1093/mnras/staf2109},
archivePrefix = {arXiv},
       eprint = {2505.22567},
 primaryClass = {astro-ph.GA},
       adsurl = {https://ui.adsabs.harvard.edu/abs/2026MNRAS.548f2109M}
}

@ARTICLE{outflows_1,
       author = {{Cooper}, Ryan A. and {Caputi}, Karina I. and {Iani}, Edoardo and {Rinaldi}, Pierluigi and {Desprez}, Guillaume and {Navarro-Carrera}, Rafael},
        title = "{High-velocity Outflows in [O III] Emitters at 2.5 < z < 9 from JWST NIRSpec Medium-resolution Spectroscopy}",
      journal = {\apj},
         year = 2025,
        month = nov,
       volume = {994},
       number = {1},
          eid = {102},
        pages = {102},
          doi = {10.3847/1538-4357/ae0580},
archivePrefix = {arXiv},
       eprint = {2502.18310},
 primaryClass = {astro-ph.GA},
       adsurl = {https://ui.adsabs.harvard.edu/abs/2025ApJ...994..102C}
}

@ARTICLE{outflows_2,
       author = {{Carniani}, Stefano and {Venturi}, Giacomo and {Parlanti}, Eleonora and {de Graaff}, Anna and {Maiolino}, Roberto and {Arribas}, Santiago and {Bonaventura}, Nina and {Boyett}, Kristan and {Bunker}, Andrew J. and {Cameron}, Alex J. and {Charlot}, Stephane and {Chevallard}, Jacopo and {Curti}, Mirko and {Curtis-Lake}, Emma and {Eisenstein}, Daniel J. and {Giardino}, Giovanna and {Hausen}, Ryan and {Kumari}, Nimisha and {Maseda}, Michael V. and {Nelson}, Erica and {Perna}, Michele and {Rix}, Hans-Walter and {Robertson}, Brant and {Del Pino}, Bruno Rodr{\'\i}guez and {Sandles}, Lester and {Scholtz}, Jan and {Simmonds}, Charlotte and {Smit}, Renske and {Tacchella}, Sandro and {{\"U}bler}, Hannah and {Williams}, Christina C. and {Willott}, Chris and {Witstok}, Joris},
        title = "{JADES: The incidence rate and properties of galactic outflows in low-mass galaxies across 3 < z < 9}",
      journal = {\aap},
         year = 2024,
        month = may,
       volume = {685},
          eid = {A99},
        pages = {A99},
          doi = {10.1051/0004-6361/202347230},
archivePrefix = {arXiv},
       eprint = {2306.11801},
 primaryClass = {astro-ph.GA},
       adsurl = {https://ui.adsabs.harvard.edu/abs/2024A&A...685A..99C}
}

@article{oiii_ratio,
    author = {Storey, P. J. and Zeippen, C. J.},
    title = {Theoretical values for the [O iii] 5007/4959 line-intensity ratio and homologous cases},
    journal = {Monthly Notices of the Royal Astronomical Society},
    volume = {312},
    number = {4},
    pages = {813-816},
    year = {2000},
    month = {03},
    issn = {0035-8711},
    doi = {10.1046/j.1365-8711.2000.03184.x},
    url = {https://doi.org/10.1046/j.1365-8711.2000.03184.x},
    eprint = {https://academic.oup.com/mnras/article-pdf/312/4/813/18634438/312-4-813.pdf},
}

@ARTICLE{sandles_bd,
       author = {{Sandles}, Lester and {D'Eugenio}, Francesco and {Maiolino}, Roberto and {Looser}, Tobias J. and {Arribas}, Santiago and {Baker}, William M. and {Bonaventura}, Nina and {Bunker}, Andrew J. and {Cameron}, Alex J. and {Carniani}, Stefano and {Charlot}, Stephane and {Chevallard}, Jacopo and {Curti}, Mirko and {Curtis-Lake}, Emma and {de Graaff}, Anna and {Eisenstein}, Daniel J. and {Hainline}, Kevin and {Ji}, Zhiyuan and {Johnson}, Benjamin D. and {Jones}, Gareth C. and {Kumari}, Nimisha and {Nelson}, Erica and {Perna}, Michele and {Rawle}, Tim and {Rix}, Hans-Walter and {Robertson}, Brant and {Del Pino}, Bruno Rodr{\'\i}guez and {Scholtz}, Jan and {Shivaei}, Irene and {Smit}, Renske and {Sun}, Fengwu and {Tacchella}, Sandro and {{\"U}bler}, Hannah and {Williams}, Christina C. and {Willott}, Chris and {Witstok}, Joris},
        title = "{JADES: Balmer decrement measurements at redshifts 4 < z < 7}",
      journal = {\aap},
         year = 2024,
        month = nov,
       volume = {691},
          eid = {A305},
        pages = {A305},
          doi = {10.1051/0004-6361/202347119},
archivePrefix = {arXiv},
       eprint = {2306.03931},
 primaryClass = {astro-ph.GA},
       adsurl = {https://ui.adsabs.harvard.edu/abs/2024A&A...691A.305S}
}

@BOOK{os_and_fer,
       author = {{Osterbrock}, Donald E. and {Ferland}, Gary J.},
        title = "{Astrophysics of gaseous nebulae and active galactic nuclei}",
         year = 2006,
       adsurl = {https://ui.adsabs.harvard.edu/abs/2006agna.book.....O}
}

@ARTICLE{cigale_2019,
       author = {{Boquien}, M. and {Burgarella}, D. and {Roehlly}, Y. and {Buat}, V. and {Ciesla}, L. and {Corre}, D. and {Inoue}, A.~K. and {Salas}, H.},
        title = "{CIGALE: a python Code Investigating GALaxy Emission}",
      journal = {\aap},
         year = 2019,
        month = feb,
       volume = {622},
          eid = {A103},
        pages = {A103},
          doi = {10.1051/0004-6361/201834156},
archivePrefix = {arXiv},
       eprint = {1811.03094},
 primaryClass = {astro-ph.GA},
       adsurl = {https://ui.adsabs.harvard.edu/abs/2019A&A...622A.103B}
}

@ARTICLE{cigale_stoch,
       author = {{Carvajal-Bohorquez}, C. and {Ciesla}, L. and {Laporte}, N. and {Boquien}, M. and {Buat}, V. and {Ilbert}, O. and {Aufort}, G. and {Shuntov}, M. and {Witten}, C. and {Oesch}, P.~A. and {Covelo-Paz}, A.},
        title = "{Stochastic star formation activity of galaxies within the first billion years probed by JWST}",
      journal = {\aap},
         year = 2025,
        month = dec,
       volume = {704},
          eid = {A290},
        pages = {A290},
          doi = {10.1051/0004-6361/202556471},
archivePrefix = {arXiv},
       eprint = {2507.13160},
 primaryClass = {astro-ph.GA},
       adsurl = {https://ui.adsabs.harvard.edu/abs/2025A&A...704A.290C}
}

@ARTICLE{BC03,
       author = {{Bruzual}, G. and {Charlot}, S.},
        title = "{Stellar population synthesis at the resolution of 2003}",
      journal = {\mnras},
         year = 2003,
        month = oct,
       volume = {344},
       number = {4},
        pages = {1000-1028},
          doi = {10.1046/j.1365-8711.2003.06897.x},
archivePrefix = {arXiv},
       eprint = {astro-ph/0309134},
 primaryClass = {astro-ph},
       adsurl = {https://ui.adsabs.harvard.edu/abs/2003MNRAS.344.1000B}
}

@ARTICLE{cloudy_13,
       author = {{Ferland}, G.~J. and {Porter}, R.~L. and {van Hoof}, P.~A.~M. and {Williams}, R.~J.~R. and {Abel}, N.~P. and {Lykins}, M.~L. and {Shaw}, G. and {Henney}, W.~J. and {Stancil}, P.~C.},
        title = "{The 2013 Release of Cloudy}",
      journal = {\rmxaa},
         year = 2013,
        month = apr,
       volume = {49},
        pages = {137-163},
          doi = {10.48550/arXiv.1302.4485},
archivePrefix = {arXiv},
       eprint = {1302.4485},
 primaryClass = {astro-ph.GA},
       adsurl = {https://ui.adsabs.harvard.edu/abs/2013RMxAA..49..137F}
}

@ARTICLE{calzetti_2000,
       author = {{Calzetti}, Daniela and {Armus}, Lee and {Bohlin}, Ralph C. and {Kinney}, Anne L. and {Koornneef}, Jan and {Storchi-Bergmann}, Thaisa},
        title = "{The Dust Content and Opacity of Actively Star-forming Galaxies}",
      journal = {\apj},
         year = 2000,
        month = apr,
       volume = {533},
       number = {2},
        pages = {682-695},
          doi = {10.1086/308692},
archivePrefix = {arXiv},
       eprint = {astro-ph/9911459},
 primaryClass = {astro-ph},
       adsurl = {https://ui.adsabs.harvard.edu/abs/2000ApJ...533..682C}
}

@ARTICLE{dicesare_2026,
       author = {{Di Cesare}, Claudia and {Matthee}, Jorryt and {Naidu}, Rohan P. and {Torralba}, Alberto and {Kotiwale}, Gauri and {Kramarenko}, Ivan G. and {Blaizot}, Jeremy and {Rosdahl}, Joakim and {Leja}, Joel and {Iani}, Edoardo and et al.},
        title = "{The slope and scatter of the star-forming main sequence at z {\ensuremath{\sim}} 5: Reconciling observations with simulations}",
      journal = {\aap},
         year = 2026,
        month = mar,
       volume = {707},
          eid = {A129},
        pages = {A129},
          doi = {10.1051/0004-6361/202557790},
archivePrefix = {arXiv},
       eprint = {2510.19044},
 primaryClass = {astro-ph.GA},
       adsurl = {https://ui.adsabs.harvard.edu/abs/2026A&A...707A.129D}
}

@ARTICLE{kramarenko_2026,
       author = {{Kramarenko}, I.~G. and {Rosdahl}, J. and {Blaizot}, J. and {Matthee}, J. and {Katz}, H. and {Di Cesare}, C.},
        title = "{H{\ensuremath{\alpha}} as a tracer of star formation in the SPHINX cosmological simulations}",
      journal = {\aap},
         year = 2026,
        month = mar,
       volume = {707},
          eid = {A184},
        pages = {A184},
          doi = {10.1051/0004-6361/202557114},
archivePrefix = {arXiv},
       eprint = {2509.05403},
 primaryClass = {astro-ph.GA},
       adsurl = {https://ui.adsabs.harvard.edu/abs/2026A&A...707A.184K}
}

@ARTICLE{theios_2019,
       author = {{Theios}, Rachel L. and {Steidel}, Charles C. and {Strom}, Allison L. and {Rudie}, Gwen C. and {Trainor}, Ryan F. and {Reddy}, Naveen A.},
        title = "{Dust Attenuation, Star Formation, and Metallicity in z {\ensuremath{\sim}} 2-3 Galaxies from KBSS-MOSFIRE}",
      journal = {\apj},
         year = 2019,
        month = jan,
       volume = {871},
       number = {1},
          eid = {128},
        pages = {128},
          doi = {10.3847/1538-4357/aaf386},
archivePrefix = {arXiv},
       eprint = {1805.00016},
 primaryClass = {astro-ph.GA},
       adsurl = {https://ui.adsabs.harvard.edu/abs/2019ApJ...871..128T}
}

@ARTICLE{cullen_2016,
       author = {{Cullen}, F. and {Cirasuolo}, M. and {Kewley}, L.~J. and {McLure}, R.~J. and {Dunlop}, J.~S. and {Bowler}, R.~A.~A.},
        title = "{Changing physical conditions in star-forming galaxies between redshifts 0 < z < 4: [O III]/H {\ensuremath{\beta}} evolution}",
      journal = {\mnras},
         year = 2016,
        month = aug,
       volume = {460},
       number = {3},
        pages = {3002-3013},
          doi = {10.1093/mnras/stw1181},
archivePrefix = {arXiv},
       eprint = {1605.04228},
 primaryClass = {astro-ph.GA},
       adsurl = {https://ui.adsabs.harvard.edu/abs/2016MNRAS.460.3002C}
}

@ARTICLE{shapely_2023,
       author = {{Shapley}, Alice E. and {Sanders}, Ryan L. and {Reddy}, Naveen A. and {Topping}, Michael W. and {Brammer}, Gabriel B.},
        title = "{JWST/NIRSpec Balmer-line Measurements of Star Formation and Dust Attenuation at z   3-6}",
      journal = {\apj},
         year = 2023,
        month = sep,
       volume = {954},
       number = {2},
          eid = {157},
        pages = {157},
          doi = {10.3847/1538-4357/acea5a},
archivePrefix = {arXiv},
       eprint = {2301.03241},
 primaryClass = {astro-ph.GA},
       adsurl = {https://ui.adsabs.harvard.edu/abs/2023ApJ...954..157S}
}

@ARTICLE{nakajima_2023,
       author = {{Nakajima}, Kimihiko and {Ouchi}, Masami and {Isobe}, Yuki and {Harikane}, Yuichi and {Zhang}, Yechi and {Ono}, Yoshiaki and {Umeda}, Hiroya and {Oguri}, Masamune},
        title = "{JWST Census for the Mass-Metallicity Star Formation Relations at z = 4-10 with Self-consistent Flux Calibration and Proper Metallicity Calibrators}",
      journal = {\apjs},
         year = 2023,
        month = dec,
       volume = {269},
       number = {2},
          eid = {33},
        pages = {33},
          doi = {10.3847/1538-4365/acd556},
archivePrefix = {arXiv},
       eprint = {2301.12825},
 primaryClass = {astro-ph.GA},
       adsurl = {https://ui.adsabs.harvard.edu/abs/2023ApJS..269...33N}
}

@ARTICLE{marszewski_2025,
       author = {{Marszewski}, Andrew and {Faucher-Gigu{\`e}re}, Claude-Andr{\'e} and {Feldmann}, Robert and {Sun}, Guochao},
        title = "{Explaining the Weak Evolution of the High-redshift Mass─Metallicity Relation with Galaxy Burst Cycles}",
      journal = {\apjl},
         year = 2025,
        month = sep,
       volume = {991},
       number = {1},
          eid = {L4},
        pages = {L4},
          doi = {10.3847/2041-8213/adf74b},
archivePrefix = {arXiv},
       eprint = {2505.22712},
 primaryClass = {astro-ph.GA},
       adsurl = {https://ui.adsabs.harvard.edu/abs/2025ApJ...991L...4M}
}

@ARTICLE{witstok_2023,
       author = {{Witstok}, Joris and {Shivaei}, Irene and {Smit}, Renske and {Maiolino}, Roberto and {Carniani}, Stefano and {Curtis-Lake}, Emma and {Ferruit}, Pierre and {Arribas}, Santiago and {Bunker}, Andrew J. and {Cameron}, Alex J. and {Charlot}, Stephane and {Chevallard}, Jacopo and {Curti}, Mirko and {de Graaff}, Anna and {D'Eugenio}, Francesco and {Giardino}, Giovanna and {Looser}, Tobias J. and {Rawle}, Tim and {Rodr{\'\i}guez del Pino}, Bruno and {Willott}, Chris and {Alberts}, Stacey and {Baker}, William M. and {Boyett}, Kristan and {Egami}, Eiichi and {Eisenstein}, Daniel J. and {Endsley}, Ryan and {Hainline}, Kevin N. and {Ji}, Zhiyuan and {Johnson}, Benjamin D. and {Kumari}, Nimisha and {Lyu}, Jianwei and {Nelson}, Erica and {Perna}, Michele and {Rieke}, Marcia and {Robertson}, Brant E. and {Sandles}, Lester and {Saxena}, Aayush and {Scholtz}, Jan and {Sun}, Fengwu and {Tacchella}, Sandro and {Williams}, Christina C. and {Willmer}, Christopher N.~A.},
        title = "{Carbonaceous dust grains seen in the first billion years of cosmic time}",
      journal = {\nat},
         year = 2023,
        month = sep,
       volume = {621},
       number = {7978},
        pages = {267-270},
          doi = {10.1038/s41586-023-06413-w},
archivePrefix = {arXiv},
       eprint = {2302.05468},
 primaryClass = {astro-ph.GA},
       adsurl = {https://ui.adsabs.harvard.edu/abs/2023Natur.621..267W}
}

@ARTICLE{hanae_2022,
       author = {{Inami}, Hanae and {Algera}, Hiddo S.~B. and {Schouws}, Sander and {Sommovigo}, Laura and {Bouwens}, Rychard and {Smit}, Renske and {Stefanon}, Mauro and {Bowler}, Rebecca A.~A. and {Endsley}, Ryan and {Ferrara}, Andrea and {Oesch}, Pascal and {Stark}, Daniel and {Aravena}, Manuel and {Barrufet}, Laia and {da Cunha}, Elisabete and {Dayal}, Pratika and {De Looze}, Ilse and {Fudamoto}, Yoshinobu and {Gonzalez}, Valentino and {Graziani}, Luca and {Hodge}, Jacqueline A. and {Hygate}, Alexander P.~S. and {Nanayakkara}, Themiya and {Pallottini}, Andrea and {Riechers}, Dominik A. and {Schneider}, Raffaella and {Topping}, Michael and {van der Werf}, Paul},
        title = "{The ALMA REBELS Survey: dust continuum detections at z > 6.5}",
      journal = {\mnras},
         year = 2022,
        month = sep,
       volume = {515},
       number = {3},
        pages = {3126-3143},
          doi = {10.1093/mnras/stac1779},
archivePrefix = {arXiv},
       eprint = {2203.15136},
 primaryClass = {astro-ph.GA},
       adsurl = {https://ui.adsabs.harvard.edu/abs/2022MNRAS.515.3126I}
}

@ARTICLE{trump_2023,
       author = {{Trump}, Jonathan R. and {Arrabal Haro}, Pablo and {Simons}, Raymond C. and {Backhaus}, Bren E. and {Amor{\'\i}n}, Ricardo O. and {Dickinson}, Mark and {Fern{\'a}ndez}, Vital and {Papovich}, Casey and {Nicholls}, David C. and {Kewley}, Lisa J. and {Brunker}, Samantha W. and {Salzer}, John J. and {Wilkins}, Stephen M. and {Almaini}, Omar and {Bagley}, Micaela B. and {Berg}, Danielle A. and {Bhatawdekar}, Rachana and {Bisigello}, Laura and {Buat}, V{\'e}ronique and {Burgarella}, Denis and {Calabr{\`o}}, Antonello and {Casey}, Caitlin M. and {Ciesla}, Laure and {Cleri}, Nikko J. and {Cole}, Justin W. and {Cooper}, M.~C. and {Cooray}, Asantha R. and {Costantin}, Luca and {Croton}, Darren and {Ferguson}, Henry C. and {Finkelstein}, Steven L. and {Fujimoto}, Seiji and {Gardner}, Jonathan P. and {Gawiser}, Eric and {Giavalisco}, Mauro and {Grazian}, Andrea and {Grogin}, Norman A. and {Hathi}, Nimish P. and {Hirschmann}, Michaela and {Holwerda}, Benne W. and {Huertas-Company}, Marc and {Hutchison}, Taylor A. and {Jogee}, Shardha and {Juneau}, St{\'e}phanie and {Jung}, Intae and {Kartaltepe}, Jeyhan S. and {Kirkpatrick}, Allison and {Kocevski}, Dale D. and {Koekemoer}, Anton M. and {Lotz}, Jennifer M. and {Lucas}, Ray A. and {Magnelli}, Benjamin and {Matharu}, Jasleen and {P{\'e}rez-Gonz{\'a}lez}, Pablo G. and {Pirzkal}, Nor and {Rafelski}, Marc and {Rose}, Caitlin and {Seill{\'e}}, Lise-Marie and {Somerville}, Rachel S. and {Straughn}, Amber N. and {Tacchella}, Sandro and {Vanderhoof}, Brittany N. and {Weiner}, Benjamin J. and {Wuyts}, Stijn and {Yung}, L.~Y. Aaron and {Zavala}, Jorge A.},
        title = "{The Physical Conditions of Emission-line Galaxies at Cosmic Dawn from JWST/NIRSpec Spectroscopy in the SMACS 0723 Early Release Observations}",
      journal = {\apj},
         year = 2023,
        month = mar,
       volume = {945},
       number = {1},
          eid = {35},
        pages = {35},
          doi = {10.3847/1538-4357/acba8a},
archivePrefix = {arXiv},
       eprint = {2207.12388},
 primaryClass = {astro-ph.GA},
       adsurl = {https://ui.adsabs.harvard.edu/abs/2023ApJ...945...35T}
}

@ARTICLE{curti_2024,
       author = {{Curti}, Mirko and {Maiolino}, Roberto and {Curtis-Lake}, Emma and {Chevallard}, Jacopo and {Carniani}, Stefano and {D'Eugenio}, Francesco and {Looser}, Tobias J. and {Scholtz}, Jan and {Charlot}, Stephane and {Cameron}, Alex and {{\"U}bler}, Hannah and {Witstok}, Joris and {Boyett}, Kristian and {Laseter}, Isaac and {Sandles}, Lester and {Arribas}, Santiago and {Bunker}, Andrew and {Giardino}, Giovanna and {Maseda}, Michael V. and {Rawle}, Tim and {Rodr{\'\i}guez Del Pino}, Bruno and {Smit}, Renske and {Willott}, Chris J. and {Eisenstein}, Daniel J. and {Hausen}, Ryan and {Johnson}, Benjamin and {Rieke}, Marcia and {Robertson}, Brant and {Tacchella}, Sandro and {Williams}, Christina C. and {Willmer}, Christopher and {Baker}, William M. and {Bhatawdekar}, Rachana and {Egami}, Eiichi and {Helton}, Jakob M. and {Ji}, Zhiyuan and {Kumari}, Nimisha and {Perna}, Michele and {Shivaei}, Irene and {Sun}, Fengwu},
        title = "{JADES: Insights into the low-mass end of the mass-metallicity-SFR relation at 3 < z < 10 from deep JWST/NIRSpec spectroscopy}",
      journal = {\aap},
         year = 2024,
        month = apr,
       volume = {684},
          eid = {A75},
        pages = {A75},
          doi = {10.1051/0004-6361/202346698},
archivePrefix = {arXiv},
       eprint = {2304.08516},
 primaryClass = {astro-ph.GA},
       adsurl = {https://ui.adsabs.harvard.edu/abs/2024A&A...684A..75C}
}

@ARTICLE{woodrum_2025,
       author = {{Woodrum}, Charity and {Shivaei}, Irene and {Witstok}, Joris and {Saxena}, Aayush and {Simmonds}, Charlotte and {Scholtz}, Jan and {Bhatawdekar}, Rachana and {Bunker}, Andrew J. and {Carniani}, St{\'e}fano and {Charlot}, Stephane and et al.},
        title = "{JADES: The Star Formation and Dust Attenuation Properties of Galaxies at 3<z<7}",
      journal = {arXiv e-prints},
         year = 2025,
        month = sep,
          eid = {arXiv:2510.00235},
        pages = {arXiv:2510.00235},
          doi = {10.48550/arXiv.2510.00235},
archivePrefix = {arXiv},
       eprint = {2510.00235},
 primaryClass = {astro-ph.GA},
       adsurl = {https://ui.adsabs.harvard.edu/abs/2025arXiv251000235W}
}

@ARTICLE{shapely_2022,
       author = {{Shapley}, Alice E. and {Sanders}, Ryan L. and {Salim}, Samir and {Reddy}, Naveen A. and {Kriek}, Mariska and {Mobasher}, Bahram and {Coil}, Alison L. and {Siana}, Brian and {Price}, Sedona H. and {Shivaei}, Irene and {Dunlop}, James S. and {McLure}, Ross J. and {Cullen}, Fergus},
        title = "{The MOSFIRE Deep Evolution Field Survey: Implications of the Lack of Evolution in the Dust Attenuation-Mass Relation to z   2}",
      journal = {\apj},
         year = 2022,
        month = feb,
       volume = {926},
       number = {2},
          eid = {145},
        pages = {145},
          doi = {10.3847/1538-4357/ac4742},
archivePrefix = {arXiv},
       eprint = {2109.14630},
 primaryClass = {astro-ph.GA},
       adsurl = {https://ui.adsabs.harvard.edu/abs/2022ApJ...926..145S}
}

@ARTICLE{maheson_2024,
       author = {{Maheson}, Gabriel and {Maiolino}, Roberto and {Curti}, Mirko and {Sanders}, Ryan and {Tacchella}, Sandro and {Sandles}, Lester},
        title = "{Unravelling the dust attenuation scaling relations and their evolution}",
      journal = {\mnras},
         year = 2024,
        month = jan,
       volume = {527},
       number = {3},
        pages = {8213-8233},
          doi = {10.1093/mnras/stad3685},
archivePrefix = {arXiv},
       eprint = {2306.00069},
 primaryClass = {astro-ph.GA},
       adsurl = {https://ui.adsabs.harvard.edu/abs/2024MNRAS.527.8213M}
}

@ARTICLE{cloudy_2025,
       author = {{Gunasekera}, C.~M. and {van Hoof}, P.~A.~M. and {Dehghanian}, M. and {Chakraborty}, P. and {Shaw}, G. and {Bianchi}, S. and {Chatzikos}, M. and {Tsujimoto}, M. and {Ferland}, G.~J.},
        title = "{The 2025 release of Cloudy}",
      journal = {\rmxaa},
         year = 2025,
        month = nov,
       volume = {61},
        pages = {120-133},
          doi = {10.22201/ia.01851101p.2025.61.03.01},
archivePrefix = {arXiv},
       eprint = {2508.01102},
 primaryClass = {astro-ph.GA},
       adsurl = {https://ui.adsabs.harvard.edu/abs/2025RMxAA..61..120G}
}

@ARTICLE{BPASS_1,
       author = {{Eldridge}, J.~J. and {Stanway}, E.~R. and {Xiao}, L. and {McClelland}, L.~A.~S. and {Taylor}, G. and {Ng}, M. and {Greis}, S.~M.~L. and {Bray}, J.~C.},
        title = "{Binary Population and Spectral Synthesis Version 2.1: Construction, Observational Verification, and New Results}",
      journal = {\pasa},
         year = 2017,
        month = nov,
       volume = {34},
          eid = {e058},
        pages = {e058},
          doi = {10.1017/pasa.2017.51},
archivePrefix = {arXiv},
       eprint = {1710.02154},
 primaryClass = {astro-ph.SR},
       adsurl = {https://ui.adsabs.harvard.edu/abs/2017PASA...34...58E}
}

@ARTICLE{Sanders_2024,
       author = {{Sanders}, Ryan L. and {Shapley}, Alice E. and {Topping}, Michael W. and {Reddy}, Naveen A. and {Brammer}, Gabriel B.},
        title = "{Direct T $_{e}$-based Metallicities of z = 2─9 Galaxies with JWST/NIRSpec: Empirical Metallicity Calibrations Applicable from Reionization to Cosmic Noon}",
      journal = {\apj},
         year = 2024,
        month = feb,
       volume = {962},
       number = {1},
          eid = {24},
        pages = {24},
          doi = {10.3847/1538-4357/ad15fc},
archivePrefix = {arXiv},
       eprint = {2303.08149},
 primaryClass = {astro-ph.GA},
       adsurl = {https://ui.adsabs.harvard.edu/abs/2024ApJ...962...24S}
}

@ARTICLE{degraaff_2024,
       author = {{de Graaff}, Anna and {Rix}, Hans-Walter and {Carniani}, Stefano and {Suess}, Katherine A. and {Charlot}, St{\'e}phane and {Curtis-Lake}, Emma and {Arribas}, Santiago and {Baker}, William M. and {Boyett}, Kristan and {Bunker}, Andrew J. and {Cameron}, Alex J. and {Chevallard}, Jacopo and {Curti}, Mirko and {Eisenstein}, Daniel J. and {Franx}, Marijn and {Hainline}, Kevin and {Hausen}, Ryan and {Ji}, Zhiyuan and {Johnson}, Benjamin D. and {Jones}, Gareth C. and {Maiolino}, Roberto and {Maseda}, Michael V. and {Nelson}, Erica and {Parlanti}, Eleonora and {Rawle}, Tim and {Robertson}, Brant and {Tacchella}, Sandro and {{\"U}bler}, Hannah and {Williams}, Christina C. and {Willmer}, Christopher N.~A. and {Willott}, Chris},
        title = "{Ionised gas kinematics and dynamical masses of z {\ensuremath{\gtrsim}} 6 galaxies from JADES/NIRSpec high-resolution spectroscopy}",
      journal = {\aap},
         year = 2024,
        month = apr,
       volume = {684},
          eid = {A87},
        pages = {A87},
          doi = {10.1051/0004-6361/202347755},
archivePrefix = {arXiv},
       eprint = {2308.09742},
 primaryClass = {astro-ph.GA},
       adsurl = {https://ui.adsabs.harvard.edu/abs/2024A&A...684A..87D}
}

@ARTICLE{umeda_2026,
       author = {{Umeda}, Hiroya and {Ouchi}, Masami and {Nakajima}, Kimihiko and {Watanabe}, Kuria and {Harikane}, Yuichi and {Isobe}, Yuki and {Nishigaki}, Moka and {Yoshiaki}, Ono and {Yajima}, Hidenobu and {Xu}, Yi},
        title = "{Gas-Phase Metallicity and Nitrogen Abundances in Low-Mass Galaxies Down to $M_\star\simeq10^{5.7}\,M_\odot$ at $z\simeq4.5$--$10.1$ from JWST Lensing Cluster Surveys}",
      journal = {arXiv e-prints},
         year = 2026,
        month = jul,
          eid = {arXiv:2607.15515},
        pages = {arXiv:2607.15515},
          doi = {10.48550/arXiv.2607.15515},
archivePrefix = {arXiv},
       eprint = {2607.15515},
 primaryClass = {astro-ph.GA},
       adsurl = {https://ui.adsabs.harvard.edu/abs/2026arXiv260715515U}
}

@ARTICLE{BPASS_2,
       author = {{Stanway}, E.~R. and {Eldridge}, J.~J.},
        title = "{Re-evaluating old stellar populations}",
      journal = {\mnras},
         year = 2018,
        month = sep,
       volume = {479},
       number = {1},
        pages = {75-93},
          doi = {10.1093/mnras/sty1353},
archivePrefix = {arXiv},
       eprint = {1805.08784},
 primaryClass = {astro-ph.GA},
       adsurl = {https://ui.adsabs.harvard.edu/abs/2018MNRAS.479...75S}
}

@ARTICLE{BPASS_3,
       author = {{Byrne}, C.~M. and {Stanway}, E.~R. and {Eldridge}, J.~J. and {McSwiney}, L. and {Townsend}, O.~T.},
        title = "{The dependence of theoretical synthetic spectra on {\ensuremath{\alpha}}-enhancement in young, binary stellar populations}",
      journal = {\mnras},
         year = 2022,
        month = jun,
       volume = {512},
       number = {4},
        pages = {5329-5338},
          doi = {10.1093/mnras/stac807},
archivePrefix = {arXiv},
       eprint = {2203.13275},
 primaryClass = {astro-ph.SR},
       adsurl = {https://ui.adsabs.harvard.edu/abs/2022MNRAS.512.5329B}
}

@ARTICLE{kroupa_2001,
       author = {{Kroupa}, Pavel},
        title = "{On the variation of the initial mass function}",
      journal = {\mnras},
         year = 2001,
        month = apr,
       volume = {322},
       number = {2},
        pages = {231-246},
          doi = {10.1046/j.1365-8711.2001.04022.x},
archivePrefix = {arXiv},
       eprint = {astro-ph/0009005},
 primaryClass = {astro-ph},
       adsurl = {https://ui.adsabs.harvard.edu/abs/2001MNRAS.322..231K}
}

@ARTICLE{solar_abundance,
       author = {{Asplund}, Martin and {Grevesse}, Nicolas and {Jacques Sauval}, A.},
        title = "{The solar chemical composition}",
      journal = {\nphysa},
         year = 2006,
        month = oct,
       volume = {777},
        pages = {1-4},
          doi = {10.1016/j.nuclphysa.2005.06.010},
archivePrefix = {arXiv},
       eprint = {astro-ph/0410214},
 primaryClass = {astro-ph},
       adsurl = {https://ui.adsabs.harvard.edu/abs/2006NuPhA.777....1A}
}

@software{hviding_unite,
  author       = {Hviding, Raphael Erik},
  title        = {unite: Unified liNe Integration Turbo Engine},
  month        = may,
  year         = 2026,
  publisher    = {Zenodo},
  version      = {v2.9.0},
  doi          = {10.5281/zenodo.20136821},
  url          = {https://doi.org/10.5281/zenodo.20136821},
}

@ARTICLE{VO87,
       author = {{Veilleux}, Sylvain and {Osterbrock}, Donald E.},
        title = "{Spectral Classification of Emission-Line Galaxies}",
      journal = {\apjs},
         year = 1987,
        month = feb,
       volume = {63},
        pages = {295},
          doi = {10.1086/191166},
       adsurl = {https://ui.adsabs.harvard.edu/abs/1987ApJS...63..295V}
}
\bibliographystyle{aasjournalv7}



\end{document}